\documentclass[12pt]{article}
\usepackage{amssymb,amsmath}
\usepackage{epsfig,psfrag}

\catcode`\@=11
\def \res{\mathop{\rm res}\nolimits}
\def \e  {\mathop{\rm e}\nolimits}
\newcommand\lr[1]{{\left({#1}\right)}}
\newcommand \widebar [1] {\overline{#1}}

\newcommand \ket [1] {|{#1}\rangle}
\newcommand \bra [1] {\langle {#1}|}
\newcommand\re[1]{(\ref{#1})}
\def \qqquad {\qquad\quad}

\newcommand{\pa}{\partial}

\newcommand{\be}{\begin{equation}}
\newcommand{\ee}{\end{equation}}
\newcommand{\bea}{\begin{eqnarray}}
\newcommand{\eaa}{\end{eqnarray}}
\newcommand{\nn}{\nonumber}

\renewcommand{\a}{\alpha}

\newcommand{\da}{{\dot\alpha}}
\newcommand{\db}{{\dot\beta}}
\newcommand{\bl}{{\tilde\lambda}}

\renewcommand{\b}{\beta}
\newcommand{\s}{\sigma}
\renewcommand{\r}{\rho}
\newcommand{\rt}{\tilde\rho}

\newcommand{\la}{\lambda}

\newcommand{\q}{\theta}

\newcommand{\ep}{\epsilon}

\newcommand{\cN}{{\cal N}}
\newcommand{\cA}{{\cal A}}

\newcommand{\p}[1]{(\ref{#1})}
\newcommand{\bt}[1]{{\bar t}}

\newcommand \vev [1] {\langle{#1}\rangle}
\newcommand \ran [1] {|{#1}\rangle}
\newcommand \lan [1] {\langle{#1}|}

\def\numberbysection{\@addtoreset{equation}{section}
                     \def\theequation{\thesection.\arabic{equation}}}
\numberbysection

\begin{document}



\thispagestyle{empty}
\null\vskip-12pt \hfill IPhT-T09/102~~~ \\
\null\vskip-12pt \hfill  LAPTH-1341/09 \\
\vskip2.2truecm
\begin{center}
\vskip 0.2truecm {\Large\bf
 Twistor transform of all\\[3mm]  tree amplitudes in   $\cN=4$ SYM theory}

\vskip 1truecm
{\bf  G.P. Korchemsky$^{*}$\footnote{On leave from Laboratoire de Physique Th\'eorique, Universit\'e de Paris XI, 91405 Orsay Cedex, France}
 and E. Sokatchev$^{**}$ \\
}

\vskip 0.4truecm
$^{*}$ {\it Institut de Physique Th\'eorique\,\footnote{Unit\'e de Recherche Associ\'ee au CNRS URA 2306},
CEA Saclay, \\
91191 Gif-sur-Yvette Cedex, France\\
\vskip .2truecm $^{**}$ {\it LAPTH\,\footnote[3]{Laboratoire d'Annecy-le-Vieux de Physique Th\'{e}orique, UMR 5108},   Universit\'{e} de Savoie, CNRS, \\
B.P. 110,  F-74941 Annecy-le-Vieux, France
                       }
  } \\
\end{center}

\vskip 1truecm 
\centerline{\bf Abstract} 

\medskip

 \noindent
We perform the twistor (half-Fourier) transform of all tree $n-$particle superamplitudes in $\cN=4$ SYM and show that it has a transparent geometric interpretation. We find that
the N${}^k$MHV amplitude is supported on a set of $2k+1$ intersecting lines in twistor space and demonstrate that the corresponding line moduli form a lightlike $(2k+1)-$gon in moduli space. This polygon is triangulated into two kinds of lightlike triangles lying in different planes. We formulate simple graphical rules for constructing the triangulated polygons, from which the analytic expressions of the N${}^k$MHV amplitudes follow directly, both in twistor and in momentum space. We also discuss the ordinary and dual conformal properties and the cancellation of spurious singularities in twistor space.

\newpage

\thispagestyle{empty}

{\small \tableofcontents}

\newpage
\setcounter{page}{1}\setcounter{footnote}{0}

\section{Introduction}

It has been known since the work of Parke and Taylor \cite{Parke:1986gb}
that the tree MHV gluon scattering amplitudes are remarkably simple~\cite{Berends:1987me,Mangano:1990by}.  These amplitudes have a
universal form in all gauge theories but, as was shown by Nair \cite{Nair:1988bq}, an additional simplification occurs in the maximally supersymmetric $\mathcal{N}=4$ Yang-Mills theory.
In an attempt to explain these properties, Witten  \cite{Witten:2003nn} has proposed to perform a half-Fourier transform of the scattering
amplitudes from momentum to twistor space. Twistor methods in field theory have attracted a lot of attention since the early work of Penrose \cite{Penrose:1967wn}. What makes them so appealing in the context of scattering amplitudes is the introduction of geometric notions. Indeed,
performing the twistor transform of the simplest, tree MHV gluon amplitude, Witten
has observed that this amplitude has support on a single line in twistor space.
He conjectured that all amplitudes should be supported on holomorphic curves of higher degree and gave some evidence based on the six-gluon next-to-MHV (NMHV) tree amplitude. Later on Bern et al \cite{Bern:2004ky}, \cite{Bern:2004bt} (see also \cite{Britto:2004tx}) extended this observation by showing that all $n-$gluon NMHV tree amplitudes are localized on three intersecting lines in twistor space.

The study of scattering amplitude initiated in  \cite{Witten:2003nn} led to the formulation of
two efficient methods for computing tree amplitudes. The CSW approach \cite{Cachazo:2004kj} (for supersymmetric extensions see \cite{Georgiou:2004by,Brandhuber:2004yw,Bianchi:2008pu}) uses Feynman diagram rules based on MHV vertices and it allows one to express an arbitrary tree amplitude as a sum of products of MHV amplitudes, each depending on the so-called `reference spinors'. This dependence drops out in the sum of all terms, but its presence makes the geometric properties of non-MHV tree amplitudes in twistor space less transparent \cite{Gukov:2004ei,Bena:2004ry}.\footnote{Attempts have also been made to elucidate the twistor space structure of one-loop amplitudes \cite{Cachazo:2004zb,Cachazo:2004by,Bena:2004xu}.}  In the second, BCFW approach \cite{Britto:2005fq}, one exploits the analytic properties of the tree amplitudes to derive recursion relations. They allow one to construct all tree amplitudes starting from the simplest, three-particle amplitude, without need for any auxiliary parameters.

Recently, a lot of progress in the understanding of the tree amplitudes has been made by  fully exploiting the supersymmetry in the maximally supersymmetric ${\cal N}=4$ super-Yang-Mills theory \cite{Drummond:2008vq,ArkaniHamed:2008gz}. ${\cal N}=4$ supersymmetry helps organize the gluon and other parton amplitudes into very compact expressions
for superamplitudes, with a much clearer structure than the gluon amplitudes themselves. 
This led to the discovery of a new dynamical symmetry of the perturbative superamplitudes, the so-called `dual superconformal' symmetry \cite{Drummond:2008vq}.\footnote{The strong coupling version of this symmetry was subsequently found in string theory \cite{Berkovits:2008ic,Beisert:2008iq}.} The combination of this dynamical symmetry with the ordinary superconformal symmetry puts very strong restrictions on  the tree amplitudes \cite{Bargheer:2009qu,Korchemsky:2009hm}. The reformulation of the superamplitudes in dual superspace, combined with the supersymmetric 
generalization of the BCFW recursive procedure \cite{Bianchi:2008pu,Brandhuber:2008pf,ArkaniHamed:2008gz,Elvang:2008na}, extended the proof of dual conformal symmetry to all tree amplitudes \cite{Brandhuber:2008pf}, and led to the full solution of these recursion relation \cite{Drummond:2008cr}. The resulting tree $\mathcal{N}=4$ superamplitudes are given by sums of different classes of dual superconformal  invariants. These invariants have a very nontrivial dependence on the particle (super)momenta and, with Witten's proposal in mind, we may expect that they
should become much simpler after a twistor transform.
One of the main goals of the present paper is to give a geometric interpretation
to the superconformal invariants in twistor space.

Another approach to the twistor
transform of the tree superamplitudes, adopted by Mason and Skinner in Ref.~\cite{Mason:2009sa} and by Arkani-Hamed et al in Ref.~\cite{ArkaniHamed:2009si}, is to formulate and try to solve the BCFW  recursion relations directly in twistor space. In this way,  one does not actually carry out  the Fourier transform of the amplitudes from momentum to twistor space, but instead one constructs them in twistor space.\footnote{With the exception of the simplest, three-point amplitudes, needed at the staring level of the BCFW recursion.}
One of the main motivations of these works was to make the conformal symmetry of the tree amplitudes manifest. Indeed, as shown in \cite{Witten:2003nn}, the conformal group acts non-locally in momentum space, with generators realized as second-order differential operators. Consequently, the direct proof of (super)conformal invariance in momentum space requires some effort.\footnote{For a recent proof in the case of the NMHV superamplitudes see \cite{Korchemsky:2009hm}.} In twistor space with split signature $(++--)$ the conformal group is $SL(4,\mathbb{R})$ and it acts linearly on the twistor variables. The latter form either the fundamental, or the anti-fundamental representation of $SL(4,\mathbb{R})$, denoted by $Z^a$ and $W_a$ (with $a=1,\ldots,4$), respectively. Conformal symmetry implies that the twistor transform of the tree amplitudes
should be functions of $SL(4,\mathbb{R})$ invariants. There exist two types of such invariants, regular ones of the form  $Z\cdot W \equiv Z^a W_a$, and singular ones of the form
$\delta^{(4)}(Z_1 - s Z_2)$ (with $s$ being a real singlet parameter). The `ambitwistor' approach  of Ref.~\cite{ArkaniHamed:2009si} (see also the earlier work by Hodges \cite{Hodges:2005bf}) employs the former invariants, while the chiral approach of Ref.~\cite{Mason:2009sa} uses only fundamental $Z-$twistors  and hence conformal invariants of the second type. Although the BCFW relations could in principle give all non-MHV amplitudes with an arbitrary number of particles, actually solving  their twistor versions, in full generality, is not an obvious task in both approaches. In \cite{Mason:2009sa} this has been done for the NMHV amplitude, and a couple of examples of NNMHV amplitudes with 7 and 8 particles are given. In Ref.~\cite{ArkaniHamed:2009si} the case of a 6-particle NMHV amplitude is treated explicitly, and graphical rules for the construction of more complicated amplitudes are formulated.

In this paper, we start from the explicit expressions for all tree superamplitudes in ${\cal N}=4$ SYM found in \cite{Drummond:2008cr}, and pursue Witten's original proposal to carry out their twistor transform. The lightlike momenta of the $n$ massless particles are represented in  terms of commuting two-component spinor variables, $p^{\a\da}_i = \la^\a_i \bl^{\da}_i$ with $\a=1,2$, $\da=1,2$ and $i =1, \ldots, n$. In Minkowski
space-time, the spinors $ \la^\a_i$ and $\bl^{\da}_i$ are complex conjugate to each other. The twistor transform is most easily done if $\la$ and $\bl$ are treated as real and independent variables. This can be achieved by working in a space-time with split signature  $(++--)$. Then, the (bosonic) twistor transform  of the
amplitude is defined as a half-Fourier transform over the $\bl$'s, with Fourier conjugate variables $\mu_{\da}$. As was
shown in  \cite{Witten:2003nn}, the result  for the MHV amplitude is (omitting a purely holomorphic factor depending only on the $\la$'s)
\begin{equation}\label{-01}
    \int d^4X \prod_1^{n}  \delta^{(2)}(\mu_{i \, \da} + \la_{i}^\a X_{\a\da})\,,
\end{equation}
where $X_{\a\da}$ is a real four-vector parameter with the scaling dimension of a coordinate in position space. The geometric interpretation of this result is very simple:
The twistor transform of the MHV amplitude is supported on configurations of $n$ points that all lie on a twistor line defined by the equation
\begin{equation}\label{liequ}
    \mu_{\da} + \la^\a X_{\a\da} = 0\,.
\end{equation}
Here $X_{\a\da}$ is the line parameter, and in \p{-01} we see the integral over the moduli space of such lines. The same approach works in the case of  Nair's MHV superamplitude, adding a fermionic analog of the line equation \p{liequ} with a new chiral Grassmann modulus $\Theta^\a_A$ (with $A=1,\ldots,4$ being an $SU(4)$ R-symmetry index).

After reviewing the MHV case in Sect.~\ref{mhvsupa}, we do the twistor transform of the NMHV superamplitudes in Sect.~\ref{nmhvsup}. A very compact form of the NMHV superamplitudes in dual superspace was found in \cite{Drummond:2008vq} (and later on rederived in \cite{Drummond:2008bq} using the generalized unitarity method in superspace, and in \cite{Drummond:2008cr} as a solution to the BCFW recursion relations). Despite the apparently complicated $\bl$ dependence of these superamplitudes, the twistor transform is easy to perform by introducing new moduli. In Sect.~\ref{nmhvsup} we show that the NMHV superamplitude is supported on three intersecting lines in twistor superspace, characterized by the moduli $X_1$, $X_2$ and $X_3$ (and their fermionic counterparts $\Theta_{1,2,3}$). The new moduli are not completely independent. They satisfy the constraint that their differences $X_{i,i+1}\equiv X_i-X_{i+1}$ (with $i+3\equiv i$)
are lightlike vectors,
\begin{equation}\label{xlili}
    X^2_{12}=X^2_{23}=X^2_{31}=0\,,
\end{equation}
in addition to the obvious relation $X_{12}+X_{23}+X_{31}=0$.  In a space-time with split signature $(++--)$,
these constraints can be solved as follows,
\begin{equation}\label{bysett}
    X^{\a\da}_{12} = \r^\a\rt^{\da}\,, \qquad X^{\a\da}_{23} = \sigma^\a\rt^{\da}\,, \qquad X^{\a\da}_{31} = \r^\a_0\rt^{\da}\,,
\end{equation}
where $\rt$ is a common antichiral spinor and $\r$, $\s$, $\r_0$ are three chiral spinors satisfying the relation $\r+\sigma+\r_0=0$. One of these spinors, $\r_0$, is identified with the twistor variable of one of the incoming particles, e.g., $\r_0 \equiv \la_n$. So, the new independent moduli are $\r$ and $\rt$ (and their fermionic counterpart $\xi_A$), in addition to $X \equiv X_3$ and $\Theta \equiv \Theta_3$ inherited from the MHV amplitude.

The explicit form of the twistor transform of the NMHV superamplitude is similar to the MHV one \p{-01}. The main differences are that, firstly, it involves
additional integrals over the new moduli (see Eq.~\re{sugg} below) and, secondly, the product of delta functions in \p{-01} breaks up into three sets, each involving the  moduli $X_1$, $X_2$ and $X_3$, respectively. Thus, the single line equation \p{liequ} for the support of the twistor transform is replaced by three such equations. This makes the geometric structure of three intersecting lines in twistor space perfectly transparent.\footnote{As mentioned earlier, this three-line structure of the NMHV gluon amplitudes was already established by Bern et al in \cite{Bern:2004ky,Bern:2004bt}. However, they had to use the collinearity and coplanarity differential operators from \cite{Witten:2003nn} to test the geometric properties of the amplitude in {\it momentum space}.}

Another property of the twistor transform of the MHV and NMHV superamplitudes studied in detail in Sects.~\ref{mhvsupa} and \ref{nmhvsup} is their ordinary and dual conformal invariance. To show ordinary (super)conformal invariance, one has to transform the moduli $X$ and $\Theta$ (and the related spinor moduli from \p{bysett}). It turns out that they transform exactly as the coordinates of some fictitious configuration superspace (not to be confused with the configuration space of the particles). But ordinary conformal symmetry is not powerful enough to determine uniquely the twistor transform of the superamplitude. We argue that adding to it dual conformal symmetry, one can fix its form up to constants (see also \cite{Korchemsky:2009hm} for a similar argument without twistors).

One must however bare in mind that ordinary conformal symmetry is not an {\it exact} symmetry of the  tree amplitudes. In fact, conformal symmetry is broken by the physical
singularities of the amplitude, corresponding to the vanishing invariant masses of several
color-adjacent particles. This effect is hard to control in momentum space, but it can be
easily identified in twistor space. We confirm the observation of Refs.~\cite{Mason:2009sa,ArkaniHamed:2009si} that the global conformal symmetry of the MHV superamplitude is broken by sign factors of the form ${\rm sgn}(\la_1^\a \la_{2\a})$. In addition, we show that for NMHV amplitudes other sign factors of the form ${\rm sgn}(s_{a...b})$ (with $s_{a...b}=(p_a + \ldots + p_{b-1})^2$) imply the breakdown of conformal symmetry even at the infinitesimal level. To avoid such effects,
we restore the infinitesimal conformal symmetry of the tree amplitudes by multiplying them by
the appropriate sign factors, and then study their properties in twistor space. Notice that
the inverse Fourier transform of the resulting expression
back to momentum space yields a function coinciding with the true amplitude only in a restricted kinematic domain, where the Mandelstam invariants $s_{a...b}$ are all positive. The original momentum space amplitude can then be obtained by analytic continuation.
Further, we examine the cancellation of the so-called `spurious' singularities of the NMHV superamplitude. This issue was also addressed in \cite{Korchemsky:2009hm}, but the twistor transform makes the analysis considerably simpler.\footnote{In a recent paper \cite{Hodges:2009hk} the absence of spurious singularities in the NMHV split-helicity amplitude was shown in the so-called `momentum-twistor' approach. We remark that this approach does not employ a Fourier transform to twistor space, but treats the spinors $\la,\bl$, related to the particle momenta, as twistor variables.  }

One of the main results of the present paper is the clarification of the geometric properties of the twistor transform of all the non-MHV amplitudes, and of the associated structure in the moduli space. In Sect.~\ref{gitt}, after a general discussion of intersecting lines in twistor space,
we propose a simple diagrammatic representation  of the twistor transform of the NMHV amplitude. It consists of a set of three intersecting twistor lines, in which a moduli space lightlike triangle is inscribed.
As we then show in Sect.~\ref{allnmhvsu}, this picture is very easy to generalize, providing a simple graphical procedure for generating all N${}^k$MHV amplitudes.
In close analogy with the NMHV case, we work out the twistor transform of the N${}^2$MHV amplitudes. We find that the amplitude is supported on five intersecting lines lying in three different planes. Three of these lines have a common intersection point. The associated line parameters $X_1, \ldots, X_5$ form a lightlike `pentagon' in moduli space. Actually, this `pentagon' is not planar but is made of three lightlike triangles, two of the type considered above and one of a different kind, with a common chiral spinor on all sides. The three triangles lie in different planes, like the twistor lines themselves. This pentagon configuration appears in two different orientations, as well as in two degenerate versions with one twistor line absent.

The experience with the case N${}^2$MHV allows us to immediately generalize to all N${}^k$MHV amplitudes. We formulate simple rules for constructing sets of $(2k+1)$ intersecting lines in twistor space and the associated inscribed lightlike $(2k+1)-$gons in moduli space. The latter are triangulated into the two types of triangles mentioned above, following a simple regular pattern. Once the relevant diagrams are drawn, it is straightforward to translate them into analytic expressions for the twistor transform. It is then equally easy to work out the inverse twistor transform, leading to the expressions in dual superspace found in \cite{Drummond:2008cr}.


\section{MHV superamplitude}\label{mhvsupa}

In this section we first review Witten's twistor transform of the MHV tree superamplitude. We then study its superconformal properties, paying attention to the violation of conformal invariance by global conformal transformations, and, finally, discuss to what extent the MHV superamplitude is fixed by the combination of ordinary and dual superconformal symmetry.

{ Throughout the paper we use the on-shell superspace description of scattering amplitudes
in $\mathcal{N}=4$ SYM theory. In this approach, all on-shell states (gluons, gluinos, scalars) are described by a single superstate and all $n-$particle
scattering amplitudes can be combined into a single object, the on-shell superamplitude
\be\label{An}
\mathcal{A}_n=\mathcal{A}(\la_1,\bl_1,\eta_1;\ldots;\la_n,\bl_n,\eta_n)\,.
\ee
Each scattered superstate is characterized by a pair of commuting two-component spinors
$\la_i$ and $\bl_i$,  defining the lightlike momenta
\be
p^{\a\da}_i = \la^\a_i \bl^{\da}_i\,,
\ee
and by a Grassmann variable $\eta_i^A$ with an $SU(4)$ index $A=1,\ldots,4$. The variables $\la,\bl,\eta$ carry helicities $-1/2,1/2,1/2$, respectively.  The expansion of $\mathcal{A}_n$ in powers of the $\eta$'s
generates scattering amplitudes for the various types of particles and it has the following
form,
\begin{align}\label{An-dec}
\mathcal{A}_n = \mathcal{A}_n^{\rm MHV} +  \mathcal{A}_n^{\rm NMHV}
+\ldots+\mathcal{A}_n^{\rm \widebar{MHV}}\,,
\end{align}
where $\mathcal{A}_n^{\rm N^kMHV}$ is a homogenous polynomial
in the $\eta$'s of degree $8+4k$, with $k=0,\ldots,n-4$.
}

The twistor transform we are going to study is a Fourier transform of \re{An} with respect to the  $\bl$'s  and $\eta$'s,
\begin{align}\label{T}
T[\mathcal{A}_n](\{\la,\mu,\psi\}) =  \int  \prod_1^n \frac{d^2 \bl_i}{(2\pi)^2}\, d^4\eta_i\,  \e^{i\sum_1^n (\mu_{i\da} \bl_i^{\da} + \psi_{i A}  \eta_i^A )}\mathcal{A}_n(\{\la,\bl,\eta\})\,,
\end{align}
so that  $\mu_{i \,\da}$ and $\psi_{i\,A}$ are Fourier conjugated to $\bl_i^{\da}$ and $\eta_{i}^A$, respectively. We recall that in split  $(++--)$ signature  the $\bl$'s are real spinors, independent from the $\la$'s. The Lorentz group in this case is $SO(2,2) \sim SL(2,\mathbb{R}) \times SL(2,\mathbb{R})$, so $\la^\a$ and $\bl^{\da}$ transform under the first and second $SL(2,\mathbb{R})$, respectively.  It is standard to use the bra-ket notation for (anti)chiral spinors,
\begin{align}
 \la^\a_i \equiv \bra{i}\,, \qquad
    \bl^\da_i \equiv |i]\,,\qquad p^{\a\da}_i \equiv |i]\bra{i}\,.
\end{align}
The two kinds of spinor indices are raised and lowered with the help of Levi-Civita tensors
\begin{align}\label{notat}
 \la_{i\, \a} = \ep_{\a\b}\la^\b_i \equiv \ket{i}\,, \qquad
\bl_{i\, \da} = \ep_{\da\db}\la^{\db}_i \equiv [i| \,,
 \qquad
 \vev{i\, j} =   \la^\a_i \la_{j\, \a}
 \,, \qquad [i \,j] = \bl_{i\, \da} \bl_{j}^{\da}
 \,.
\end{align}

\subsection{Twistor transform}\label{mmhhvv}

{The MHV  superamplitude  in \re{An-dec} is a homogenous polynomial of degree 8. At tree level, it has the following form \cite{Nair:1988bq}}
\begin{equation}\label{mhvsu}
    \cA_n^{\rm MHV} = \frac{i(2\pi)^4}{\prod_1^n \vev{i\, i+1}}\ \delta^{(4)}(\sum_1^n \ran{i} [i|)\ \delta^{(8)}(\sum_1^n \ran{i}\eta_i)\,,
\end{equation}
with the periodicity condition $ n+1\equiv 1$.
{To do the twistor transform \re{T},} we follow \cite{Witten:2003nn} and replace the two delta functions in \re{mhvsu} by their Fourier integrals,
\begin{eqnarray}
  (2\pi)^4\delta^{(4)}(\sum_1^n \ran{i} [i|) &=& \int d^4X\, \e^{i \sum_1^n \lan{i} X |i]}\,, \nn\\
  \delta^{(8)}(\sum_1^n \ran{i}\eta_i) &=& \int d^8\Theta\, \e^{i \sum_1^n \vev{i\, \Theta} \eta_i}\, , \label{foexp}
\end{eqnarray}
thus introducing {two} new integration variables, a real four-vector $X^{\a\da}$ and a chiral anticommuting spinor $\Theta^\a_A$.
{After this the Fourier integrals in \re{T} are immediately done and} the twistor transform of the MHV superamplitude \p{mhvsu} is given by
\begin{equation}\label{01}
    T\left[\cA_n^{\rm MHV}\right] = \frac{i}{\prod_1^n \vev{i\, i+1}}\ \int d^4X d^8\Theta\   \prod_1^{n}  \delta^{(2)}(\mu_i + \lan{i} X)\  \delta^{(4)}(\psi_i + \vev{i\, \Theta})\,.
\end{equation}
{Regarded as a function in the supertwistor space with coordinates $(\lambda,\mu,\psi)$, it is localized on the line defined by the equations  (for a detailed discussion of twistor lines see Sect.~\ref{ltsp})}
\begin{equation}\label{twleqs}
    \mu_i + \lan{i} X= 0\,, \qquad \psi_{i\, A} + \vev{i\, \Theta_A} = 0\,,
\end{equation}
We remark that the parameters $X$ and $\Theta$ have the same scaling dimensions as the coordinates of {\it configuration superspace}. Indeed, it follows from \p{foexp} that, in mass units, the dimension of $X$ is $(-1)$ and {that } of $\Theta$ is $(-1/2)$. However, $X$ and $\Theta$ have nothing to do with the particle coordinates
and we must think of them as defining the {\it moduli} of the line in twistor space.
In what follows we shall call the chiral superspace with coordinates $X^{\a\da}$ and $\Theta^{\a}_A$  the {\it moduli superspace}. Below we will see that this moduli space acquires an interesting geometric structure for non-MHV superamplitudes.

The integrals in \p{01} can be easily computed using two of the delta functions
\cite{Mason:2009sa,ArkaniHamed:2009si}, e.g., those for $i=1,2$. They fix $X$ and $\Theta$ as follows,
\begin{equation}\label{exprxth}
    X^{\a\da} = \frac{\la^\a_1\mu^{\da}_2 - \la^\a_2\mu^{\da}_1}{\vev{12}}\,, \qquad \Theta^{\a}_A = \frac{\la^\a_1\psi_{2\, A} - \la^\a_2\psi_{1\, A}}{\vev{12}}\,,
\end{equation}
after which \p{01} becomes
\begin{equation}\label{01express}
    T\left[\cA_n^{\rm MHV}\right] = i  \frac{\vev{12}^2}{\prod_1^n \vev{i\, i+1}}\  \prod_3^{n}  \delta^{(2)}\bigg(\mu_i + \frac{\vev{i1}}{\vev{12}} \mu_2 +  \frac{\vev{2i}}{\vev{12}} \mu_1\bigg)\  \delta^{(4)}\bigg(\psi_i + \frac{\vev{i1}}{\vev{12}} \psi_2 +  \frac{\vev{2i}}{\vev{12}} \psi_1\bigg)\ .
\end{equation}
For our purposes it is however preferable to keep the integrals over the moduli undone.
As we shall see, the twistor transform of the non-MHV amplitudes involves new bosonic and fermionic moduli. Although they can be integrated out with the help of the delta functions, the resulting expressions loose most of their clarity.

\subsection{Superconformal properties}

Let us now examine the transformation properties of the MHV superamplitude under the ordinary $\cN=4$ conformal supersymmetry transformations. After the twistor transform \re{T}, they are generated by {\it linear} differential
operators acting on the twistor space coordinates:
\begin{eqnarray}
  && q^{A\, \a} = \la^\a \frac{\pa}{\pa\psi_A}\,, \qqquad \bar q_A^{\da} = \psi_A \frac{\pa}{\pa\mu_{\da}}\,, \qqquad  p^{\a\da} = \la^\a \frac{\pa}{\pa\mu_{\da}}\,, \nn \\
  &&  s_{A\, \a} = \psi_A \frac{\pa}{\pa\la^\a}\,, \qqquad \bar s^A_{\da} = \mu_{\da} \frac{\pa}{\pa\psi_A}\,, \qqquad  k_{\a\da} = \mu_{\da} \frac{\pa}{\pa\la^\a}\,, \label{ordcsusygen}
\end{eqnarray}
which form, together with the $SU(4)$ rotations, the Lorentz group $SO(2,2)$ and dilatations, the superconformal algebra $SL(4|4)$.

In addition, the twistor variables are ascribed a `helicity' weight under rescaling with a real parameter.\footnote{In a space with Minkowski metric. where $\bl = \la^*$, this rescaling is a phase factor.} Conventionally, the helicity weights are $\la \ \to \ a^{-1/2}\la$, $\bl \ \to \ a^{1/2}\bl$ and $\eta \ \to \ a^{1/2}\eta$, and, consequently, $\mu \ \to \ a^{-1/2}\mu$ and $\psi \ \to \ a^{-1/2}\psi$.  It is this helicity scaling which turns the twistor space into a projective space. Indeed, the superamplitude \p{mhvsu}  is a homogeneous function of $(\la_i, \bl_i,\eta_i)$ with unit helicity  weight at each point $i=1,\ldots,n$.
As a result,
its twistor transform \p{01} is a homogeneous function of $(\la_i, \mu_i,\psi_i)$ of vanishing weight.

{The extension of the generators \p{ordcsusygen} to the space of the moduli  $X$ and $\Theta$ can be obtained by requiring that the arguments of the delta functions in \p{01},
\begin{equation}\label{modtwiva}
    \hat\mu_i = \mu_i + \lan{i} X\,,\qqquad \hat\psi_i = \psi_i + \vev{i\, \Theta}\,,
\end{equation}
remain invariant or transform into each other. Thus, the induced action of translations ($p$) and Poincar\'e supersymmetry ($q$ and $\bar q$) in moduli space is as follows,
\begin{equation}\label{sus}
    \begin{array}{lllll}
       p^{\b\db} \mu_{i\, \da} = \delta^{\db}_{\da}\la^\a_{i} & \rightarrow & p^{\b\db} X_{\a\da} = -   \delta^{\b}_{\a}\delta^{\db}_{\da}  & \rightarrow & p^{\b\db} \hat\mu_{i\, \da} =0 
       \\[2.5mm]
        q^{B\, \b}\psi_{i\, A} = \la^\b_{i}\delta^B_A & \rightarrow &  q^{B\, \b} \Theta_{\a\,A} = -\delta^{\b}_{\a}\delta^B_A  & \rightarrow &  q^{B\, \b} \hat\psi_{i\, A} =0 \\[2mm]
        {\bar q}^{\db}_B \mu_{i\, \da} = \delta^{\db}_{\da}\psi_{i\, B}  & \rightarrow & {\bar q}^{\db}_B X_{\a\da} = \delta^{\db}_{\da}\Theta_{\a\,B}  & \rightarrow & {\bar q}^{\db}_B \hat\mu_{i\, \da} =\delta^{\db}_{\da}\hat\psi_{i\, B}
     \end{array}
\end{equation}
(all other variations vanish). These are the standard transformations of the coordinates of points in {\it chiral position superspace}. It is then clear that \re{01} is invariant under
the transformations \re{sus} (the $\bar q-$transformation of $\hat\mu_i$ vanishes due to $\delta^{(4)}(\hat\psi_i)$).

{Similarly, a conformal $k-$transformation in twistor space induces the standard conformal transformation of the moduli $X$ and $\Theta$, as if they were coordinates in position superspace,
\begin{equation}\label{cotr}
    k_{\b\db} \la^\a_i = \delta^{\b}_{\a}\mu_{i\, \db}  \ \quad \rightarrow\quad \ k_{\b\db} X^{\a\da} =  X_\b^{\da} X^\a_{\db}
 \,,\qquad    k_{\b\db} \Theta_{\a\,A} =  \Theta_{\b\,A}X_{\a\db}\,,
\end{equation}
so that
\begin{equation}\label{trhatmps}
    k_{\b\db} \hat\mu_{i\, \da} =  X_{\b\da} \hat\mu_{i\, \db}\,, \qquad k_{\b\db} \hat\psi_{i\, A} = \Theta_{A\, \b} \hat\mu_{i\, \db}\,.
\end{equation}
Finally, special conformal supersymmetry ($s$ and $\bar s$) yields
\begin{equation}\label{actonhat}
    \quad s_{B\, \b}  \hat\mu_{i\, \da} = X_{\b\da} \hat\psi_{i\, B}\,, \qquad s_{B\, \b}   \hat\psi_{i\, A} = \Theta_{A\, \b} \hat\psi_{i\, B}\,, \qquad \bar s^B_{\db} \hat\psi_{i\, A} = \delta^B_A \hat\mu_{i\, \db}\,,
\end{equation}
together with the standard transformations of $X$ and $\Theta$.

Let us now examine how \re{01} transforms under the conformal $k-$transformations.
Applying \re{cotr}, we find that the integration measure $\int d^4 X d^8\Theta$ is invariant. Further, from \p{trhatmps} it follows that the variations of the
fermionic delta functions $\delta^{(4)}(\hat\psi_i)$ are suppressed by the bosonic deltas $\delta^{(2)}(\hat\mu_i)$. It remains to consider the infinitesimal transformations of the bosonic factors in \re{01},
\begin{align} \notag
& \delta_k  \delta^{(2)}(\hat\mu_i) \equiv (b \cdot k) \delta^{(2)}(\hat\mu_i) =-2(b\cdot X)  \delta^{(2)}(\hat\mu_i)\,,
\\[2mm] \label{matchfact}
&  \delta_k \vev{ij} = [\mu_i|b\ran{j} -  [\mu_j|b\ran{i}= -2(b\cdot X) \vev{ij}\,,
\end{align}
where in the second relation we have used $\delta^{(2)}(\hat\mu_i)$ to replace $\mu_i$ by $-\lan{i}X$. We see that under {\it infinitesimal} conformal transformations the weight factors cancel in the ratios $\delta^{(2)}(\hat\mu_i)/\vev{i\,i+1}$, so that the MHV superamplitude \re{01} is invariant. However, in the next subsection  we will show
that this ceases to be true  under {\it global} conformal transformations.}}

\subsection{Conformal inversion in twistor space and breaking of conformal invariance}\label{ciits}

Conformal inversion in twistor space is a discrete element of the group $SL(4,\mathbb{R})$ whose square is the identity. It swaps $\la$ and $\mu$, but leaves $\psi$  unchanged,
\begin{align}\label{confinv}
I[\la_\a] = \mu_{\da}\,, \qquad I[\mu_{\da}] = \la_\a\,,\qquad  I[\la^\a] = - \mu^{\da}\,, \qquad I[\mu^{\da}] = -\la^\a \,,\qquad I[\psi_A] = \psi_A\,,
\end{align}
where we took into account that $ I[\la^\a] = I[\ep^{\a\b}\la_\b] = \ep^{\db\da}\mu_{\db}=- \mu^{\da}$.
{As before,} the line parameters $X$ and $\Theta$ have the standard transformations of
coordinates in position superspace,
\begin{equation}\label{stasuco}
    I[X_{\a\da}] = X^{-1}_{\a\da}\,, \qquad I[\Theta_\a] = \Theta^\a (X^{-1})_{\a\da}\,.
\end{equation}
Combining inversion with the infinitesimal super-Poincar\'e transformations \p{sus}, we can obtain the rest of the superconformal algebra. For instance, consider the action of the special conformal supersymmetry generator $\bar s = IqI$ on $\psi$:
\begin{equation}\label{invpss}
   \psi_A \ \stackrel{I}{\longrightarrow} \ \psi_A  \ \stackrel{q^B_\a}{\longrightarrow} \ \la_\a \delta^B_A \ \stackrel{I}{\longrightarrow} \ \mu_{\da}  \delta^B_A\ ,
\end{equation}
exactly as in \p{ordcsusygen}.

Let us see how the superamplitude \p{01} transforms under inversion. First of all, the arguments of the bosonic delta functions are covariant,
\begin{equation}\label{argbode}
    I[\hat\mu_{i\, \da}] = I[\mu_{i\, \da} + \la^\a_i X_{\a\da}] = \la_{i\, \a} - \mu_{i}^{\db} X^{-1}_{\a\db} = \hat\mu_{i\, \db}(X^{-1})^{\db}{}_\a\,,
\end{equation}
so that each $\delta^{(2)}(\hat\mu_i)$ produces the same weight factor,
\begin{equation}\label{eadepro}
    I[\delta^{(2)}(\hat\mu_i)] = |X^2|\ \delta^{(2)}(\hat\mu_i)
\end{equation}
(notice the absolute value of $X^2$, due to the properties of the bosonic delta function).
Further, in the presence of $\delta^{(2)}(\hat\mu_i)$ the arguments of the fermionic delta functions remain invariant,
\begin{equation}\label{inthefe}
    I[\hat\psi_i] = I[\psi_i+\la^\a_i \Theta_\a] = \psi_i -\mu^{\da}_i \Theta^\a X^{-1}_{\da\a} = \psi_i + \la^\b_iX_\b{}^{\da}X^{-1}_{\da\a}\Theta^\a = \hat\psi_i\,.
\end{equation}
Next, each angular bracket in the denominator of \re{01} gives (again, in virtue of $\delta^{(2)}(\hat\mu_i)$)
\begin{equation}\label{eachinde}
    I[\vev{i\ i+1}] = I[\la_{i}^\a \la_{i+1\, \a}] = -\mu^{\da}_i \mu_{\da\, i+1} = X^2 \vev{i\ i+1}\,.
\end{equation}
Finally, the complete integrand in the amplitude \p{01} picks the factor $(\mbox{sgn} (X^2))^n$, while the measure $\int d^4X d^8\Theta$ is invariant under inversion. Thus,
the twistor transformed MHV amplitude \re{01} is {invariant under {\it global} conformal transformations only for an even number of particles}.

{Here we are facing the phenomenon of broken global conformal invariance already observed in Refs.~\cite{Mason:2009sa,ArkaniHamed:2009si}. For an odd number of particles $n$, it can be repaired by replacing one of the angular brackets in
the denominator of \re{mhvsu} and \re{01} by its modulus,
 e.g., $\vev{12}$  by $|\vev{12}|$. However, the resulting object will certainly not be the physical amplitude. The important point is that the amplitude \p{mhvsu} has physical singularities when two adjacent particles become collinear, i.e. $\la_i \sim \la_{i+1}$. Changing $\vev{12}$ into $|\vev{12}|$ in the denominator in \p{mhvsu} modifies the behavior of the amplitude in the collinear limit $\la_1\sim\la_2$. This is a common and basic problem of all twistor approaches. As we shall see in Sect.~\ref{nmhvsup}, the problem becomes even worse in the case of the NMHV superamplitude. There similar sign factors break conformal invariance even at the infinitesimal level!}

We can observe the same phenomenon using the form \p{01express} of the amplitude, where the line parameters have been integrated out and expressed in terms of the basis twistors with $i=1,2$. Applying the inversion rules \p{confinv} for $\la$ and $\mu$, it is easy to show that $X$ and $\Theta$ from \p{exprxth} transform exactly as in \p{stasuco}. In this case
\begin{equation}\label{x2expr}
    X^2 =   \frac{[\mu_1\mu_2]}{\vev{12}}\,,
\end{equation}
so the sign flip responsible for the breakdown of conformal invariance occurs when either  $[\mu_1\mu_2]$, or $\vev{12}$ changes sign.

As a direct illustration of this effect, consider the twistor transform of the simplest, three-point MHV amplitude,\footnote{Three-particle amplitudes do not exist in a spacetime with Minkowski metric, because three real lightlike momenta cannot sum up to zero (unless they are collinear). However, the split signature $(++--)$ allows such amplitudes.} a special case of \p{01express},
\begin{equation}\label{exerc}
T[\mathcal{A}_3^{\rm MHV}]=   i\  \frac{\delta^{(2)}(\mu_1 \vev{23} +\mu_2 \vev{31} +\mu_3 \vev{12} )}{\vev{12}\vev{23}\vev{31}}\ \delta^{(4)}(\psi_1 \vev{23} +\psi_2 \vev{31} +\psi_3 \vev{12} )\,.
\end{equation}
Performing inversion using the rules \p{confinv}, we obtain
\begin{equation}\label{exercinv}
I[T[\mathcal{A}_3^{\rm MHV}]]=  i\  \frac{\delta^{(2)}(\la_1 [\mu_2 \mu_3] +\la_2 [\mu_3 \mu_1] +\la_3 [\mu_1 \mu_2])}{[\mu_1 \mu_2][\mu_2 \mu_3][\mu_3 \mu_1]}\ \delta^{(4)}(\psi_1 [\mu_2 \mu_3] +\psi_2 [\mu_3 \mu_1] +\psi_3 [\mu_1 \mu_2])\,.
\end{equation}
 The bosonic delta functions in \p{exerc} and \p{exercinv} have support on the same surface (see Sect.~\ref{cctsp}).   This leads  to the relation
\begin{align}\label{comp2del}
T[\mathcal{A}_3^{\rm MHV}] \, \text{sgn} (\vev{12})=
I[T[\mathcal{A}_3^{\rm MHV}]]\,  \text{sgn}([\mu_1\mu_2])= I\left[T[\mathcal{A}_3^{\rm MHV}] \,\text{sgn} (\vev{12})\right]\,,
\end{align}
which allows us to establish the equivalence of \p{exerc} and \p{exercinv}, up to the sign factor discussed above. This observation is in accord with the results of Refs.~\cite{Mason:2009sa,ArkaniHamed:2009si}. There one can find manifestly conformally invariant twistor transforms of a three-point amplitude, which differs from the true amplitude by $\mbox{sgn}(\vev{12})$.

\subsection{Uniqueness of the MHV superamplitude}\label{unimhv}

In the previous section, we applied the twistor transform to demonstrate that
the MHV superamplitude is invariant under (infinitesimal) superconformal $SL(4|4)$ transformations.
It is well known that all tree superamplitudes in $\mathcal{N}=4$ SYM have another, dynamical {\it  dual superconformal symmetry} \cite{Drummond:2008vq,Brandhuber:2008pf,Drummond:2008cr}. In Ref.~\cite{Korchemsky:2009hm} we argued that
the combination of the two symmetries, conventional and dual superconformal, fixes the form of the tree MHV superamplitude, up to a constant factor. In this section, we repeat the same analysis, this time in twistor space.

{The ordinary and dual symmetries are  difficult to study simultaneously because there exists no formulation of the amplitudes in which both of them have a simple, local realization. We can use dual superspace \cite{Drummond:2008vq} to make the dual symmetry manifest, but there the conventional conformal symmetry acts non-locally, with second-order generators \cite{Witten:2003nn}. We can instead use twistor space, where the conventional conformal symmetry is local, but then dual conformal symmetry becomes non-local. This problem is related to the fact that the closure of the two symmetries is infinite dimensional, having a Yangian structure \cite{Drummond:2009fd}.

The strategy we adopt here is to first examine the consequences of ordinary superconformal symmetry in twistor space. We will see that it leaves considerable freedom in the form of the MHV superamplitude. Then we transform the amplitude back to momentum space, change variables to dual coordinates, and impose dual conformal symmetry. The last step restricts the freedom down to a constant factor. Inversely, we can start with the dual description of the amplitude, make modifications to it compatible with dual conformal symmetry, then transform to twistor space and show that ordinary superconformal symmetry forbids these modifications.}

In twistor space, the conformal group acts linearly on the twistors $Z^a=(\la^\a,\mu_{\da})$, transforming them according to the fundamental representation of  $SL(4,\mathbb{R})$.\footnote{In \p{sus} and \p{cotr} we have shown the action of the off-diagonal block of an $SL(4,\mathbb{R})$ matrix corresponding to translations and conformal boosts. The diagonal blocks correspond to the Lorentz group $SO(2,2) \sim SL(2,\mathbb{R}) \times SL(2,\mathbb{R})$ and to dilatations.}
This suggests that the natural invariants of the conformal group have the form $Z^a W_a$ with $W_a$ belonging to anti-fundamental representation of $SL(4,\mathbb{R})$.
However, our description of the amplitudes is chiral (holomorphic) and we do not make use of $W_a$.\footnote{Such objects do appear in the so-called ambitwistor approach (see \cite{Hodges:2005bf}, \cite{ArkaniHamed:2009si}).}
Another possibility could be to construct an invariant  out of four fundamental spinors $Z^a$, in the form of a determinant,
\begin{equation}\label{twiin}
    \ep_{abcd} Z^a_1 Z^b_2 Z^c_3 Z^d_4 = \vev{\la_1\la_2}[\mu_3\mu_4] + \mbox{permutations} \,.
\end{equation}
However, for this purpose we need four {\it linearly independent} twistors. This is clearly not the case of the MHV amplitude  \p{01}, since all points $Z_i$  lie on a line and
the invariants \re{twiin} just vanish. But the very special form of the MHV amplitude in twistor space allows us to form another, exceptional type of invariant.
Indeed, we see from \p{matchfact} that all contractions of chiral twistor variables $\vev{ij}$ transform in exactly the same way under conformal transformations. Thus, any ratio $\vev{ij}/\vev{kl}$ will be conformally invariant. In order not to modify the helicity weights of the amplitude, we need to consider the helicity-free cross-ratio
\begin{equation}\label{neutcrr}
    \frac{\vev{ij}\vev{kl}}{\vev{ik}\vev{jl}}\,.
\end{equation}
In fact, it is not only a conformal, but a complete superconformal invariant. Thus, multiplying the amplitude \p{01} by an arbitrary function of such cross-ratios will not modify its superconformal properties.\footnote{Of course, a realistic amplitude must have other properties, like cyclic symmetry, correct behavior in the singular collinear limit, etc. which may not be compatible with any function of the conformal invariants \p{neutcrr}. Here we restrict ourselves to discussing only the implications of the two superconformal symmetries.}

Let us now take into account the other, {\it dual conformal} symmetry of the superamplitude \p{mhvsu}. It becomes manifest in the dual description of the amplitude obtained by changing variables from momenta to dual coordinates, $p_i = x_i - x_{i+1}$. To examine the dual conformal properties of the modified twistor transform considered above, we need to carry out the inverse twistor transform back to momentum space.  This does not affect the  $\la-$variables and, therefore, the cross-ratios \p{neutcrr} and their functions are carried over to the dual picture unaltered. At this stage dual conformal symmetry steps in and forbids any such cross-ratio. Indeed, the contractions $\vev{ij}$ are dual conformal only if $|i-j|=1$, but this cannot hold for all contractions in \p{neutcrr}. Thus, we are led to the conclusion that the combined action of the two symmetries fixes the MHV superamplitude up to a constant factor.

We can reverse this argument and start from the dual conformal description of the MHV amplitude,
\begin{equation}\label{mhvsudual}
    \cA_n^{\rm MHV} = \frac{i(2\pi)^4}{\prod_1^n \vev{i\, i+1}}\ \delta^{(4)}(x_1-x_{n+1})\ \delta^{(8)}(\theta_1-\theta_{n+1})\,,
\end{equation}
where the dual coordinates have been introduced through the change of variables\footnote{The dual coordinates $x$ and $\q$ should not be confused with the twistor line parameters $X$ and $\Theta$.}
\begin{eqnarray}
  p_i^{\a\da} &=& \la^\a_i \bl^{\da}_i = (x_i - x_{i+1})^{\a\da}\,, \nn\\[2mm]
  q^\a_{i\, A} &=& \la^\a_i \eta_{i\, A} = (\q_i - \q_{i+1})^{\a}_A\,. \label{daulcoo}
\end{eqnarray}
The role of the delta functions in \p{mhvsudual} is to identify the end points of the cycle, yielding (super)momentum conservation $\sum_1^n p_i = \sum_1^n q_i = 0$. The product of the two delta functions in \re{mhvsudual} is manifestly invariant under dual conformal transformations, while the denominator involves contractions of
adjacent spinors only, and is therefore covariant.

How can we modify the superamplitude \re{mhvsudual} in a way consistent with its dual conformal symmetry? We can form two types of dual conformal invariants. The first one is the ratio $
    {\vev{i\ i+1}}/{[i\ i+1]} = {\vev{i\ i+1}^2}/{x^2_{i\ i+2}}\,,
$
and the other is the standard cross-ratio\footnote{Due to the lightlike separation of adjacent points, such cross-ratios only exist if $n \geq 6$.}
\begin{equation}\label{stancrora}
    u_{ijkl} =\frac{x^2_{ij}x^2_{kl}}{x^2_{ik}x^2_{jl}}\,.
\end{equation}
The difference is that the former has non-vanishing helicity, while the latter is helicity neutral. So, the only modification of \p{mhvsudual} consistent with dual conformal symmetry and with the helicity weights of the amplitude is
\begin{equation}\label{mhvsudualmod}
     \frac{i(2\pi)^4}{\prod_1^n \vev{i\, i+1}}\ \delta^{(4)}(x_1-x_{n+1})\ \delta^{(8)}(\theta_1-\theta_{n+1})\ f(u_{ijkl})\,,
\end{equation}
{with some arbitrary function $ f(u_{ijkl})$ of the cross-ratios.}

One can immediately argue that the function $ f(u_{ijkl})$ is not invariant under the dual Poincar\'e supersymmetry with generator
\begin{equation}\label{dualbarq}
    \bar Q^A_{\da} = \sum_1^n \q^{A\, \a} \frac{\pa}{\pa x^{\a\da}}
\end{equation}
(as explained in \cite{Drummond:2008vq}, it coincides with the ordinary special conformal supersymmetry generator $\bar s= \bar Q$).
Indeed, {the  variation $\bar Q^A_{\da}f(u_{ijkl})$} is proportional to $\q$, and there is nothing in the amplitude \p{mhvsudualmod} which can cancel or suppress such variations. Yet, here we would like to give an equivalent argument based on the twistor transform, because it is more suitable for generalizations. Let us repeat the steps leading to the twistor transform \p{01}, but this time with an additional bosonic factor  $f(u_{ijkl})$, as in \p{mhvsudualmod}. We can represent the latter through its half-Fourier transform
\begin{equation}\label{halfff}
    f(u_{ijkl})= \int \prod_1^n \frac{d^2\mu_i}{(2\pi)^2}\ e^{-i\sum_1^n[\mu_i\ i]} F(\la,\mu)\,.
\end{equation}
After a few obvious steps, the result is
\begin{equation}\label{01obvious}
     \frac{i}{\prod_1^n \vev{i\, i+1}}\ \int d^4X d^8\Theta\   F(\la, \hat\mu) \ \prod_1^{n} \delta^{(4)}(\hat\psi_i )\,,
\end{equation}
where we have used the notation \p{modtwiva}. Now, let us perform an $\bar s \equiv \bar Q$ transformation {in  \re{01obvious}}. According to \p{actonhat}, each $\delta^{(4)}(\hat\psi_i )$ gives
\begin{equation}\label{varbars}
    \bar s \,\delta^{(4)}(\hat\psi_i ) \sim (\hat\psi_i)^3 \hat\mu_i\,,
\end{equation}
so that the invariance of \re{01obvious} under $\bar s-$transformations would require $\int d^4X \, \hat\mu_i\, F(\la, \hat\mu)=0$. This is only possible if $F(\la, \hat\mu) \sim \delta^{(2)}(\hat\mu_i)$, and we find
\begin{equation}\label{wefindf}
    F(\la, \hat\mu) = \phi(\la) \prod_1^n \delta^{(2)}(\hat\mu_i)\,.
\end{equation}
{Substituting this expression into \re{01obvious} gives back the twistor transform \p{01}, except for the extra holomorphic function $\phi(\la)$.} Then, undoing the twistor transform
of \re{01obvious} and demanding dual conformal invariance, we find that the only solution is $\phi(\la)=\mbox{const}$.

\section{NMHV superamplitude}\label{nmhvsup}

In this section we carry out the twistor transform of the simplest of the non-MHV superamplitudes in \re{An}, the next-to-MHV (NMHV) amplitude $\cA_n^{\rm NMHV}$. We show that $T[\cA_n^{\rm NMHV}]$ is given by an integral over the moduli space which is enlarged by additional bosonic and fermionic moduli. We also study the superconformal properties of the twistor transform and show that sign factors lead to the breakdown of the conformal invariance of  $T[\cA_n^{\rm NMHV}]$
even under infinitesimal transformations. We argue that dual and ordinary superconformal symmetry fix the form of the amplitude, up to constant factors,
and demonstrate the cancellation of the so-called `spurious' singularities in the twistor transform.

The NMHV tree superamplitude admits a very simple and manifestly dual superconformal formulation \cite{Drummond:2008vq},
\begin{equation}\label{nmhvsu}
    \cA_n^{\rm NMHV}  = \sum_{3 \leq a+1 < b \leq n-1} \cA_{nab}\,,\qquad
    \cA_{nab}
    = \cA_n^{\rm MHV}\ R_{nab}\,.
\end{equation}
The main building block  is the dual superconformal invariant\footnote{Compared to the form given in \cite{Drummond:2008vq}, here we use the inverse matrices $x^{-1}_{ab}$, which is helpful for doing the Fourier transform.}
\begin{equation}\label{mbubl}
    R_{nab} = \frac{\vev{a-1\, a} \vev{b-1\, b}\ \delta^{(4)}(\sum_1^{a-1} \lan{n} x_{nb} x^{-1}_{ba}\ran{i}  \eta_i + \sum_1^{b-1} \lan{n} x_{na} x^{-1}_{ab}\ran{i} \eta_i)}{x^2_{ab}\lan{n} x_{nb} x^{-1}_{ba}\ran{a-1}\lan{n} x_{nb} x^{-1}_{ba}\ran{a} \lan{n} x_{na} x^{-1}_{ab}\ran{b-1}\lan{n} x_{na} x^{-1}_{ab}\ran{b}}\,,
\end{equation}
where the dual $x-$variables were introduced in \p{daulcoo}, so that $x_{rs} =\sum_r^{s-1}p_i$.
It is easy to check that the numerator in  \re{mbubl} vanishes identically unless the indices $a,b$ satisfy the inequalities indicated in \p{nmhvsu}. The choice of the first index of $\cA_{nab}$ in the double sum  \p{nmhvsu} to be $n$ is not essential
because of the following identity among superinvariants \cite{Drummond:2008vq}:
\begin{equation}\label{iderinv}
    \sum_{3 \leq a+1 < b \leq n-1} R_{nab} = \sum_{4 \leq a+1 < b \leq n} R_{1ab}\ .
\end{equation}
This identity ensures that the double sum in  \p{nmhvsu}
is invariant under a cyclic shift of the indices of the incoming particles.

\subsection{Twistor transform}\label{ttnmhv}

The twistor transform of the MHV superamplitude in Sect.~\ref{mmhhvv} was easy to carry out, due to the almost holomorphic nature of this amplitude (the only dependence on $\bl$ comes through the momentum conservation delta function).
Comparing \re{nmhvsu}, \p{mbubl} with \re{mhvsu}, it may seem that the half-Fourier transform of the NMHV amplitude is an impossible task, because of the very non-trivial dependence on $\bl$ in \p{mbubl}.  However, we can apply the following trick. First, we define the spinors
\begin{equation}\label{defr}
 \lan{\rho_0} \equiv \lan{n}\,, \qquad \lan{\rho} = \lan{n} x_{nb} x^{-1}_{ba}\,, \qquad  \lan{\sigma} = \lan{n} x_{na} x^{-1}_{ab}\,,
\end{equation}
satisfying the relation
\begin{equation}\label{linrel}
    \lan{\rho_0} + \lan{\rho}+ \lan{\sigma} = 0\,.
\end{equation}
{Then, we use the Faddeev-Popov approach and introduce the
spinors \re{defr} into \p{mbubl} via delta function integrals. Choosing, e.g.,  $\r$ to be  the independent spinor from \p{linrel}, we have
\begin{eqnarray}
  f(\lan{n} x_{nb} x^{-1}_{ba}) &=& \int d^2\rho\  f(\r)\ \delta^{(2)}\left(\lan{\rho} - \lan{n} x_{nb} x^{-1}_{ba}\right) \nn\\
  &=& |x^2_{ab}| \int d^2\rho \, d^2\rt \ f(\r)\ \exp\left\{-i \left(\lan{\rho} x_{ba} |\rt] + \lan{n} x_{nb}  |\rt] \right)\right\}\,. \label{trick}
\end{eqnarray}
}
We see that the factor $|x^2_{ab}|$ pulled out of the delta function `almost' cancels the analogous factor $x^2_{ab}$ in the denominator in \p{mbubl}, producing $\mbox{sgn}(x^2_{ab})$. This extra sign factor has a dramatic effect on the Fourier transform, as discussed in detail in Appendix \ref{dramatic}. For the time being, we replace $x^2_{ab}$ in denominator of  \p{mbubl} with
$|x^2_{ab}|$ and perform the twistor transform \re{T} of the resulting expression for
$\cA_{nab}$.
 In other words, we will not be doing the twistor transform of the true NMHV tree amplitude, but of another function which coincides with $ \cA_n^{\rm NMHV}$ only in the kinematic region where all the kinematic invariants $x^2_{ab}= (\sum_a^{b-1}p_i)^2$ in the sum in \p{nmhvsu} have the same sign.

It is convenient to replace the Grassmann delta function in \p{mbubl} by its Fourier integral over the auxiliary odd variable $\xi_A$,
\begin{equation}\label{grde}
    \int d^4\xi \ \exp \left\{i\xi_A \left(-\sum_1^{a-1} \vev{\r_0\, i} \eta^A_i + \sum_a^{b-1} \vev{\sigma\, i} \eta^A_i\right)  \right\}\,,
\end{equation}
where we have used \p{linrel}. Finally,
we treat the delta functions in the MHV factor in \p{nmhvsu} as  in Sect.~\ref{mmhhvv} and, recalling that $x_{ab} = \sum_a^{b-1} \ran{i} [i|$ and $x_{nb} = \sum_n^{b-1} \ran{i} [i|$, we can immediately do the complete twistor transform of
\re{nmhvsu}. In this way, the result for the twistor transform of the partial amplitude $\cA_{nab}$ reads
\begin{equation}\label{sugg}
T\left[\cA_{nab}\right] =   i\int d^4X d^8\Theta\int d^2\rho\, d^2\tilde\rho\, d^4\xi \ \frac{ (\prod)_{nab}}{ \Delta_{nab}}\,,
\end{equation}
where
\begin{equation}\label{delta0}
    \Delta_{nab} = \vev{12}\ldots \vev{a-1\, \rho} \vev{\rho\, a} \ldots  \vev{b-1\, \sigma} \vev{\sigma\, b}  \ldots \vev{n1}\,,
\end{equation}
and
\begin{eqnarray}
   (\prod)_{nab} &=& \prod_{1}^{a-1} \delta^{(2)}(\mu_i + \lan{i} X_1)\ \delta^{(4)}(\psi_i + \vev{i\, \Theta_1})\nn\\
  &\times& \prod_{a}^{b-1} \delta^{(2)}(\mu_i + \lan{i} X_2)\ \delta^{(4)}(\psi_i + \vev{i\, \Theta_2})\nn\\
  &\times& \prod_{b}^{n}\delta^{(2)}(\mu_i + \lan{i} X_3)\ \delta^{(4)}(\psi_i + \vev{i\, \Theta_3}) \label{2}
\end{eqnarray}
with
\begin{equation}\label{arr}
    \begin{array}{lcl}
       X_1 = X - \r_0 \rt\,,   & \qquad & \Theta_1 = \Theta - \r_0\xi\,,   \\
       X_2 = X +\sigma\rt\,,  & \qquad & \Theta_2 = \Theta +\sigma\xi\,,   \\
       X_3 = X\,,  & \qquad & \Theta_3 = \Theta  \ .
     \end{array}
\end{equation}
Note that the term with $i=n$ in the right-hand side of \re{2} can equally well be attributed to the first cluster of deltas, since
\begin{equation}\label{nand1}
    \bra{n}X_1=\bra{n}X_3\,, \qquad \bra{n}\Theta_1=\bra{n}\Theta_3\,.
\end{equation}
We observe that the relation \re{sugg} has a structure similar to that of the MHV superamplitude \p{01}, with the essential difference that instead of the single cluster of delta functions we now have three such clusters. As we explain in detail in Sect.~\ref{gitt}, they correspond to three lines in twistor space.
Accordingly, the single set of line parameters $(X,\Theta)$ for the MHV superamplitude is replaced by three sets $(X_u,\Theta_u)$ with $u=1,2,3$.

It is very instructive to rewrite \p{arr} in terms of the differences of the line parameters,
\begin{equation}\label{arr2}
    \begin{array}{lcl}
       X_{12} = \r\rt\,, & \qquad & \Theta_{12} = \r\xi\,, \\
       X_{23} = \sigma\rt\,, & \qquad & \Theta_{23} = \sigma\xi\,,\\
       X_{31} = \r_0 \rt \,,& \qquad & \Theta_{31} = \r_0\xi\,,
    \end{array}
\end{equation}
where we have used \p{linrel}. By definition,
\begin{equation}\label{sum3x}
    X_{12}+X_{23}+X_{31}=0\,, \qquad  \Theta_{12}+\Theta_{23}+\Theta_{31}=0\,,
\end{equation}
and, most importantly, the three vectors $X_{12}$, $X_{23}$ and $X_{31}$ separating the points $X_1$, $X_2$ and $X_3$ in moduli space are {\it lightlike}. These properties
are strongly reminiscent of those of the dual superspace \p{daulcoo} but there
are two important differences. Firstly, the dual coordinates $x_i^{\a\da}$ (with $i=1,\ldots,n$) have the scaling dimension of momenta, while the moduli $X_u^{\a\da}$ (with $u=1,2,3$) have the dimension of position space coordinates. Secondly, in dual space the NMHV amplitude \p{mbubl} depends on $n$ points (equal to the number of particles), while the moduli space only involves three points. For a generic N${}^k$MHV superamplitude (see Sect.~\ref{allnmhvsu}), the latter number is in fact $2k+1$.

\subsection{Superconformal properties}\label{scpro}

One of the advantages of the twistor transform \re{sugg}, related to the presence of super-line moduli, is its geometric
clarity  (see Sect.~\ref{gitt}).
However, unlike other approaches (see \cite{Mason:2009sa} and \cite{ArkaniHamed:2009si}), where the superconformal symmetry of the twistor transform is manifest, here one needs to show it explicitly. In particular, one needs to work out the transformation properties of the moduli.
In this subsection we show that the NMHV twistor transform \re{sugg} is invariant under Poincar\'e supersymmetry and conformal inversion, which, as explained in  Sect.~\ref{ciits}, is sufficient to establish full superconformal invariance.

The action of translations ($p$) and chiral supersymmetry ($q$) on the common line parameters $(X,\Theta)$, and consequently on $(X_u,\Theta_u)$ (with $u=1,2,3$), is the same as in the MHV case (see \p{sus}). The antichiral generator $\bar q$ has to be supplemented with the rule
\begin{equation}\label{suupwi}
    \bar q^{\da}_A\, \rt_{\db} = \delta^{\da}_{\db}\,\xi_A\,,
\end{equation}
so that the line parameters $X_u$ and $\Theta_u$, Eq.~\re{arr}, transform in the standard way \p{sus}.

{We now consider inversion and take into account the standard transformation  of $X_{uv}$,}
\begin{equation}\label{wekno}
    I[X_{uv}] = - X^{-1}_u X_{uv} X^{-1}_v\,.
\end{equation}
We shall try to arrange the transformation properties of the moduli in \p{arr2} to match \p{wekno}. This is certainly possible, because conformal inversion maps a lightlike vector into another lightlike vector. Let us start with $X_{31}$, where we already know the transformation of $\r_0=\la_n$. Indeed, the latter belongs to the line with parameter $X\equiv X_3$, so repeating the steps from Sect.~\ref{ciits} we get
\begin{equation}\label{wealreakn}
    I[\la^\a_n] = -\mu^{\da}_n = (\lan{n}X_3)^{\da}\,.
\end{equation}
Then, for compatibility with \p{wekno} we need to impose
\begin{equation}\label{neetoimp}
    I[\rt] = -(X^2_1 X^2_3)^{-1}\ X_1|\rt] = -(X^2_1 X^2_3)^{-1}\ X_2|\rt]= -(X^2_1 X^2_3)^{-1}\ X_3|\rt]\,,
\end{equation}
where the three equivalent forms follow from the fact that the three lightlike vectors in \p{arr2} share the same antichiral spinor $\rt$. {In the same way, the transformation of $\r$  can be derived from those of $X_{12}$ and of $\rt$.} We find
\begin{equation}\label{fromthatof}
    I[\r] = -\frac{X^2_3}{X^2_2} \ \lan{\r}X_1  = -\frac{X^2_3}{X^2_2} \ \lan{\r}X_2\ .
\end{equation}
Finally, from the transformation of $X_{23}$ we find that of $\sigma$,
\begin{equation}\label{fromthatof'}
    I[\sigma] = -\frac{X^2_1}{X^2_2} \ \lan{\sigma}X_2  = -\frac{X^2_1}{X^2_2} \ \lan{\sigma}X_3\ .
\end{equation}
Since $\sigma$ is not independent (recall \p{defr}), \p{fromthatof'} should also follow from \p{wealreakn} and \p{fromthatof}, which can be checked directly.

In the fermionic sector, from the standard transformation \p{stasuco} of $\Theta_u$ and from, e.g., the expression for $\Theta_{31}$ in \p{arr2} we can derive the transformation of $\xi$,
\begin{equation}\label{dertraxi}
    I[\xi] = \frac{\xi}{X^2_1} - \frac{\lan{\Theta_3}X_3|\rt]}{X^2_1X^2_3}\,,
\end{equation}
{which can be rewritten in several equivalent forms.}

We are  ready to examine the properties of the NMHV twistor transform  \p{sugg}
under inversion. We can repeat the same steps as in the MHV case. The main difference is the transformation of the spinor contractions in the denominator of \p{sugg} involving the new moduli $\r$ and $\sigma$:
\begin{align}
&I[\vev{a-1\ \r}] = \frac{X^2_3 X^2_1}{X^2_2}  \vev{a-1\ \r}\,,  & & I[\vev{\r\ a}]  = X^2_3 \vev{\r\ a}\nn \\
& I[\vev{b-1\ \sigma}] = X^2_1 \vev{b-1\ \sigma}\,, & & I[\vev{\sigma\ b}]  = \frac{X^2_1 X^2_3}{X^2_2} \vev{\sigma\ b} \ . \label{newcontr}
\end{align}
{These additional weight factors cancel against those of the new integration measure $\int d^2\r d^2\rt d^4\xi$, up to sign factors,
\begin{equation}\label{sigfa}
    [\text{sgn}(X_1^2)]^{a-2} [\text{sgn}(X_2^2)]^{b-a-1} [\text{sgn}(X_3^2)]^{n-b+1}\,,
\end{equation}
as shown in Appendix \ref{jacobi}.} Notice that the total number of sign factors coincides with the number of particles, but their distribution depends on the labels $a,b$. Once again, global conformal invariance is broken by sign factors. Here we must recall that \p{sugg} is not the twistor transform of the true amplitude \p{nmhvsu}, \p{mbubl}, but of another function differing from it by $\text{sgn}(x^2_{ab})$. The latter sign factor leads to the breakdown of conformal symmetry even at the infinitesimal level, as explained in Appendix B.

We can now understand the structure of the denominator \p{delta0} in the NMHV
twistor transform \re{sugg}. {As far as conformal symmetry is concerned,}
the line parameter $\r$ transforms in a way suitable to form a covariant contraction with any point from the cluster $[1, a-1]$, and with any point from the cluster $[a, b-1]$. Then, why is the new spinor $\r$ inserted precisely  between the  points $a-1$ and $a$?  It is clear that the ordinary conformal symmetry cannot give the explanation. It comes from the other, dual conformal symmetry of the amplitude. If we undo the twistor transform, we should get back the momentum space expression \p{mbubl}, with its characteristic angular brackets in the denominator, where $\r$ is replaced by $\lan{n} x_{nb} x^{-1}_{ba}$.  As explained in \cite{Drummond:2008vq}, only the brackets $\lan{n} x_{nb} x^{-1}_{ba}\ran{a-1}$ and $\lan{n} x_{nb} x^{-1}_{ba}\ran{a}$ are dual conformally covariant, while the combination of $\r$ with any other points from the two clusters would not be. Similarly, we can argue why the new spinor $\sigma$ appears precisely between the points $b-1$ and $b$.

\subsection{Uniqueness of the NMHV superamplitude}

Let us repeat the argument given at the end of Sect.~\ref{unimhv} about the restrictions imposed by the two superconformal symmetries on the possible form of  the NMHV superamplitude. As before, we can try to modify each invariant \p{mbubl} by an arbitrary dual conformally invariant bosonic factor $f(\la,\bl) \equiv f(u_{ijkl})$.  Such a
factor will inevitably break the special conformal $\bar s-$supersymmetry
of the partial amplitude $\cA_{nab}$ in \re{nmhvsu}, but one might suspect that the invariance
would be restored in the sum \p{nmhvsu}, for some special choice of the function $f(u_{ijkl})$. In this subsection, employing the twistor transform, we rule out this possibility.

We start by introducing the compact notation (cf. \p{modtwiva})
\begin{equation}\label{compnot}
    \hat\mu_{u;i} = \mu_i + \lan{i} X_u\,, \qquad \hat\psi_{u;i} = \psi_i + \vev{i\, \Theta_u}\,, \qquad u=1,2,3\,,
\end{equation}
{where the index $i$ labels the particles and $u$ labels the three
lines in twistor space.}
It is straightforward to verify that the `hatted' twistor variables, $\hat\mu_{u;i}$ and
$\hat\psi_{u;i}$,  have the same superconformal transformation properties as in the MHV case \p{actonhat}, with the moduli $(X_u, \Theta_u)$ relevant for each of the three lines. With this notations, the relation \p{sugg}  becomes
\begin{equation}\label{suggerer}
i\int d^4X d^8\Theta \int \frac{d^2\rho d^2\tilde\rho d^4\xi}{\Delta_{nab}} \ \prod_1^{a-1} \delta^{(2)}(\hat\mu_{1;i})\delta^{(4)}(\hat\psi_{1;i})\ \prod_a^{b-1} \delta^{(2)}(\hat\mu_{2;i})\delta^{(4)}(\hat\psi_{2;i})\ \prod_b^{n} \delta^{(2)}(\hat\mu_{3;i})\delta^{(4)}(\hat\psi_{3;i}) \ .
\end{equation}
{Now, suppose that we have modified each $\cA_{nab}$ in \re{nmhvsu} by  an arbitrary bosonic factor, $\cA_{nab}\to \cA_{nab}  f_{nab}(\{u\})$, with}
\begin{equation}\label{modbyarbfa}
    f_{nab}(\{u\})   = \int \prod_1^n \frac{d^2\mu_i}{(2\pi)^2}\,\e^{-i\sum_1^n[\mu_i\ i]} F_{nab}(\la,\mu)\,.
\end{equation}
Repeating the steps leading to the twistor transform \p{suggerer}, we find
\begin{equation}\label{repstep}
 i\int \frac{d^4X d^2\rho d^2\tilde\rho d^8\Theta d^4\xi}{\Delta_{nab}} \  F_{nab}(\la_i,\hat\mu_{u;i})\ \prod_1^{a-1} \delta^{(4)}(\hat\psi_{1;i})\ \prod_a^{b-1} \delta^{(4)}(\hat\psi_{2;i})\ \prod_b^{n} \delta^{(4)}(\hat\psi_{3;i}) \ ,
\end{equation}
where the second argument in $F_{nab}(\la,\hat\mu)$ takes one of the forms \p{compnot} according to the sector it belongs to. Then, applying an $\bar s$ transformation to \re{repstep}, we find that each $\delta^{(4)}(\hat\psi_{u;i})$ produces a factor $\hat\mu_{u;i}$ which must annihilate the function $F_{nab}(\la_i,\hat\mu_{u;i})$. It is clear that the different terms in the sum over all $\cA_{nab}$ in \p{nmhvsu} cannot help each other, since they have different Grassmann structures. Thus, we are lead to the conclusion that the $\hat\mu$ dependence of $F_{nab}(\la_i,\hat\mu_{u;i})$ is given by products of delta functions, as in \p{suggerer}. The remaining $\la$ dependence can be reduced to a constant by undoing the twistor transform and imposing dual conformal symmetry.

In the argument above the decisive input was the known fermionic structure of $R_{nab}$ in \p{mbubl} (including the detailed knowledge of the argument of the Grassmann delta function). This, together with  $\bar s$  supersymmetry, effectively determines the bosonic dependence of the twistor transform. How do we know that there is no Grassmann structure other than \p{mbubl} suitable for the NMHV amplitude? For instance, we might imagine an $SU(4)$ invariant not of the simple delta function type $\delta^{(4)}(\hat\psi_{u;i})$, but a combination of four different $\hat\psi$'s. The $\bar s$ supersymmetry will again force us to have the $\mu$ dependence in the form $\delta^{(2)}(\hat\mu_{u;i})$. After that, we can impose $\bar q$ supersymmetry, which transforms $\hat\mu$ into $\hat\psi$ (see \p{sus}). This in return will require the presence of $\delta^{(4)}(\hat\psi_{u;i})$. So, although this is not a rigorous formal proof, we can see that the combination of dual with ordinary superconformal symmetry fixes the form of the invariants $R_{nab}$, up to an overall factor.

\subsection{Spurious singularities}

According to \re{sugg}, the twistor transform of $\mathcal{A}_{nab} = \mathcal{A}_n^{\rm MHV}  R_{nab}$ involves a two-dimensional integral over the real spinor $\rho^\a$. The close examination of \re{sugg} reveals that
this integral is not well defined due to the vanishing of the denominator $\Delta$ from \re{delta0},
\begin{align}\label{spur}
T[\mathcal{A}_{nab}] \sim \int\frac{d^2\rho}{\Delta_{nab}} \sim \int\frac{d^2\rho}{\vev{a-1\rho} \vev{\rho a} \vev{b-1\sigma}\vev{\sigma b}}\,,
\end{align}
with $\ket{\sigma} = -\ket{\rho} -\ket{n}$. Each factor in the right-hand side of this relation
produces a pole in $\rho^\alpha$ and, as a consequence, the twistor transform of $\mathcal{A}_{nab}$ depends on the prescription employed to deform the integration contour around the pole.  The choice of the prescription is
ambiguous but the scattering amplitude $\mathcal{A}_n$ should not depend on it. In other words, the four poles in the right-hand side of \re{spur}
are spurious and they should cancel in the sum of $\mathcal{A}_{nab}$ defining
the scattering amplitude. As was shown in \cite{Korchemsky:2009hm}, this requirement, in combination with dual and ordinary superconformal symmetry, allows us to unambiguously
reconstruct the tree expression for NMHV superamplitude. In this subsection,
we demonstrate the cancellation of the spurious poles in the twistor transform.

The twistor transform  \re{sugg} and \re{spur} has four spurious poles at
\begin{align}\label{poles}
\vev{a-1\,\rho} = \vev{a\,\rho} = \vev{b-1\,\sigma} = \vev{b\,\sigma} = 0\,.
\end{align}
Let us first examine the spurious pole at $\vev{a\rho}= 0$.
{In terms of $\ket{\rho}$ it corresponds to the kinematic configuration where
$\ket{\rho}\sim \ket{a}$. Without loss of generality, we can choose $\ket{\rho}= \ket{a}$, because  the integral in \p{spur} is invariant under rescaling of $\r$. As we showed in the previous subsections, the twistor
transform of $\mathcal{A}_{nab}$ is characterized by the three moduli $X_1$, $X_2$ and $X_3$ with lightlike separations, see \p{arr2}. For $\ket{\rho}= \ket{a}$ these separations take the form
\begin{align}\label{tri1}
T[\mathcal{A}_{nab}] \quad  \stackrel{\vev{a\rho}=0}{\Longrightarrow} \quad X_{12} = \lambda_a \rt\,,\quad X_{23} = -(\lambda_a+\lambda_n)\rt\,,\quad X_{31} = \lambda_n\rt\,.
\end{align}
This configuration defines the residue of the integrand of $T[\mathcal{A}_{nab}]$ at the spurious pole at $\vev{a\rho}=0$,
\begin{align}\label{res1}
\res_{\vev{a\rho}=0}\left[\frac{ (\prod)_{nab}}{ \Delta_{nab}}\right]=
 \frac{\vev{b-1\, b}}{\vev{b-1,a+n}\vev{a+n,b}}
  \frac{(\prod)_{nab}}{ \prod_1^n \vev{i \, i+1}}\ .
\end{align}

Now, let us examine the spurious poles of the partial amplitude $T[\mathcal{A}_{abn}]$. It can be obtained from $T[\mathcal{A}_{nab}]$ through a cyclic shift of the indices. As a result, $T[\mathcal{A}_{abn}]$ also has four spurious poles whose
positions however are different from \re{poles}. In particular, the pole  $\vev{b\sigma} = 0 $ in \re{poles} is now located at $\vev{n\sigma}=0$ with $\ket{\sigma}=-\ket{\rho}-\ket{a}$.
Examining the corresponding configuration in the moduli space we find that the three moduli are given by
\begin{align}\label{tri2}
T[\mathcal{A}_{abn}] \quad  \stackrel{\vev{n\sigma}=0}{\Longrightarrow} \quad X_{31} = \lambda_a \rt\,,\quad X_{12} = -(\lambda_a+\lambda_n)\rt\,,\quad X_{23} = \lambda_n\rt\,.
\end{align}
Comparing \re{tri1} and \re{tri2}, we notice that the two configurations in fact coincide after relabeling the moduli as follows, $X_1\to X_3, X_2\to X_1, X_3\to X_2$. This procedure becomes quite natural
after we identify the indices of the particles belonging to the three lines in twistor space. Namely, for $T[\mathcal{A}_{nab}]$ the particles  $1,\ldots,a-1$ lie on
the line with modulus $X_1$, while for $T[\mathcal{A}_{abn}]$ the same points lie on the line with modulus $X_3$.
The same identification applies to the two remaining lines. Then, the residue of the
integrand of  $T[\mathcal{A}_{abn}]$ at the spurious pole $\vev{n\sigma}=0$ reads
 \begin{align}\label{res2}
\res_{\vev{n\sigma}=0}\left[\frac{ (\prod)_{abn}}{ \Delta_{abn}}\right]= -
 \frac{\vev{b-1\, b}}{\vev{b-1,a+n}\vev{a+n,b}}
  \frac{(\prod)_{abn}}{ \prod_1^n \vev{i \, i+1}}\,.
\end{align}
Due to the identical moduli space configurations \re{tri1} and \re{tri2}, the products of delta functions entering \re{res1} and \re{res2} coincide, $ (\prod)_{abn}= (\prod)_{nab}$.
Therefore, we conclude that the sum of the residues of the integrands of $T[\mathcal{A}_{nab}]$ at $\vev{a\rho}=0$, and of $T[\mathcal{A}_{abn}]$ at $\vev{n\sigma}=0$, vanishes.

This result is in agreement with Ref.~\cite{Korchemsky:2009hm} where it was found that
all spurious poles of $\mathcal{A}_{nab}$ cancel in the following linear combination
of partial amplitudes
\begin{align}\label{master}
\mathcal{A}_{nab} + (\mathcal{A}_{abn}+\mathcal{A}_{bna}- \mathcal{A}_{a-1\, bn}-\mathcal{A}_{b-1\, na})\,.
\end{align}
Obviously, the same property should hold after the twistor transform. Indeed, in this subsection we have demonstrated that the spurious pole  of $T[\mathcal{A}_{nab}]$ at $\vev{a\rho}=0$ cancels against a similar pole in $T[\mathcal{A}_{abn}]$ because both amplitudes are characterized by the same moduli space configuration. The same applies
to the remaining three spurious poles of $T[\mathcal{A}_{nab}]$ in \re{poles}.
In particular, the spurious poles of  $T[\mathcal{A}_{abn}]$ at $\vev{b\sigma}=0$,
$\vev{a-1\rho}=0$ and $\vev{b-1\sigma}=0$ cancel, respectively,
against the following spurious poles: $\vev{n\rho}=0$ of $T[\mathcal{A}_{bna}]$,
$\vev{n\sigma}=0$ of $T[\mathcal{A}_{a-1bn}]$ and $\vev{n\rho}=0$ of $T[\mathcal{A}_{b-1na}]$. In all three cases, the underlying kinematic configurations in moduli space are identical,
and the relative signs between the partial amplitudes are determined by the relative signs of the residues of the $\Delta-$factors.} {Making use of \re{master} and going along the same lines as in Ref.~\cite{Korchemsky:2009hm}, we can
show that the requirement of cancellation of spurious poles fixes
the relative coefficients in front of $\cA_{nab}$ in the sum \re{nmhvsu} and, therefore,
determines the form of tree NMHV superamplitude.}

\section{Geometric interpretation of the twistor transform of the NMHV superamplitude}\label{gitt}

In this section we show that the twistor transform \p{sugg} of the NMHV superamplitude has support on a set of three intersecting lines in twistor space. This result itself is not new, its bosonic version was first established in Refs.~\cite{Bern:2004ky}, \cite{Bern:2004bt}, \cite{Britto:2004tx} by applying Witten's collinearity and coplanarity differential operators \cite{Witten:2003nn}. More recently, it was also confirmed in twistor space \cite{Mason:2009sa}. {In the preceding section we found a new form of the twistor transform of the NMHV superamplitude in which this line structure becomes manifest.} In addition, below we reveal a transparent geometric structure in the moduli space {of the twistor transform}, which will allow us, in the next section, to propose a general  construction for all non-MHV tree amplitudes.

\subsection{Lines in twistor space}\label{cctsp}

{Upon the twistor transform \re{T}, the superamplitude depends on $n$ sets of bosonic $(\lambda_i^\a,\mu_{i\da})$ and fermionic $\psi_{i\,A}$ variables. To simplify the discussion, we ignore for the time being the fermions and consider $T[\cA_n]$
as a function on the twistor space ${\mathbb R}{\mathbb P}^3$ with signature $(++--)$. Its projective coordinates $(\la^\a, \mu_{\da})$ are two-component commuting real spinors.
To each particle corresponds a `twistor', i.e. a vector in the projective space   $Z_i^a=(\la_i^\a,\mu_{i\,\da})$ (with $a=1,\ldots,4$), transforming under the fundamental representation of the conformal group $SL(4,{\mathbb R})$. }
As explained in \cite{Witten:2003nn}, the condition for any three  points
to lie on a line (i.e., on a ${\mathbb R}{\mathbb P}^1 \subset {\mathbb R}{\mathbb P}^3$) is that the three corresponding vectors $Z^a_i$ (with $i=1,2,3$) {have to} be linearly dependent, i.e.,
\begin{equation}\label{concoll}
    \epsilon_{abcd} Z^b_1 Z^c_2 Z^d_3 =0\,.
\end{equation}
This equation has a free index $a=(\a,\da)$, so we can consider its two components, chiral and antichiral,
\begin{eqnarray}
  [\mu_1\mu_2]\la_3 + [\mu_2\mu_3]\la_1+ [\mu_3\mu_1]\la_2 &=& 0\,,  \label{chicons} \\[2mm]
  \vev{12}\mu_3 + \vev{23}\mu_1 + \vev{31}\mu_2 &=& 0\,. \label{achicons}
\end{eqnarray}
Each of these relations, considered separately, simply states the fact that three two-component spinors are always linearly dependent. Let us assume, for example, that
{the contractions} $\vev{12} \neq 0$ and $[\mu_1 \mu_2] \neq 0$, i.e., that the corresponding pairs of spinors are linearly independent. Then we can express $\la_3$ from \p{chicons} and $\mu_3$ from \p{achicons}, as linear combinations of $\la_1,\la_2$ and $\mu_1,\mu_2$, respectively. What makes the two conditions non-trivial is the fact that the coefficients of the two linear combinations are the same. Indeed, projecting either  \p{chicons} with $\la_1$ and $\la_2$, or   \p{achicons} with $\mu_1$ and $\mu_2$, we find the relations
\begin{equation}\label{wefire}
   \frac{\vev{13}}{\vev{12}} = \frac{[\mu_1\mu_3]}{[\mu_1\mu_2]} \equiv t\,, \qquad \frac{\vev{23}}{\vev{12}} = \frac{[\mu_2\mu_3]}{[\mu_1\mu_2]} \equiv s \,.
\end{equation}
Putting this back into \p{chicons} and \p{achicons}, we obtain the following linear relation between the three twistors:
\begin{equation}\label{msk}
    Z_3 + s Z_1 - t Z_2  = 0\,.
\end{equation}
Here the scalar parameters $s$ and $t$ are inert under conformal $SL(4,{\mathbb R})$
transformations {but they undergo helicity rescaling}, since the three vectors in \p{msk} scale independently. The linear, manifestly $SL(4,{\mathbb R})$ covariant relation \re{msk} was exploited in Ref.~\cite{Mason:2009sa} in order to exhibit the line structure of the twistor transform of the amplitudes.

Alternatively, we may write down the solution to \p{chicons} and \p{achicons} in  the form
\begin{equation}\label{tleq0}
    \mu_{i\, \da} + \la^\a_i X_{\a\da}=0\,, \qquad i=1,2,3,
\end{equation}
where $X^{\a\da}$ is  a real four-vector parameter.
{This  relation is just the (bosonic) twistor line equation obtained by half-Fourier transforming the MHV amplitude in Sect.~\ref{mmhhvv}.  The vector line parameter $X$ can be expressed in terms of the projective coordinates of two points as in \p{exprxth}.} Unlike the conformally invariant scalar parameters $s$ and $t$, Eq.~\p{wefire}, the parameter $X^{\a\da}$ transforms under $SL(4,{\mathbb R})$ as if it were the coordinate of a point in configuration space, Eqs.~\re{cotr} and \re{stasuco}. {However, in contrast with $s$ and $t$, it has vanishing helicity.}

{The somewhat abstract notion of twistor line
becomes more intuitive by rewriting its equation $\mu_{\da} + \la^\a X_{\a\da}=0$}
in ${\mathbb R}^3$, as explained in \cite{Witten:2003nn}. Throwing away, for example, the set $\la^2 \neq 0$ in ${\mathbb R}{\mathbb P}^3$, we can describe the rest of ${\mathbb R}{\mathbb P}^3$ by the affine (homogeneous) coordinates $z_1 = \mu_{\dot 1}/\la^2$,  $z_2 = \mu_{\dot 2}/\la^2$, $z_3 = \la^1/\la^2$. They parametrize a copy of ${\mathbb R}^3$, in which the twistor line equation takes the form
\begin{eqnarray}
  z_1 + z_3 X_{1\dot 1}+ X_{2\dot 1} &=& 0\,, \nn\\
  z_2 + z_3 X_{1\dot 2}+ X_{2\dot 2} &=& 0\ . \label{twline}
\end{eqnarray}
These relations define a straight line in ${\mathbb R}^3$  passing through the point $(-X_{2\dot 1}, -X_{2\dot 2}, 0)$ in the direction specified by the tangent vector $(-X_{1\dot 1}, -X_{1\dot 2}, 1)$.

\subsection{Intersecting lines in twistor space}\label{ltsp}

Let us consider two twistor lines with parameters $X_1$ and $X_2$ and examine the condition for them to intersect, i.e. to have a common point ${\cal O}(\la,\mu)$,
\begin{eqnarray}
  \mu_{\da} + \la^\a (X_{1})_{\a\da} &=& 0\,, \nn \\
  \mu_{\da} + \la^\a (X_{2})_{\a\da} &=& 0\,, \label{twoteq}
\end{eqnarray}
{or equivalently (with $X_{12}=X_1-X_2$)}
\begin{equation}\label{hencee}
    \la^\a (X_{12})_{\a\da} = 0\,.
\end{equation}
This equation has a  non-trivial solution $\la^\a \neq 0$ if and only if $\mbox{det}\|X_{12}\| =(X_{12})^2=0$, i.e. if the difference of the two line parameters is a lightlike vector,    $(X_{12})_{\a\da} = \r_\a \rt_{\da}$, {with $\r$ and $\rt$ being some chiral and antichiral real spinors.} Then, the solution to \p{twoteq} {is given  (up to an inessential scale) by
$\la^\a=\r^\a$  and $\mu = -\bra{\r}X_1 = -\bra{\r}X_2$.} It is often convenient to use translation invariance to set, e.g., $X_2=0$, {so that the two lines intersect at the point $\la^\a=\r^\a$ and $\mu_\da=0$.}\footnote{Exactly the same argument applies to the fermionic sector: $\psi_A = \vev{\la \Theta_A} = \vev{\la \Theta'_A}$ implies $\Theta_A - \Theta'_A = \la\xi_A$ with some arbitrary odd variable $\xi_A$. Using supersymmetry to set, e.g., $\Theta'_A=0$, we obtain $\psi_A = 0$ at the intersection point. }

\begin{figure}[h]
\psfrag{X}[cc][cc]{$X$}
\psfrag{X1}[cc][cc]{$X_1$}
\psfrag{X2}[cc][cc]{$X_2$}
\psfrag{rr}[cc][cc]{$\scriptstyle \r\rt$}
\psfrag{M}[cc][cc]{$\scriptstyle \cal{O} $}
\centerline{\includegraphics[height=45mm]{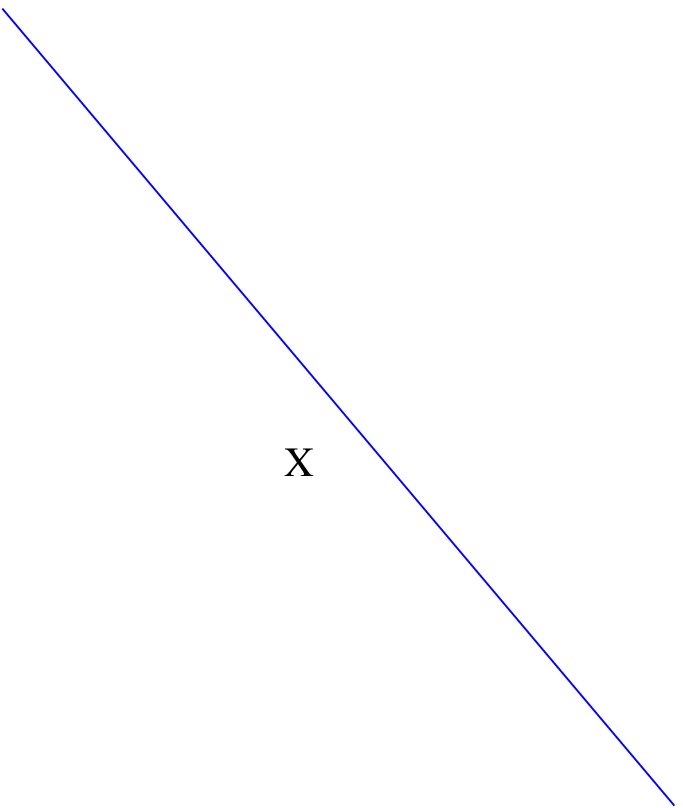} \qquad\qquad\qquad \includegraphics[height=50mm]{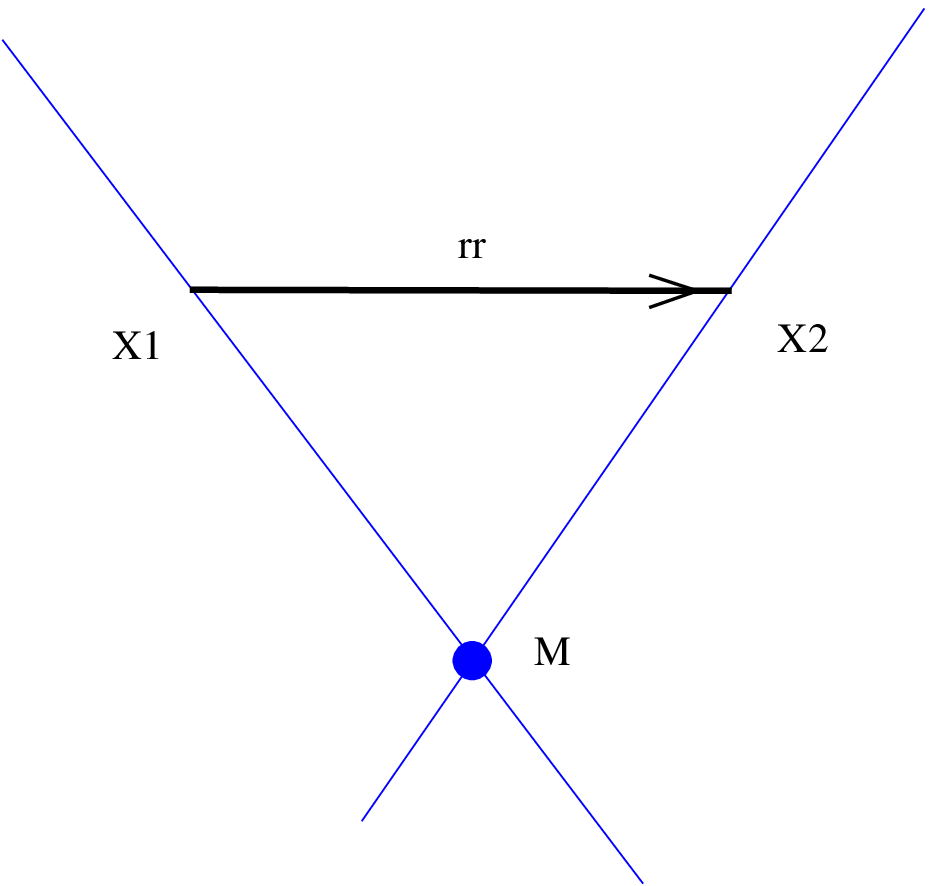}}
\caption{\small A single line and two intersecting lines in twistor space. {The arrowed line denotes the lightlike vector $X_{21}=\rho\rt$ in moduli space.}
 }\label{1and2lin}
\end{figure}

In what follows we shall intensively use a graphical representation of the twistor line configurations, as well as of the associated moduli $X-$space points. The simplest examples in Fig.~\ref{1and2lin} are those of a single line and of {two lines intersecting at the point $\mathcal{O}(\la,\mu)$.}  In this figure we have also shown, symbolically, the line parameters $X_i$,   and used the arrowed line to denote the lightlike vector $\r\rt$ between $X_1$ and $X_2$. One should not think of the $X$'s as of points on the twistor space lines (indeed, they are points in a different space, the moduli space), their positions just indicate which line they are associated with.

Now, suppose that not just two, but three twistor lines intersect at the common point ${\cal O}(\la,\mu)$,
\begin{equation}\label{3lines}
    \mu = -\bra{\la}X_1 = -\bra{\la}X_2 = -\bra{\la}X_3\,.
\end{equation}
Compatibility of these equations requires that
\begin{equation}\label{com3eq}
    X_{12}^{\a\da} = \r^\a\rt_1^\da\,, \qquad X_{23}^{\a\da} = \r^\a\rt_2^\da\,, \qquad X_{31}^{\a\da} = \r^\a\rt_3^\da\,,
\end{equation}
with a {\it common chiral} spinor $\r^\a$. Moreover, since $X_{12}+X_{23}+X_{31}=0$,\footnote{Note that in a space with Minkowski metric this would not be possible for three {\it real lightlike} vectors. The split signature $(++--)$ that we are using here allows it.} the three antichiral spinors are not linearly independent,
\begin{equation}\label{3notlin}
    \rt_1^\da+\rt_2^\da+\rt_3^\da=0\,.
\end{equation}
Clearly,  the same condition applies to any number of lines with a common intersection - the differences of the line parameters {of any pair of intersecting lines}
must be lightlike, with a common chiral spinor $\r$.\footnote{We are grateful to David Skinner for a discussion of this point.}

In general, the three intersecting lines {with the moduli \re{com3eq}} do not lie in a single plane. {As explained in \cite{Witten:2003nn}, four points in twistor space lie  in a plane (i.e., in a ${\mathbb R}{\mathbb P}^2 \subset {\mathbb R}{\mathbb P}^3$) iff the corresponding vectors $Z_i^a=(\la_i^\a,\mu_{i\,\da})$ (with $i=1,\ldots,4$) are linearly dependent.}
In other words, the condition for coplanarity is
\begin{equation}\label{concoplan}
    \det \|Z^a_i\| = \epsilon_{abcd} Z^a_1 Z^b_2 Z^c_3 Z^d_4 =0\,.
\end{equation}
Rewritten in terms of the twistor coordinates,  this condition reads (recall \p{twiin})
\begin{equation}\label{copla}
    \vev{12} [\mu_3\mu_4] + \text{[5 permutations]} = 0\,.
\end{equation}
{Coming back to the configuration of three intersecting lines with the moduli \re{com3eq},
we may test it for coplanarity \re{concoplan} by choosing, e.g., point $Z_1$ on line 1, point $Z_2$ on line 2  and points $Z_{3,4}$ on line 3. To simplify the analysis, }
we may use translation invariance to set, e.g., $X_3=0$. Together with \re{com3eq}, this implies $X_1=-\r\rt_3$ and $X_2=\r\rt_2$. In addition, the points  $Z_3$ and $Z_4$ from line 3 have vanishing antichiral coordinates, $\mu_3=\mu_4=0$. Substituting these relations into \p{copla}, we see that the only non-vanishing {term in the left-hand side of \p{copla}} is $\vev{34} [\mu_1\mu_2] =   \vev{34} \vev{1|X_1 X_2 |2 }=-\vev{34} \vev{1\r} [\rt_3 \rt_2]\vev{\r \,2} =0$. {Since $\ket{i}\equiv \lambda_i$ are arbitrary real spinors,  the condition for coplanarity is satisfied only if $\rt_2 \sim \rt_3$. }Taking into account \p{3notlin}, we conclude that three twistors lines with a common intersection point lie in a single plane iff the differences of their line parameters are {\it collinear} lightlike vectors, $X_{12}^{\a\da}\sim X_{23}^{\a\da}\sim X_{31}^{\a\da}$.

A different configuration of three twistor lines occurs when they intersect, but only pairwise. Let us study {the properties of the corresponding moduli. As before,} to each pair of intersecting lines {there} corresponds a lightlike vector,
\begin{equation}\label{toeach}
   X_{12}^{\a\da} = \r_1^\a\rt_1^\da \,, \qquad X_{23}^{\a\da} = \r_2^\a\rt_2^\da\,, \qquad X_{31}^{\a\da} = \r_3^\a\rt_3^\da\,,
\end{equation}
subject to the condition
\begin{equation}\label{sublinr}
   X_{12}^{\a\da}+X_{23}^{\a\da}+X_{31}^{\a\da}=\r_1^\a\rt_1^\da+\r_2^\a\rt_2^\da+\r_3^\a\rt_3^\da =0\,,
\end{equation}
with $\a=1,2$ and $\da=\dot 1, \dot 2$.
These four linear equations allow us to express either two of the $\r$'s in terms of the third $\r$, or two of the $\rt$'s in terms of the third $\rt$. The first solution is equivalent to  \p{com3eq} and it brings us back to the case of three lines with a common point. {The second solution is}
\begin{equation}\label{altsol}
   X_{12}^{\a\da} = \r_1^\a\rt^\da\,, \qquad X_{23}^{\a\da} = \r_2^\a\rt^\da\,, \qquad X_{31}^{\a\da} = \r_2^\a\rt^\da\,,
\end{equation}
where
\begin{equation}\label{conrh}
    \r_1^\a+\r_2^\a+\r_3^\a=0\,.
\end{equation}
Repeating the coplanarity argument above, we see that {the  three lines
with the moduli \re{altsol}} also lie in a plane. This is natural, since the vectors $X_{12}$, $X_{23}$ and $X_{31}$
form a `triangle' in {moduli space}.

\begin{figure}[h]
\psfrag{X1}[cc][cc]{$X_1$}
\psfrag{X2}[cc][cc]{$X_2$}
\psfrag{X3}[cc][cc]{$X_3$}
\psfrag{rr1}[cc][cc]{$\scriptstyle \r\,\rt_1$}
\psfrag{rr2}[cc][cc]{$\scriptstyle \r\,\rt_2$}
\psfrag{rr3}[cc][cc]{$\scriptstyle \r\,\rt_3$}
\psfrag{M}[cc][cc]{$\scriptstyle \mathcal{O}$}
\centerline{\includegraphics[height=60mm]{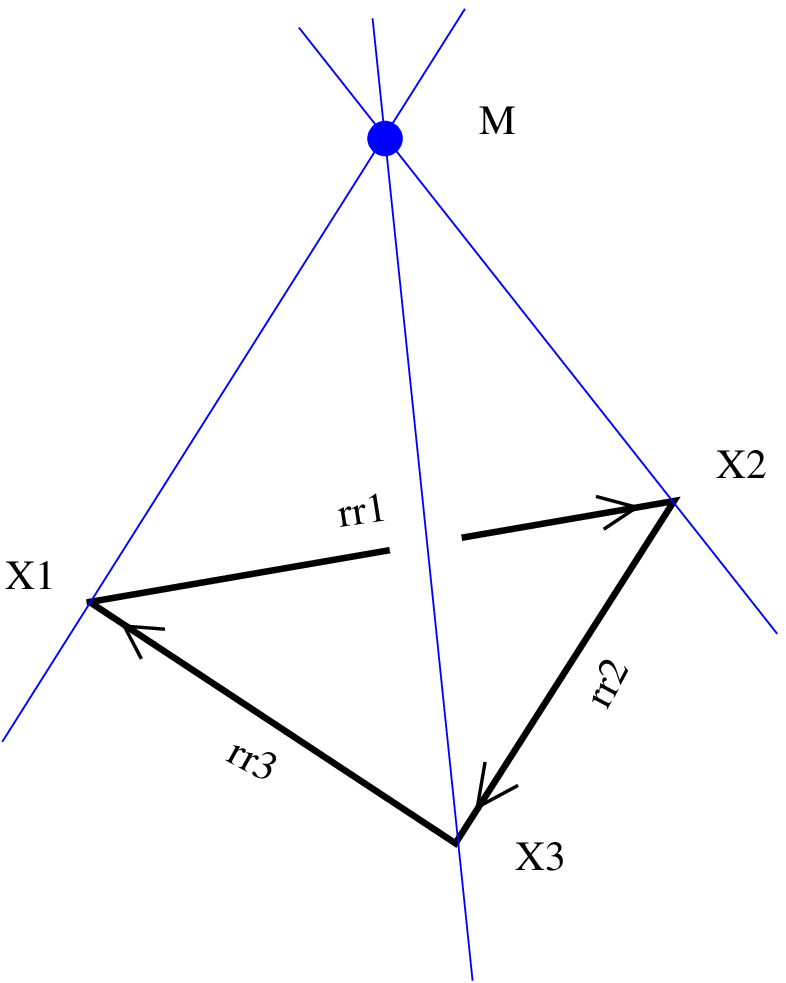} \qquad\qquad\qquad
\psfrag{rr1}[cc][cc]{$\scriptstyle \r_1\rt$}
\psfrag{rr2}[cc][cc]{$\scriptstyle \r_2\rt$}
\psfrag{rr3}[cc][cc]{$\scriptstyle \r_3\rt$}
\psfrag{M1}[cc][cc]{$\scriptstyle \mathcal{O}_1$}
\psfrag{M2}[cc][cc]{$\scriptstyle \mathcal{O}_2$}
\psfrag{M3}[cc][cc]{$\scriptstyle \mathcal{O}_3$}
\includegraphics[height=60mm]{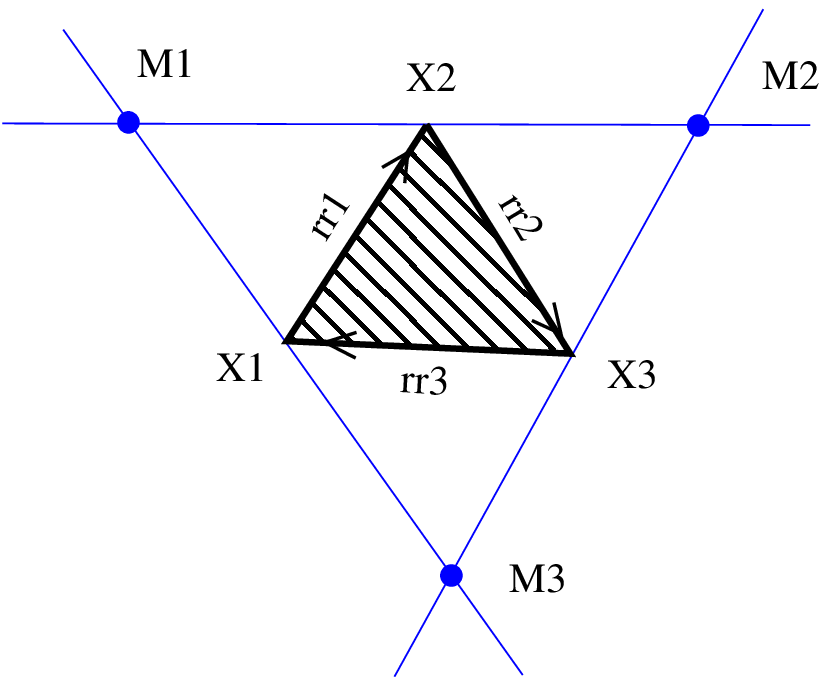}}

\caption{\small The two basic three-line configurations, planar (right-hand side part) and three-dimensional  (left-hand side part) . Blue lines are in twistor space, black lines are in moduli space.}\label{twotri}
\end{figure}

The two types of three-line configurations considered above admit the graphical representation {shown} in Fig.~\ref{twotri}. The {left-hand side part} depicts the first, non-planar configuration {in which the three twistor lines intersect} at a common point ${\cal O}$. Generically, they do not lie in a plane, but rather form a `pyramid' whose apex is at point ${\cal O}$ and whose `base' is the moduli space triangle of the first kind \p{com3eq}, i.e.  with a common chiral spinor $\r$. The lightlike separations between the moduli are shown {by arrowed lines}.

{The right-hand side part} in Fig.~\ref{twotri} describes the planar configuration {in which the three intersecting twistor lines lie in the plane of the figure.}  We  see the corresponding points $X_i$ in the moduli space,  forming a lightlike triangle of their own. In this case  the  lightlike separations of the moduli, $X_{12}$, $X_{23}$ and $X_{31}$, Eq.~\re{altsol}, share a common antichiral spinor $\rt$, as shown in the figure. The shading of the triangle indicates that it is of the second kind \p{altsol}.

\subsection{The twistor line support of the NMHV superamplitude}\label{tlsnm}

{We are now ready to elucidate the geometric interpretation of the twistor transform of the NMHV superamplitude obtained in Sect.~\ref{ttnmhv}. We recall that the MHV superamplitude analyzed in Sect.~\ref{mmhhvv} is localized on  a single supertwistor line, that is, all points $(\la_i,\mu_i,\psi_i)$ in the (super)twistor space belong to the same line
parameterized by the moduli $X^{\a\da}$ and $\Theta^{\a}_A$. According to \p{nmhvsu} and \re{sugg},  each term in the expression for the NMHV superamplitude has support on a set of {\it three} supertwistor lines, with the $n$ points distributed over them. Like in the MHV case, the three lines have a common set of moduli, $X^{\a\da}$ and $\Theta^{\a}_A$, which are needed to ensure the vanishing of the total momentum and supercharge of the superamplitude. A novel feature of the NMHV superamplitude is that its twistor
transform depends on the new even variables $\r$, $\tilde\rho$ and on the new odd variable $\xi$. They appear in the expressions for moduli of the three lines. Namely, one of the lines  has the moduli $X^{\a\da}_{1} = X^{\a\da} -\r_0^\a\tilde\rho^{\da}$ and $\Theta^\a_{1\, A} = \Theta^{\a}_A - \r_0^\a\xi_A $ (with $\r_0=\lambda_n$), and it contains
the cluster of particles with indices $i\in [1,a-1]$ (in fact, $[n,a-1]$, see \p{nand1}. The second line
has the moduli $X^{\a\da}_{2} = X^{\a\da} +\sigma^\a\tilde\rho^{\da}$ and $\Theta^\a_{2\, A} = \Theta^{\a}_A + \sigma^\a\xi_A $ (with $\sigma=-\r-\r_0$), and it
contains the cluster of particles $[a,b-1]$. Finally, the third line is parametrized by the MHV moduli $X^{\a\da}_{3} = X^{\a\da}$ and $\Theta^{\a}_{3\, A} = \Theta^{\a}_A$,
and it contains the cluster of particles $[b,n]$. Notice that the particle with index $n$
belongs to two lines simultaneously, and that the sum over the indices in \re{nmhvsu} is such that each line contains
at least two points. }

Another important observation is that the three lines above intersect pairwise and, hence, lie in a plane, as follows from our discussion in Sect.~\ref{ltsp} (see \p{altsol}). Looking at \p{arr2}, we see that this is precisely the case for the three lines supporting the term $\cA_{nab}$ of the NMHV superamplitude \p{nmhvsu}. The (bosonic) twistor space coordinates  $(\la,\mu)$ of the three intersection points are as follows:\footnote{Here we have used the translation invariance to fix the common line parameter at $X=X_3=0$.}
\begin{eqnarray}
  (\mbox{line 1})\ \times\ (\mbox{line 2}) &=& (\r, -\vev{\r_0\r}\rt)\,,  \nn\\
  (\mbox{line 2})\ \times\ (\mbox{line 3}) &=& (\sigma, 0)\,, \nn\\
  (\mbox{line 3})\ \times\ (\mbox{line 1}) &=& (\r_0,0)\,. \label{intersec}
\end{eqnarray}
{As was just mentioned,
lines 1 and 3 intersect at the point corresponding to the $n-$th particle.}
At the same time, the intersection of lines 1 and 2 and of 2 and 3 are `moving points' with chiral twistor coordinates $\r$ and $\sigma$, respectively. They do not coincide with any of the incoming particle momenta. Instead, they serve as `moduli', i.e. they are integrated over in the twistor transform \p{sugg}.
The set of integration parameters $(X^{\da\a}$, $\rho^\a$, $\tilde\rho^{\da}$, $\Theta^{\a}_A$, $\xi_A)$ constitute the `moduli space' of the amplitude in twistor space.

\begin{figure}[!h]
\psfrag{X1}[cc][cc]{$X_1$}\psfrag{X}[cc][cc]{$X$}
\psfrag{X2}[cc][cc]{$X_2$}
\psfrag{X3}[cc][cc]{$X_3$}
\psfrag{a-1}[cc][cc]{$\scriptstyle a-1$}\psfrag{a}[cc][cc]{$\scriptstyle a$}
\psfrag{b-1}[cc][cc]{$\scriptstyle b-1$}\psfrag{b}[cc][cc]{$\scriptstyle b$}
\psfrag{1}[cc][cc]{$\scriptstyle 1$}\psfrag{n}[cc][cc]{$\scriptstyle n$}
\psfrag{n-1}[cc][cc]{$\scriptstyle n-1$}
\centerline{\includegraphics[width=170mm]{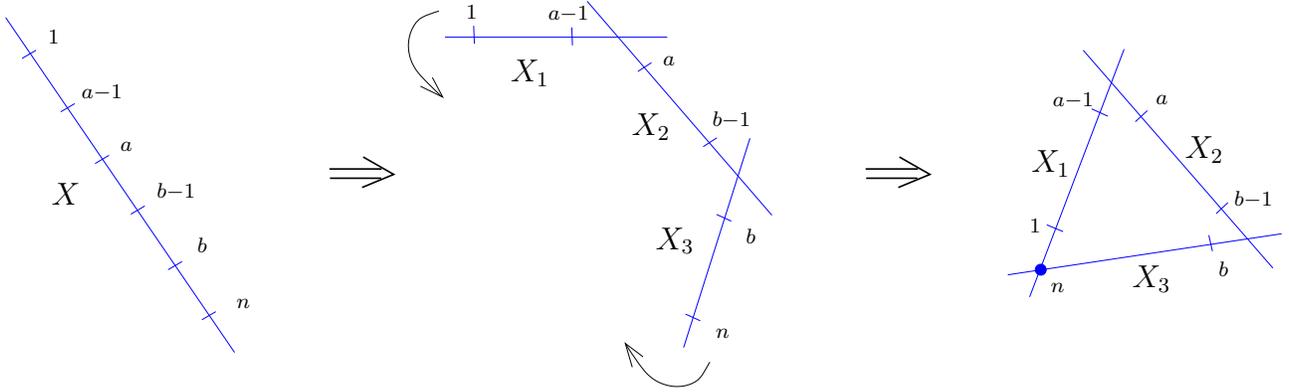}}
\caption{\small Transforming a single line into three lines. }
\label{3-lines}
\end{figure}

The twistor transform of the NMHV amplitude \re{sugg} admits a very simple and suggestive graphical description, as shown in Fig.~\ref{3-lines}. Imagine the $n$ particles of the MHV amplitude as points on a single straight line.  Next, we choose four points labeled by $a-1$, $a$, $b-1$ and $b$, so that we split the line into three segments. Then we  `bend' the segments $1 = [1, a-1]$ and $3 = [b,n]$ in such a way that they remain in the same plane with the line carrying the segment $2 = [a, b-1]$ and that
the intersection point of lines 1 and 3 is  the point with chiral twistor coordinate $\r_0\equiv \la_n$. Thus, we have transformed the initial line into a triangle. The amount of `flexing' at each bending point is determined by the lightlike vectors $X_{12}$ and $X_{23}$, while the `angle' at point $n$ is given by the third lightlike vector $X_{13} = X_{12} + X_{23}$.

\begin{figure}[!h]
\psfrag{X1}[cc][cc]{$X_3$}
\psfrag{X2}[cc][cc]{$X_1$}
\psfrag{X3}[cc][cc]{$X_2$}
\psfrag{a1}[cc][cc]{$a$}\psfrag{a}[cc][cc]{$a-1$}
\psfrag{b1}[cc][cc]{$b$}\psfrag{b}[cc][cc]{$b-1$}
\psfrag{1}[cc][cc]{$1$}\psfrag{n}[cc][cc]{$n$}
\psfrag{rr1}[cc][cc]{$\sigma\rt$}
\psfrag{rr2}[cc][cc]{$\r_0\rt$}
\psfrag{rr3}[cc][cc]{$\r\rt$}
\centerline{\includegraphics[width=80mm]{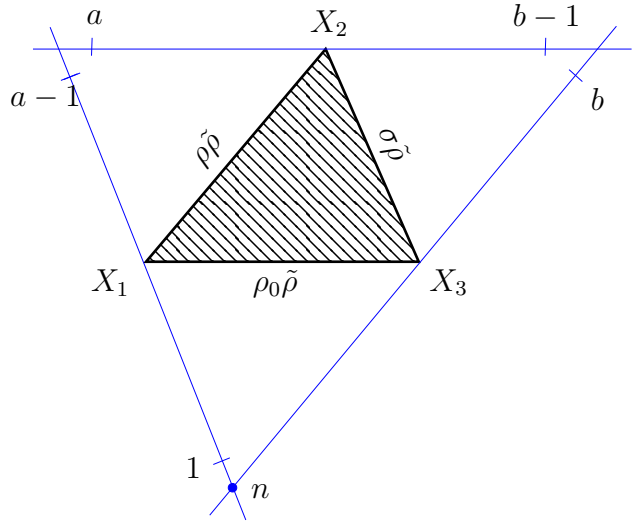}}
\caption{\small NMHV: The three twistor lines lie in a plane. The inscribed moduli space figure is a shaded triangle.  }
\label{basictri}
\end{figure}

Note that the points can be situated anywhere on a given line, not necessarily on the sides of the triangle (as we have indicated in the figure), but also outside. The actual `ordering' of the points along the line is irrelevant.\footnote{According to \p{twline}, ordering the points would require ordering the free parameters $z_3$, which would imply restrictions on the chiral twistor variables $\la_i$.}  The fact that we show point $a-1$ after (clockwise) point 1 on line 1, or point $a$ before $b-1$ on line 2, or point $b$ before point $n$ on line 3 is just a convention which reminds us that $1<a<b < n$. What matters here and in the subsequent pictures is that some set of points are localized on a given line.

In Fig.~\ref{basictri} we have shown the three intersecting twistor lines, together with the line parameters $X_{1,2,3}$. These points in moduli space form a triangle of their own, with lightlike sides sharing the same antichiral spinor $\rt$, as indicated in the figure. This is an example of the second basic three-line configuration from Fig.~\ref{twotri}.

The above geometric picture has to be complemented with a prescription how to modify the MHV denominator. The new denominator $\Delta_{nab}$ in \p{sugg} and \p{delta0} looks like an MHV denominator in which we have inserted two extra points, $\rho$ between $a-1$ and $a$ and $\sigma$ between $b-1$ and $b$. The justification of this insertion was given in Sect.~\ref{scpro} where we studied the conformal properties of the amplitude.  The insertion is somewhat reminiscent of the `reference spinor' procedure of CSW \cite{Cachazo:2004kj}, but the crucial difference here is that the Lorentz covariance is not broken as we {\it integrate} over the moduli spinor $\rho$.

As we show in the next section, the repeated application of this simple procedure generates all the terms in all tree amplitudes of the N${}^k$MHV type. The generalization of the picture in Fig.~\ref{basictri} is a set of $2k+1$ intersecting lines in twistor space, in which a lightlike  moduli space $(2k+1)-$gon  is inscribed. The latter is {\it triangulated} into lightlike triangles of both kinds described in Sect.~\ref{ltsp}, following a set of very simple rules.

\section{All non-MHV superamplitudes}\label{allnmhvsu}

The step from the twistor transform of the MHV superamplitude to that of the NMHV one was succinctly summarized by the very simple geometric idea formulated in Sect.~\ref{tlsnm}: We broke the single MHV line up into three lines, forming a triangle. In this section we show that repeating this procedure, i.e. making three lines out of one of the lines of the preceding configuration, we can recursively construct the whole sequence of N${}^k$MHV superamplitudes. The result of this geometric construction is in one-to-one correspondence with the solution of the supersymmetric BCFW recursion relations found in \cite{Drummond:2008cr}.

\subsection{N${}^2$MHV superamplitude}\label{nnmhvsupp}

{As follows from \re{An-dec}},
a characteristic feature of the N${}^k$MHV superamplitudes is that they are homogeneous polynomials of degree $4k+8$ in the Grassmann variables $\eta$.
Writing the amplitude in a factorized form with the MHV superamplitude as a prefactor (see \p{nmhvsu}),
\be\label{nnmmhhvv}
\cA_n^{\rm N^kMHV} =  \cA_n^{\rm MHV} R_n^{\rm N^kMHV}\,,
\ee
we find that the `ratio' $R_n^{\rm N^kMHV}$ is a homogeneous {polynomial in $\eta$} of degree $4k$. In the NMHV case ($k=1$), this polynomial is the sum of all superinvariants $R_{nab}$, Eq~\p{mbubl}. In the N${}^2$MHV case ($k=2$) it is given by a sum of products of  superinvariants {of a slightly modified form, as compared to
$R_{nab}$.  In fact, the expression for the N${}^2$MHV tree superamplitude is given by
a sum of four terms \cite{Drummond:2008cr},}
\begin{align}\label{4-terms}
\cA_n^{\rm N^2MHV} = \cA_n^{\rm (A)} + \cA_n^{\rm (B)} +  \cA_n^{\rm (A), deg} + \cA_n^{\rm (B),deg}\,,
\end{align}
{where  the last two terms can be viewed as degenerate cases of the first two.
Below we describe each of these terms and their twistor transforms.}

\subsubsection{Twistor transform of the term $\cA_n^{\rm (A)}$}\label{ttnnsu}

{The term $\cA_n^{\rm (A)}$  has the form }
\begin{equation}\label{defAnab}
    \cA_n^{\rm (A)} = \sum_{3 \leq a_1+1 < b_1 \leq n-1}\quad \sum_{a_1 +2 \leq a_2+1 < b_2 \leq b_1-1} \ \cA_{na_1 b_1a_2 b_2}^{\rm (A)}\,,
\end{equation}
where
\begin{equation}\label{twoR}
  \cA_{na_1 b_1a_2 b_2}^{\rm (A)} = \cA_n^{\rm MHV}\times  R_{n a_1 b_1} \times R_{na_1 b_1a_2 b_2}\,,
\end{equation}
with $R_{n a_1 b_1}$ defined in \re{mbubl}, and
the new superinvariant given by
\begin{equation}\label{newR}
    R_{na_1 b_1a_2 b_2} = \frac{\vev{a_2-1\, a_2} \vev{b_2-1\, b_2}\ \delta^{(4)}(\sum_{a_1}^{a_2-1} \vev{\r_2\, i}  \eta_i + \sum_{a_1}^{b_2-1} \vev{\sigma_2\, i} \eta_i)}{x^2_{a_2 b_2}\vev{\r_2\, a_2-1}\vev{\r_2\, a_2}\vev{\sigma_2\, b_2-1}\vev{\sigma_2\, b_2}}\,.
\end{equation}
Here we have used the new spinors $\r_2$ and $\sigma_2$ which are defined in a way similar to the old $\r$ and $\sigma$ from \p{defr}:
\begin{eqnarray}
\lan{\r_0} = \bra{n} &&  \nn \\
  \lan{\r_1} = \lan{n} x_{n b_1} x^{-1}_{b_1 a_1}\,, && \lan{\sigma_1} = \lan{n} x_{n a_1} x^{-1}_{a_1 b_1}\,, \qquad  \lan{\r_1} + \lan{\sigma_1} + \lan{\r_0} = 0 \nn \\
   \lan{\r_{2}} = \lan{\r_1}x_{a_1 b_2} x^{-1}_{b_2a_2}\,, &&  \lan{\sigma_{2}} = \lan{\r_1}x_{a_1 a_2} x^{-1}_{a_2 b_2}\,, \qquad    \lan{\r_{2}} + \lan{\sigma_{2}} + \lan{\r_{1}} = 0\,. \label{defr2}
\end{eqnarray}

The close similarity between \p{newR} and the NMHV superinvariant \p{mbubl} allows us to carry out the twistor transform \re{T}  of each term in the sum \p{twoR} in exactly the same way as before.\footnote{As in the NMHV case, we replace the factor $x^2_{a_2 b_2}$ in the denominator of \p{newR} by $|x^2_{a_2 b_2}|$. This means restricting the original amplitude to a special kinematic region, away from the physical singularities.} The definitions of $\r_1$ and $\r_2$ from \p{defr2} are introduced via delta functions, which are cast into Fourier form with the help of the conjugate variables $\rt_1$ and $\rt_{2}$. The Grassmann delta functions from $R_{na_1 b_1}$ and from $R_{na_1 b_1a_2 b_2}$ are cast into Fourier form with conjugate variables $\xi_1$ and $\xi_{2}$. After this the Fourier integration gives rise to a product of delta functions. So, the contribution of each term in the sum \p{twoR} is
\begin{equation}\label{sugg'}
 T\left[\cA_{na_1 b_1a_2 b_2}^{\rm (A)}\right]   = i    \int d^4X d^2\r_1 d^2\rt_1 d^2\r_{2} d^2\rt_{2} d^8\Theta d^4\xi_1 d^4\xi_{2} \ \frac{(\prod)^{\rm (A)}}{\Delta^{\rm (A)}}\,,
\end{equation}
where
\begin{eqnarray}
  \Delta^{\rm (A)} &=& \vev{12}\ldots \vev{a_1-1\, \r_1} \vev{\r_1\, a_1} \ldots \vev{a_2-1\, \r_{2}} \vev{\r_{2}\, a_2} \ldots  \nn\\
  & \times & \vev{b_2-1\, \sigma_{2}} \vev{\sigma_{2}\, b_2} \ldots  \vev{b_1-1\, \sigma_1} \vev{\sigma_1\, b_1}  \ldots \vev{n1} \,,   \label{delta2}
\end{eqnarray}
and
\begin{eqnarray}
   (\prod)^{\rm (A)} &=& \prod_{1}^{a_1-1} \delta^{(2)}(\mu_i + \lan{i} X_1)\ \delta^{(4)}(\psi_i + \vev{i\, \Theta_1})\nn\\
  &\times& \prod_{a_1}^{a_2-1} \delta^{(2)}(\mu_i + \lan{i} X_2)\ \delta^{(4)}(\psi_i + \vev{i\, \Theta_2})\nn\\
  &\times& \prod_{a_2}^{b_2-1} \delta^{(2)}(\mu_i + \lan{i} X_3)\ \delta^{(4)}(\psi_i + \vev{i\, \Theta_3})\nn\\
  &\times& \prod_{b_2}^{b_1-1} \delta^{(2)}(\mu_i + \lan{i} X_4)\ \delta^{(4)}(\psi_i + \vev{i\, \Theta_4})\nn\\
  &\times& \prod_{b_1}^{n}\delta^{(2)}(\mu_i + \lan{i} X_5)\ \delta^{(4)}(\psi_i + \vev{i\, \Theta_5}) \label{2other}
\end{eqnarray}
with
\begin{equation}\label{arr22}
    \begin{array}{lcl}
       X_1 = X - \r_0 \rt_1 & \quad & \Theta_1 = \Theta - \r_0\xi_1 \\
       X_2 = X +\sigma_1\rt_1 - \r_1 \rt_{2} & \quad & \Theta_2 = \Theta +\sigma_1\xi_1- \r_1\xi_{2} \\
       X_3 = X +\sigma_1\rt_1 +\sigma_{2}\rt_{2} & \quad & \Theta_3 = \Theta +\sigma_1\xi_1 + \sigma_{2}\xi_{2} \\
       X_4 = X +\sigma_1\rt_1 & \quad & \Theta_4 = \Theta +\sigma_1\xi_1 \\
       X_5 = X & \quad & \Theta_5 = \Theta \,.
     \end{array}
\end{equation}

\subsubsection{Geometric interpretation of the twistor transform}

Comparing \p{2other} with \p{2}, we see that the amplitude is now supported on five lines with parameters \p{arr22}, involving the new independent moduli $\r_2$, $\rt_2$ and $\xi_2$. Next, as we did in the NMHV case, we rewrite \p{arr22} in terms of differences,
\begin{equation}\label{arr23}
    \begin{array}{lcl}
       X_{12} = \r_1(\rt_1+\rt_2) & \quad & \Theta_{12} = \r_1(\xi_1+\xi_2) \\
       X_{23} = \r_2\rt_2 & \quad & \Theta_{23} = \r_2\xi_2\xi\\
       X_{34} = \sigma_2 \rt_2 & \quad & \Theta_{34} = \sigma_2\xi_2 \\
       X_{45} = \sigma_1 \rt_1 & \quad & \Theta_{45} = \sigma_1\xi_1 \\
       X_{51} = \r_0 \rt_1 & \quad & \Theta_{51} = \r_0\xi_1 \ .
     \end{array}
\end{equation}
We verify that $\sum_{u=1}^5 X_{u\, u+1}=\sum_{u=1}^5 \Theta_{u\, u+1}=0$, as a corollary of the linear relations \p{defr2} between $\r$ and $\sigma$. Once again, we see that the separations between the adjacent points in moduli space are lightlike, $X_{u\, u+1}^2=0$. This means that the five twistor lines intersect pairwise, that is, line 1 with 2, line 2 with 3, etc.

\begin{figure}[h]
\psfrag{X1}[cc][cc]{$X_5$}
\psfrag{X2}[cc][cc]{$X_1$}
\psfrag{X3}[cc][cc]{$X_2$}
\psfrag{X4}[cc][cc]{$X_3$}\psfrag{X5}[cc][cc]{$X_4$}
\psfrag{a1}[cc][cc]{$\scriptstyle a_1$}\psfrag{a2}[cc][cc]{$\scriptstyle a_2$}
\psfrag{b1}[cc][cc]{$\scriptstyle b_1$}\psfrag{b2}[cc][cc]{$\scriptstyle b_2$}
\psfrag{1}[cc][cc]{$\scriptstyle 1$}\psfrag{n}[cc][cc]{$\scriptstyle n$}
\psfrag{rr1}[cc][cc]{$\scriptstyle \r_0\rt_1$}
\psfrag{rr2}[cc][cc]{$\scriptstyle \rho_1(\rt_1+\rt_2)$}
\psfrag{rr3}[cc][cc]{$\scriptstyle \r_2\rt_2$}
\psfrag{rr4}[cc][cc]{$\scriptstyle \sigma_2\rt_2$}
\psfrag{rr5}[cc][cc]{$\scriptstyle \sigma_1\rt_1$}
\psfrag{rr6}[cc][cc]{$\scriptstyle \r_1\rt_2$}
\psfrag{rr7}[cc][cc]{$\scriptstyle \r_1\rt_1$}
\psfrag{M}[cc][cc]{$\scriptstyle \mathcal{O}$}
\centerline{\includegraphics[width=90mm]{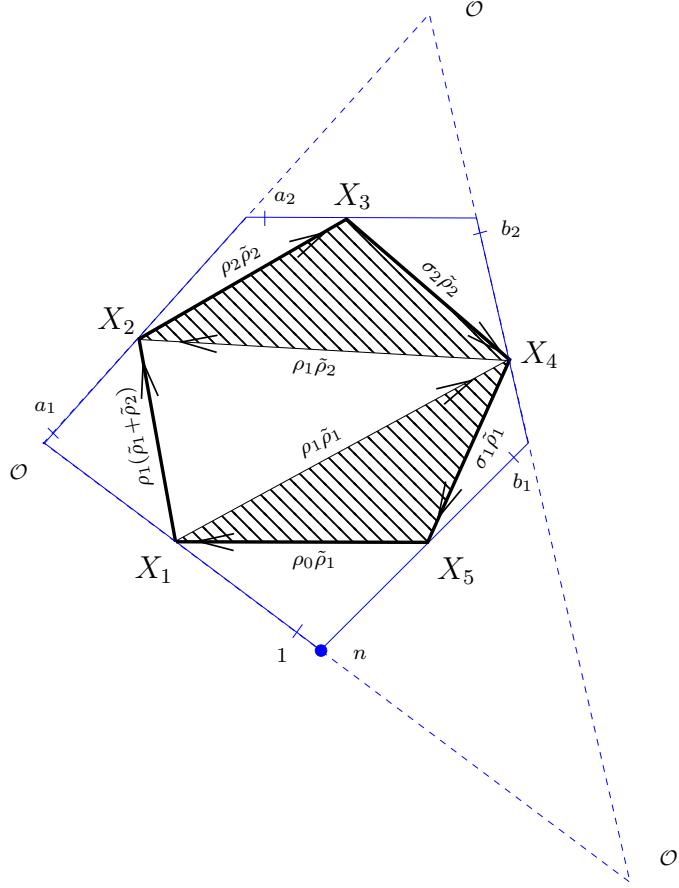}}
\caption{\small N${}^2$MHV: {Two-dimensional image of the pentagon corresponding
to the term  (A) in Eq.~\re{4-terms}.}
 }
\label{5a2d}
\end{figure}

{The new line configuration} corresponding to \re{arr23} is depicted in Fig.~\ref{5a2d} in the form of a `pentagon' in twistor space. {We can equivalently describe the same configuration
as  another `pentagon' in moduli space, whose vertices are identified with the line parameters $X_u$.}
{All the sides of the moduli space pentagon are lightlike, as well as two of its diagonals, $X_{14} = \r_1\rt_1$ and $X_{42}= \r_1\rt_2$.} Thus, the moduli space pentagon is `triangulated' into two types of triangles. Each triangle is made of three lightlike vectors summing up to zero. As explained in Sect.~\ref{gitt}, such vectors admit two alternative representations in terms of pairs of spinors, with a common chiral or antichiral spinor. The triangle $X_1 X_2 X_4$  in Fig.~\ref{5a2d} is of the first type, with the common chiral spinor $\r_1$. The two other triangles, $X_1 X_4 X_5$ and $X_2 X_3 X_4$, are of the second type, having the common spinors $\rt_1$ and $\rt_2$, respectively.  Like the similar triangle $X_1 X_2 X_3$ in Fig.~\ref{basictri}, they are shaded in the figure.

The fact that the moduli space points $X_1$ and $X_4$, as well as points $X_2$ and $X_4$, are lightlike separated, implies that the twistor space lines 1 and 4, as well as lines 2 and 4 intersect, as indicated in the figure by dashed lines. These lines pass through the vertices
{of the triangle $X_1X_2X_4$, i.e.  a triangle of the first type with a common chiral spinor $\r_1$.} As explained in Sect.~\ref{gitt}, this implies that the three lines 1, 2 and 4 intersect at the same point $\mathcal{O}$ with twistor coordinates $(\lambda = \r_1, \mu=0)$ {(here, as usual, we have used translation invariance to set the common modulus at $X \equiv X_5=0$).}

This shows that the two-dimensional picture in Fig.~\ref{5a2d} is not exact. {Indeed,
we now understand that the three  moduli space triangles $X_1X_2X_4$, $X_2X_3X_4$ and $X_1X_4X_5$ lie in three different planes.\footnote{To see that the moduli space `pentagon' is not flat, it sufficient to show that any three of its sides are defined by linearly independent vectors, hence they are not coplanar. For instance, the sides $X_{12}$, $X_{23}$ and $X_{34}$ are made of the three linearly independent vectors $\r_1\rt_1$, $\r_2\rt_2$ and $\r_1\rt_2$.} The same is true for the five twistor lines,
and the three intersection points denoted by $\mathcal{O}$  in Fig.~\ref{5a2d} have to
be identified.}

\begin{figure}
\psfrag{X1}[cc][cc]{$X_1$}
\psfrag{X2}[cc][cc]{$X_2$}
\psfrag{X3}[cc][cc]{$X_3$}
\psfrag{X4}[cc][cc]{$X_4$}\psfrag{X5}[cc][cc]{$X_5$}
\psfrag{a1}[cc][cc]{$\scriptstyle a_1$}\psfrag{a2}[cc][cc]{$\scriptstyle a_2$}
\psfrag{b1}[cc][cc]{$\scriptstyle b_1$}\psfrag{b2}[cc][cc]{$\scriptstyle b_2$}
\psfrag{1}[cc][cc]{$\scriptstyle 1$}
\psfrag{M}[cc][cc]{$\scriptstyle \mathcal{O}$}
\centerline{\includegraphics[width=90mm]{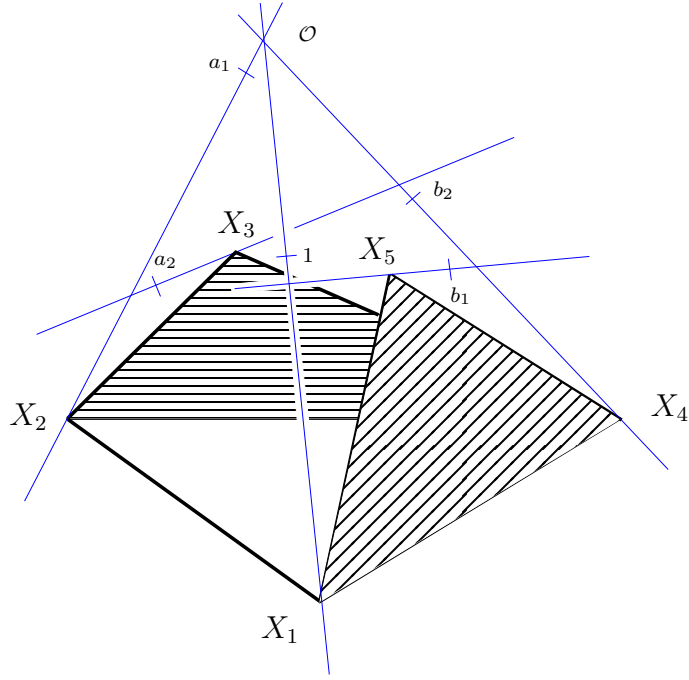}}
\caption{\small N${}^2$MHV:  {Three-dimensional image of the pentagon shown in Fig.~\ref{5a2d}.}}
\label{A3d1}
\end{figure}

A more accurate, three-dimensional representation of the pentagon configuration \re{arr23} is shown in Fig.~\ref{A3d1}. In it we clearly see the three twistor lines intersecting at point $\mathcal{O}$. This point is the tip of a pyramid whose base is the white triangle $X_1 X_2 X_4$ (compare with the first diagram in Fig.~\ref{twotri}). The two shaded triangles are now shown on two of the faces of the pyramid (the  upward or downward orientation is  purely conventional). We can think of the two-dimensional Fig.~\ref{5a2d} as being obtained by opening up the pyramid at the tip and slicing down along the edges, and then unfolding the three faces until they lie in the plane of the white triangle $X_1 X_2 X_4$. Thus, the common intersection point of the three lines splits up into three points shown in Fig.~\ref{5a2d}. The three-dimensional pictures become more difficult to draw in the cases N${}^k$MHV with $k>2$, so we shall systematically use the simplified two-dimensional representation. It can be generated, for any value of $k$, following a set of very simple rules, as we shall see in Sect.~\ref{geca}.

To conclude the discussion of the  term \p{twoR}, we comment on the structure of the denominator \p{delta2}. We see that the new variables $\r_2$ and $\sigma_2$ are inserted between the points $a_2-1, a_1$ and $b_2-1,b_2$, respectively. As in the NMHV case, this is determined by the combination of ordinary and dual conformal invariance. The ordinary conformal properties of these spinors are correlated with those of the end points of the moduli space segments $X_{23}$ and $X_{34}$ (see Fig.~\ref{5a2d}). Thus, $\r_2$ shares the same conformal properties with all points on the twistor lines 2 and 3, and $\sigma_2$ with the lines 3 and 4. Then dual conformal invariance requires that these spinors be contracted with the spinors at the adjacent points $a_2-1, a_1$ and $b_2-1,b_2$, respectively.

\subsubsection{Remaining terms}

Besides the term (A) {defined in}   \p{twoR}, the N${}^2$MHV superamplitude \p{nnmmhhvv} involves three other terms. Term (B) {is} a modified version of \p{twoR}, where the second superinvariant is of the same type as the first, but with different labels,
\begin{equation}\label{defBnab}
    \cA_n^{\rm (B)} = \sum_{3 \leq a_1+1 < b_1 \leq n-1} \quad \sum_{b_1 +2 \leq a_2+1 < b_2 \leq n-1}\ \cA^{\rm (B)}_{na_1 b_1a_2 b_2}\,,
\end{equation}
with
\begin{equation}\label{twoR'}
     \cA^{\rm (B)}_{na_1 b_1a_2 b_2} = \cA_{MHV}\times  R_{n a_1 b_1} \times R_{n a_2 b_2}\,.
\end{equation}
As in Sect.~\ref{ttnnsu}, we introduce the independent spinor variables
\begin{equation}\label{auva3}
    \lan{\r_1} = \lan{n} x_{n b_1} x^{-1}_{b_1 a_1}\,, \qquad \lan{\r_{2}} = \lan{n}x_{nb_2} x^{-1}_{b_2 a_2}\,,
\end{equation}
as well as their complements $\sigma_1 = -\r_1-\r_0$, $\sigma_2 = -\r_2-\r_0$. Then we find the twistor transform
\begin{equation}\label{sugg5}
T\left[\cA^{\rm (B)}_{na_1 b_1a_2 b_2}\right] = i\int d^4X d^2\r_1 d^2\rt_1 d^2\r_{2} d^2\rt_{2} d^8\Theta d^4\xi_1 d^4\xi_{2} \ \frac{(\prod)^{\rm (B)}}{\Delta^{\rm (B)} }
\end{equation}
with
\begin{eqnarray}
  \Delta^{\rm (B)} &=& \vev{12}\ldots \vev{a_1-1\, \r_1} \vev{\r_1\, a_1} \ldots \vev{b_1-1\, \sigma_1} \vev{\sigma_1\, b_1} \ldots  \nn\\
  & \times & \vev{a_2-1\, \r_{2}} \vev{\r_{2}\, a_2} \ldots \vev{b_2-1\, \sigma_{2}} \vev{\sigma_{2}\, b_2} \ldots   \vev{n1} \,,   \label{delta5}
\end{eqnarray}
and
\begin{eqnarray}
  (\prod)^{\rm (B)} &=& \prod_{1}^{a_1-1} \delta^{(2)}(\mu_i + \lan{i} X_1)\ \delta^{(4)}(\psi_i + \vev{i\, \Theta_1}) \nn \\
   &\times& \prod_{a_1}^{b_1-1} \delta^{(2)}(\mu_i + \lan{i} X_2)\ \delta^{(4)}(\psi_i + \vev{i\, \Theta_2}) \nn \\
  &\times&  \prod_{b_1}^{a_2-1} \delta^{(2)}(\mu_i + \lan{i} X_3)\  \delta^{(4)}(\psi_i + \vev{i\, \Theta_3}) \nn\\
  &\times&  \prod_{a_2}^{b_2-1} \delta^{(2)}(\mu_i + \lan{i} X_4)\  \delta^{(4)}(\psi_i + \vev{i\, \Theta_4}) \nn\\
  &\times& \prod_{b_2}^{n}\delta^{(2)}(\mu_i + \lan{i} X_5)\ \delta^{(4)}(\psi_i + \vev{i\, \Theta_5})\ ,
\end{eqnarray}
where
\begin{equation}\label{arr'3}
    \begin{array}{lcl}
       X_1 = X - \r_0 (\rt_1 + \rt_{2}) & \quad & \Theta_1 = \Theta - \r_0(\xi_1 + \xi_{2}) \\
       X_2 = X +\sigma_1\rt_1 - \r_0 \rt_{2} & \quad & \Theta_2 = \Theta +\sigma_1\xi_1- \r_0\xi_{2} \\
       X_3 = X - \r_0 \rt_{2} & \quad & \Theta_3 = \Theta  - \r_0\xi_{2} \\
       X_4 = X  +\r_{2}\rt_{2} & \quad & \Theta_4 = \Theta  + \r_{2}\xi_{2} \\
       X_5 = X & \quad & \Theta_5 = \Theta\ .
     \end{array}
\end{equation}

\begin{figure}[h]
\psfrag{X1}[cc][cc]{$X_5$}
\psfrag{X2}[cc][cc]{$X_1$}
\psfrag{X3}[cc][cc]{$X_2$}
\psfrag{X4}[cc][cc]{$X_3$}\psfrag{X5}[cc][cc]{$X_4$}
\psfrag{a1}[cc][cc]{$\scriptstyle a_1$}\psfrag{a2}[cc][cc]{$\scriptstyle a_2$}
\psfrag{b1}[cc][cc]{$\scriptstyle b_1$}\psfrag{b2}[cc][cc]{$\scriptstyle b_2$}
\psfrag{1}[cc][cc]{$\scriptstyle 1$}
\psfrag{rr1}[cc][cc]{$\scriptstyle \rho_0(\rt_1+\rt_2)$}
\psfrag{rr2}[cc][cc]{$\scriptstyle \r_1\rt_1$}
\psfrag{rr3}[cc][cc]{$\scriptstyle \sigma_1\rt_1$}
\psfrag{rr4}[cc][cc]{$\scriptstyle \r_2\rt_2$}
\psfrag{rr5}[cc][cc]{$\scriptstyle \sigma_2\rt_2$}
\psfrag{rr6}[cc][cc]{$\scriptstyle \r_0\rt_1$}
\psfrag{rr7}[cc][cc]{$\scriptstyle \r_0\rt_2$}
\psfrag{M}[cc][cc]{$\scriptstyle \mathcal{O}$}
\centerline{\includegraphics[width=160mm]{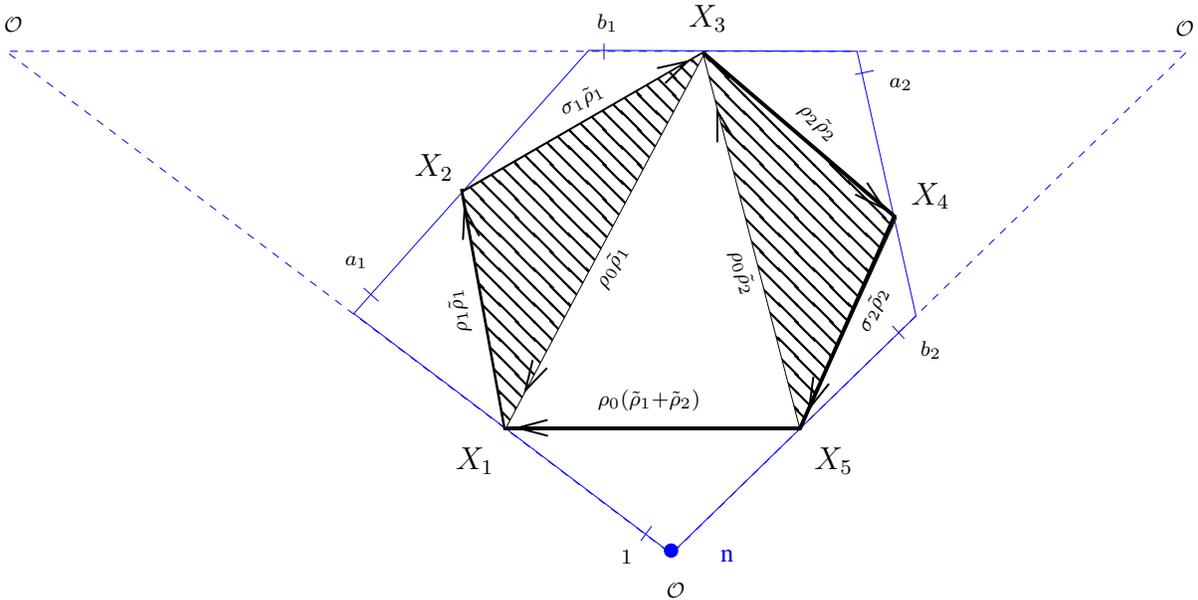}}
\caption{\small N${}^2$MHV: Pentagon (B) in two dimensions. }
\label{5a2dB}
\end{figure}

The geometric representation of this contribution to the N${}^2$MHV superamplitude is shown in Fig.~\ref{5a2dB}. It resembles very closely  Fig.~\ref{5a2d}, up to a rotation of the inscribed moduli space pentagon. Once again, the twistor lines 1, 3 and 5 intersect at the same point $\mathcal{O}$ {with the twistor coordinates}
 $(\lambda=  \la_n, \mu=0)$, as follows from the presence of the common spinor $\r_0\equiv \la_n$ in the white triangle.

\begin{figure}[h]
\psfrag{X1}[cc][cc]{$X_5$}
\psfrag{X2}[cc][cc]{$X_1$}
\psfrag{X3}[cc][cc]{$X_2$}
\psfrag{X4}[cc][cc]{$X_3$}\psfrag{X5}[cc][cc]{$X_4$}
\psfrag{a1}[cc][cc]{$\scriptstyle a_1$}\psfrag{a2}[cc][cc]{$\scriptstyle a_2$}
\psfrag{b1}[cc][cc]{$\scriptstyle b_1$}\psfrag{b2}[cc][cc]{$\scriptstyle b_2$}
\psfrag{1}[cc][cc]{$\scriptstyle 1$}\psfrag{n}[cc][cc]{$\scriptstyle n$}
\psfrag{rr1}[cc][cc]{$\scriptstyle\rho_0 \rt_1 $}
\psfrag{rr2}[cc][cc]{$\scriptstyle\r_1(\rt_1+\rt_2)$}
\psfrag{rr3}[cc][cc]{$\scriptstyle\r_2\rt_2$}
\psfrag{rr4}[cc][cc]{$\scriptstyle\sigma_2\rt_2$}
\psfrag{rr5}[cc][cc]{$\scriptstyle\sigma_1\rt_1$}
\psfrag{rr6}[cc][cc]{  $\scriptstyle\r_1\rt_2$}
\psfrag{rr7}[cc][cc]{$\scriptstyle\r_1\rt_1$}
\psfrag{bb1}[cc][cc]{$\scriptstyle b_1=b_2$}
\centerline{\includegraphics[width=160mm]{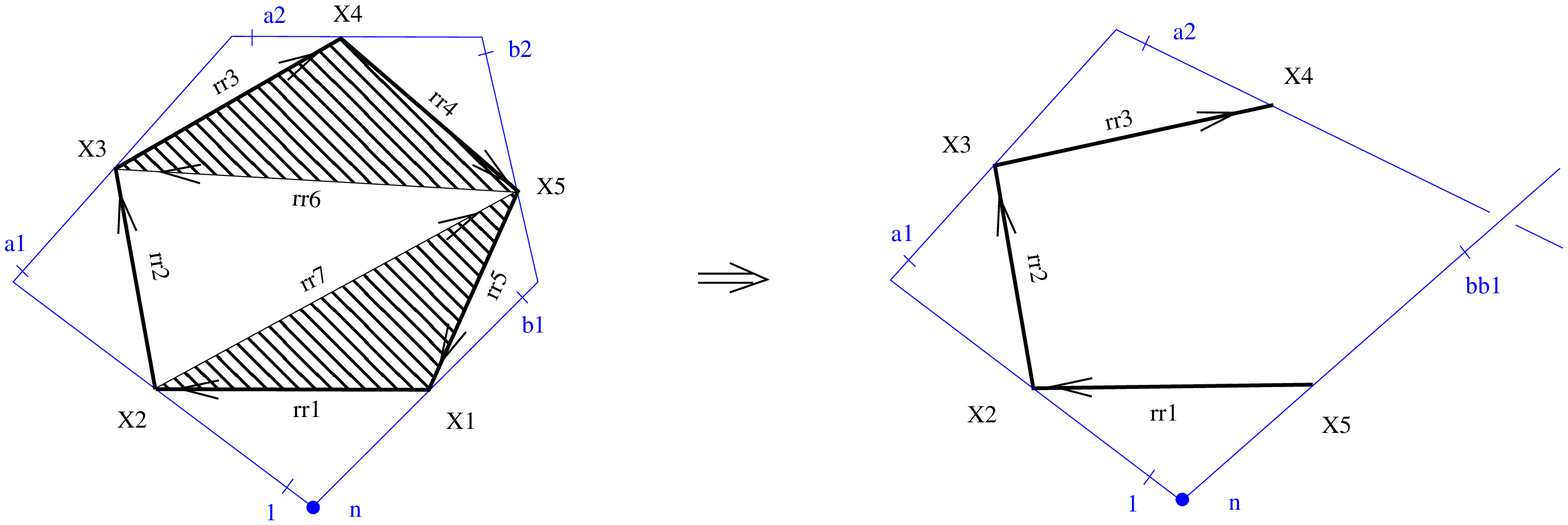}}
\caption{\small N${}^2$MHV: Degenerate pentagon (A): $b_1=b_2$}
\label{Adeg}
\end{figure}

\begin{figure}[h]
\psfrag{X1}[cc][cc]{$X_5$}
\psfrag{X2}[cc][cc]{$X_1$}
\psfrag{X3}[cc][cc]{$X_2$}
\psfrag{X4}[cc][cc]{$X_3$}\psfrag{X5}[cc][cc]{$X_4$}
\psfrag{a1}[cc][cc]{$\scriptstyle a_1$}\psfrag{a2}[cc][cc]{$\scriptstyle a_2$}
\psfrag{b1}[cc][cc]{$\scriptstyle b_1$}\psfrag{b2}[cc][cc]{$\scriptstyle b_2$}
\psfrag{1}[cc][cc]{$\scriptstyle 1$}\psfrag{n}[cc][cc]{$\scriptstyle n$}
\psfrag{rr1}[cc][cc]{$\scriptstyle \rho_0(\rt_1+\rt_2)$}
\psfrag{rr2}[cc][cc]{$\scriptstyle \r_1\rt_1$}
\psfrag{rr3}[cc][cc]{$\scriptstyle \sigma_1\rt_1$}
\psfrag{rr4}[cc][cc]{$\scriptstyle \r_2\rt_2$}
\psfrag{rr5}[cc][cc]{$\scriptstyle \sigma_2\rt_2$}
\psfrag{rr6}[cc][cc]{  $\scriptstyle \r_0\rt_1$}
\psfrag{rr7}[cc][cc]{$\scriptstyle \r_0\rt_2$}
\psfrag{aa2}[cc][cc]{$\scriptstyle b_1 = a_2$}

\centerline{\includegraphics[width=160mm]{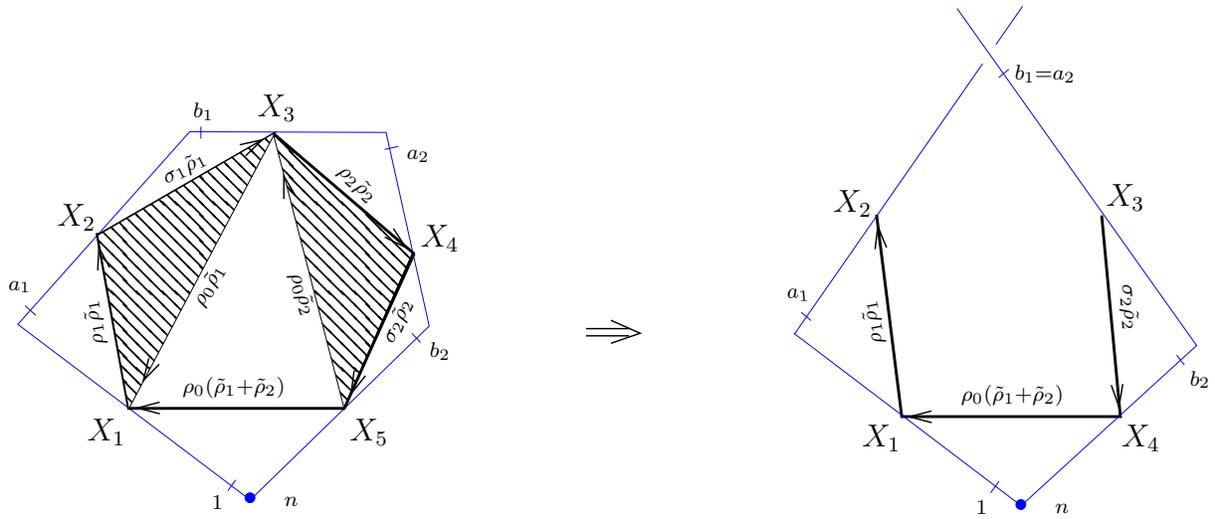}}
\caption{\small N${}^2$MHV: Degenerate pentagon (B): $b_1 = a_2$ }
\label{Bdeg}
\end{figure}

The remaining two contributions to the N${}^2$MHV superamplitude found in \cite{Drummond:2008cr} (named ``boundary terms" there) are degenerate cases of the first two. Instead of repeating the whole procedure of the twistor transform, we immediately give the corresponding diagrams, from which it is easy to read off the analytic expressions. The pentagon configuration of Fig.~\ref{5a2d} degenerates when $b_2=b_1$, as shown in Fig.~\ref{Adeg}. This leaves no points on the twistor line 4, so it disappears. The new lines 3 and 4 do not intersect anymore. The inscribed moduli space pentagon degenerates into three lightlike segments (the segment $X_{34}$ is not lightlike). The denominator \p{delta2} shrinks to
\begin{equation}\label{delta2shrink}
\Delta^{\rm (A),deg} = \vev{12}\ldots \vev{a_1-1\, \r_1} \vev{\r_1\, a_1} \ldots \vev{a_2-1\, \r_{2}} \vev{\r_{2}\, a_2} \ldots  \vev{b_1-1\, \sigma_{2}} \vev{\sigma_{2}\,  \sigma_1} \vev{\sigma_1\, b_1}  \ldots \vev{n1} \,.
\end{equation}
Similarly, the picture in Fig.~\ref{5a2dB} degenerates into that in Fig.~\ref{Bdeg} when $b_1 = a_2$.


\subsection{General case}\label{geca}

The experience we have gained in considering the N${}^2$MHV superamplitudes is sufficient to formulate the general rules which allow us to construct the twistor transform of any tree superamplitude.

{Let us first recall how the NMHV triangular configuration shown in Fig.~\ref{basictri}
was obtained from the MHV one described by a single line.}
The key step was breaking the line up into three lines, containing the cluster of particles   $[1,a_1-1]$, $[a_1, b_1-1]$ (with $b_1-a_1 \geq 2$) and $[b_1,n]$, respectively. Then we bent the first and the third lines in a way to form a triangle. Effectively, we created three lines instead of the original single line. We can reformulate this by saying that we have inserted the segment $[a_1, b_1-1]$ into the MHV line and have broken the line up around the insertion.

Let us now reexamine the construction of the two pentagon configurations (A) and (B) in the N${}^2$MHV case. {As can easily be seen from  Figs.~\ref{5a2d} and \ref{5a2dB},
it follows the same pattern as the transition from MHV to NMHV. Namely, going from NMHV to N${}^2$MHV,} we inserted the new segment $[a_2, b_2-1]$ (with $b_2-a_2 \geq 2$) into one of the lines of the NMHV triangle in Fig.~\ref{basictri}. We could do this in three ways, inserting the new segment into lines 1, 2 or 3. However, it is easy to see that the first and the third options are equivalent, up to a relabeling of the insertion points. So, to avoid double counting, we adopt the rule $a_1 < a_2$, i.e., the new insertion has to appear after (clockwise) the beginning of the preceding insertion.\footnote{We recall once again that there is no ordering of the points along the twistor lines. The fact that we show point $a_i$ before (clockwise) $b_i$ is just a convention which reminds us that $a_i<b_i$.} Thus, we restrict ourselves to two insertions, in which the new segment $[a_2, b_2-1]$ appears:
\begin{description}
  \item[(A)]  inside the existing segment $[a_1, b_1-1]$ (Fig.~\ref{5a2d});
  \item[(B)] after (clockwise) the existing segment $[a_1, b_1-1]$ (Fig.~\ref{5a2dB}).
\end{description}
We then break up the line into which the insertion takes place, in such a way that its two segments and the inserted segment form a triangle in twistor space. Thus, we pass from a 3-line configuration (triangle) to a 5-line one (`pentagon'). One should remember that the five lines are not in the same plane, so the pentagon is just the two-dimensional projection of the three-dimensional pyramidal configuration of Fig.~\ref{A3d1}.

This construction is complemented by drawing a figure in the associated moduli space. Now, based on the pentagon example,  we can formulate a set of simple rules which can immediately be generalized to any N${}^k$MHV superamplitude (with $k=1,2,3,\ldots$):
\begin{itemize}

  \item The insertion of a new segment amounts to transforming the twistor space $(2k-1)-$gon of the N${}^{k-1}$MHV superamplitude into the $(2k+1)-$gon of the N${}^{k}$MHV superamplitude. The inscribed moduli space polygon has $(2k+1)$ vertices, with  lightlike separations between the adjacent points.

  \item The new segment $[a_k, b_k-1]$ (with $b_k-a_k \geq 2$) is inserted into the sides of the existing twistor space $(2k-1)-$gon in all possible ways, respecting the rule $a_1  < \ldots <a_{k-1} < a_k$, i.e. the new segment has to appear after (clockwise) all existing points $a_1, \ldots, a_{k-1}$.

  \item Two of the twistor lines sides of the $(2k+1)-$gon have a fixed intersection point with coordinates $(\la=\r_0,\mu=0)$ (if we use translations to set $X=0$). By convention, it always faces the moduli space segment formed by the first modulus $X_1$ and the last one $X_{2k+1} \equiv X$.

  \item The moduli of the three newly created twistor lines,  $X_u$, $X_{u+1}$ and $X_{u+2}$, form a shaded triangle. Here the label $u$ indicates the position of the new insertion on the old $(2k-1)-$gon (in Fig.~\ref{5a2d} point $X_2$ shows that the insertion was done on line 2 of the triangle, in Fig.~\ref{5a2dB} point $X_3$ shows that it was done on line 3). In drawing the triangle, we follow the rule that its `base' $X_{u+1} X_{u+2}$ faces the twistor point $b_k$, while its `tip' $X_u$ appears immediately before (clockwise) the matching point $a_k$.

  \item The new triangle is of the shaded type, i.e. its sides have a common antichiral spinor, the new modulus $\rt_k$ (in the pentagon example this is $\rt_2$).

  \item The three lightlike vectors forming the new shaded triangle $X_u X_{u+1} X_{u+2}$ are oriented clockwise and sum up to zero (`conservation law'). The side $X_{u} X_{u+1}$ carries the vector  $\r_k\rt_k$, while  $X_{u+1} X_{u+2}$ carries $\sigma_k \rt_k$. The third side of the triangle $X_{u+2} X_{u}$ is determined by looking at the white figure adjacent to the new shaded triangle. For instance, in Fig.~\ref{5a2d} this white figure is the triangle $X_1 X_2 X_4$. It is formed by lightlike vectors of the other type, i.e. with a common chiral spinor. Which precisely chiral spinor, is easy to find out: The sides of the old shaded triangle $X_1 X_4 X_5$ have not changed, in particular, it has the side $X_{14} = \r_1\rt_1$, common with the white triangle. This shows that the chiral spinor `running along' the white triangle is $\r_1$.  The `conservation' of the three sides of the shaded triangle yields the linear relation between the new moduli $r_k$, $\sigma_k$ and one of the old $\r$ (see, e.g., \p{linrel} and \p{defr2}). The third side $X_{12}$ of the white triangle is also determined from the `conservation' of its three sides.
  \item In general, depending on the orientation of the shaded triangles, the white figure may be a planar polygon (up to a $(k+1)-$gon), with all sides and all diagonals proportional to the same chiral spinor. The white polygon is itself triangulated into lightlike triangles of the first type (see the examples in Figs.~\ref{A12d} -- \ref{B12} below).
  \item Thus, the complete  moduli space $(2k+1)-$gon is triangulated into a number of shaded and white triangles whose sides are lightlike vectors summing up to zero. Tracing all of these `conservation laws' throughout the figure, it is easy to work out all the lightlike distances.
  \item The three twistor lines passing through the vertices of a white triangle (or, more generally, all twistor lines passing through the vertices of a white figure) have a common intersection point. In the example of Fig.~\ref{5a2d} this is made visible by redrawing the diagram in three dimensions, see Fig.~\ref{A3d1}.
\end{itemize}

The drawing procedure described above may seem recursive, i.e., one needs to start from the NMHV triangle, make insertions to obtain the two N${}^2$MHV pentagons, etc. until reaching the $(2k+1)-$gons of the N${}^k$MHV amplitude. In fact, this is not necessary, any diagram can be drawn independently. One starts by drawing a $(2k+1)-$gon in twistor space, together with the inscribed moduli space polygon. Then one marks the points $a_1, \ldots, a_k$ and  $b_1, \ldots, b_k$ on the twistor lines, according to the particular distribution one has chosen.\footnote{As pointed out in \cite{Drummond:2008cr}, the number of such distributions equals the Catalan number $C_k = \frac{(2k)!}{(k+1)! k!}$. It is amusing that the original definition of the Catalan number has to do with Euler's problem of triangulating an $m-$sided polygon, by considering all orientations independent. The number of triangulations is $C_{m-2}$. In our case we are triangulating a $(2k+1)-$gon, but we do not keep all possible orientations.  Thus, we see a much smaller number, $C_k$, of diagrams.   } The next step is to draw the shaded triangles following the rules above. This determines the type of the enclosed white figures. Finally, starting from the basic shaded triangle with common $\rt_1$, one
employs  the  `conservation laws' for the various adjacent triangles {in order to fix}
the positions of all the vertices of the moduli space polygon.

We recall that the  $n-$particle superamplitude \p{An-dec} is given by the sum over
all N${}^k$MHV amplitudes with $k$ ranging from 0 to $n-4$. The first half of them (more precisely, for $0 \leq k \leq [n/2]-2$) are naturally described in our chiral on-shell superspace, while the remaining half correspond to the so-called `googly' versions of the PCT conjugate amplitudes.  In twistor space, for $k \leq [n/2]-2$  each of these amplitudes is represented by a $(2k+1)-$gon and the
corresponding inscribed moduli space  configuration is  $(2k+1)-$gon. The latter consists of $k$ shaded and $(k-1)$ white triangles. The white ones can appear distinctly, or some of them can merge together to form planar rectangles, pentagons, etc., up to a $(k+1)-$gon. Each such planar polygon serves as the base of pyramid. Its apex is the intersection point of the twistor lines passing through the vertices of the planar polygon.

However, if $k > [n/2]-2$, it seems that the number of such twistor diagrams continues to grow, while the number of {superinvariants in the expression for  the }corresponding  N${}^k$MHV amplitudes starts decreasing, until it reaches 1 in the extreme case $k_{\rm max}=n-4$ (corresponding to the `googly' $\overline{\rm MHV}$ amplitude). What can explain this apparent mismatch? The answer is that for sufficiently large values of $k$ some of the line configurations will be prohibited, due to the limitations on the values of the indices $a_i, b_i$, namely, $a_i < a_{i+1}$ and $b_i - a_i \geq 2$. For instance, consider the `googly' $\overline{\rm MHV}$ amplitude, for which we would expect $\ell_{\rm max}=2k_{\rm max}+1= 2n-7$ lines. For $n=4$ we get $\ell_{\rm max}=1$, and this single line indeed describes the self-conjugate amplitude ${\cal A}^{\overline{\rm MHV}}_4$. For $n=5$ and $\ell_{\rm max}=3$, the only three-line configuration possible  corresponds to $a_1=2$, $b_1=4$, and it describes the amplitude ${\cal A}^{\overline{\rm MHV}}_5$. For $n=6$ and $\ell_{\rm max}=5$, trying to meet the restrictions on the indices, we find that the only solution is $a_1=2$, $a_2=3$, $b_1=b_2=5$. This is the degenerate four-line configuration from Fig.~\ref{Adeg}. One can easily check that for any value of $n$ there is a unique (degenerate for $n>5$) twistor diagram describing ${\cal A}^{\overline{\rm MHV}}_n$. It is the most `densely packed' configuration labeled by $a_1=2, a_2=3, \ldots, a_{n-4}=n-3$ and $b_1=b_2=\ldots = b_{n-4}= n-1$, and consists of $n-2$ lines.  Similar exceptions will also take place for all values $k > [n/2]-2$. They will reduce the naively expected number of diagrams $C_k$ to the smaller number of superinvariants in the `googly' amplitudes. It would be interesting to find out the explicit equivalence relation between the conventional and `googly'  descriptions of the same amplitude in twistor space.

{Given a twistor diagram with the inscribed lightlike polygon,} it is easy to write down the integrand of the twistor transform. The numerator (denoted by $(\prod)$ in Eqs.~\p{sugg}, \p{sugg'} and \p{sugg5} above) is just a collection of delta functions. It consists of $2k+1$ clusters, corresponding to each line in twistor space. The moduli of these lines are determined from the diagram, starting with $X_{2k+1} \equiv X$ and adding to it lightlike vectors until we reach  the modulus of any given twistor line (and similarly for the fermionic moduli). The denominator $\Delta$ is of the MHV type with a number of insertions, each $\r_u$ going in between $a_u-1$ and $a_u$, and each $\sigma_u$ going in between $b_u-1$ and $b_u$. Finally, one integrates over all independent moduli, that is, $X$,  $\r_1, \ldots, \r_k$, $\rt_1, \ldots, \rt_k$ and their fermionic counterparts $\Theta$, $\xi_1, \ldots, \xi_k$.

Since this twistor transform of the superamplitude is essentially given by a
product of delta functions, it is straightforward to do the inverse twistor transform. In the process the moduli $\r$ and $\sigma$ get expressed in terms of the momenta (see, e.g., \p{defr} and \p{defr2}) {and we obtain the expressions for tree superamplitudes
which are} in perfect agreement with the explicit solution to the BCFW recursion relations found in \cite{Drummond:2008cr}.

\subsection{Further examples}

\begin{figure}[h]
\psfrag{X1}[cc][cc]{$X_1$}
\psfrag{X2}[cc][cc]{$X_2$}
\psfrag{X3}[cc][cc]{$X_3$}
\psfrag{X4}[cc][cc]{$X_4$}
\psfrag{X5}[cc][cc]{$X_5$}
\psfrag{X6}[cc][cc]{$X_6$}
\psfrag{X7}[cc][cc]{$X_7$}
\psfrag{a1}[cc][cc]{$\scriptstyle a_1$}
\psfrag{a2}[cc][cc]{$\scriptstyle a_2$}
\psfrag{a3}[cc][cc]{$\scriptstyle a_3$}
\psfrag{b1}[cc][cc]{$\scriptstyle b_1$}
\psfrag{b2}[cc][cc]{$\scriptstyle b_2$}
\psfrag{b3}[cc][cc]{$\scriptstyle b_3$}
\psfrag{1}[cc][cc]{$\scriptstyle 1$}\psfrag{n}[cc][cc]{$\scriptstyle n$}
\psfrag{rr1}[cc][cc]{$\scriptstyle \r_0\rt_1$}
\psfrag{rr2}[cc][cc]{$\scriptstyle \r_1(\rt_1+\rt_2)$}
\psfrag{rr3}[cc][cc]{$\scriptstyle \r_2(\rt_2+\rt_3)$}
\psfrag{rr4}[cc][cc]{$\scriptstyle \r_3 \rt_3$}
\psfrag{rr5}[cc][cc]{$\scriptstyle \sigma_3\rt_3$}
\psfrag{rr6}[cc][cc]{$\scriptstyle \sigma_2\rt_2$}
\psfrag{rr7}[cc][cc]{$\scriptstyle \sigma_1\rt_1$}
\psfrag{ss1}[cc][cc]{$\scriptstyle \r_2\rt_3$}
\psfrag{ss2}[cc][cc]{$\scriptstyle \r_2\rt_2$}
\psfrag{ss3}[cc][cc]{$\scriptstyle \r_1\rt_2$}
\psfrag{ss4}[cc][cc]{$\scriptstyle \r_1\rt_1$}
\psfrag{M1}[cc][cc]{$\scriptstyle \mathcal{O}_1$}
\psfrag{M2}[cc][cc]{$\scriptstyle \mathcal{O}_2$}
\psfrag{M3}[cc][cc]{$\scriptstyle \mathcal{O}_3$}
\begin{center}
{\includegraphics[width=70mm]{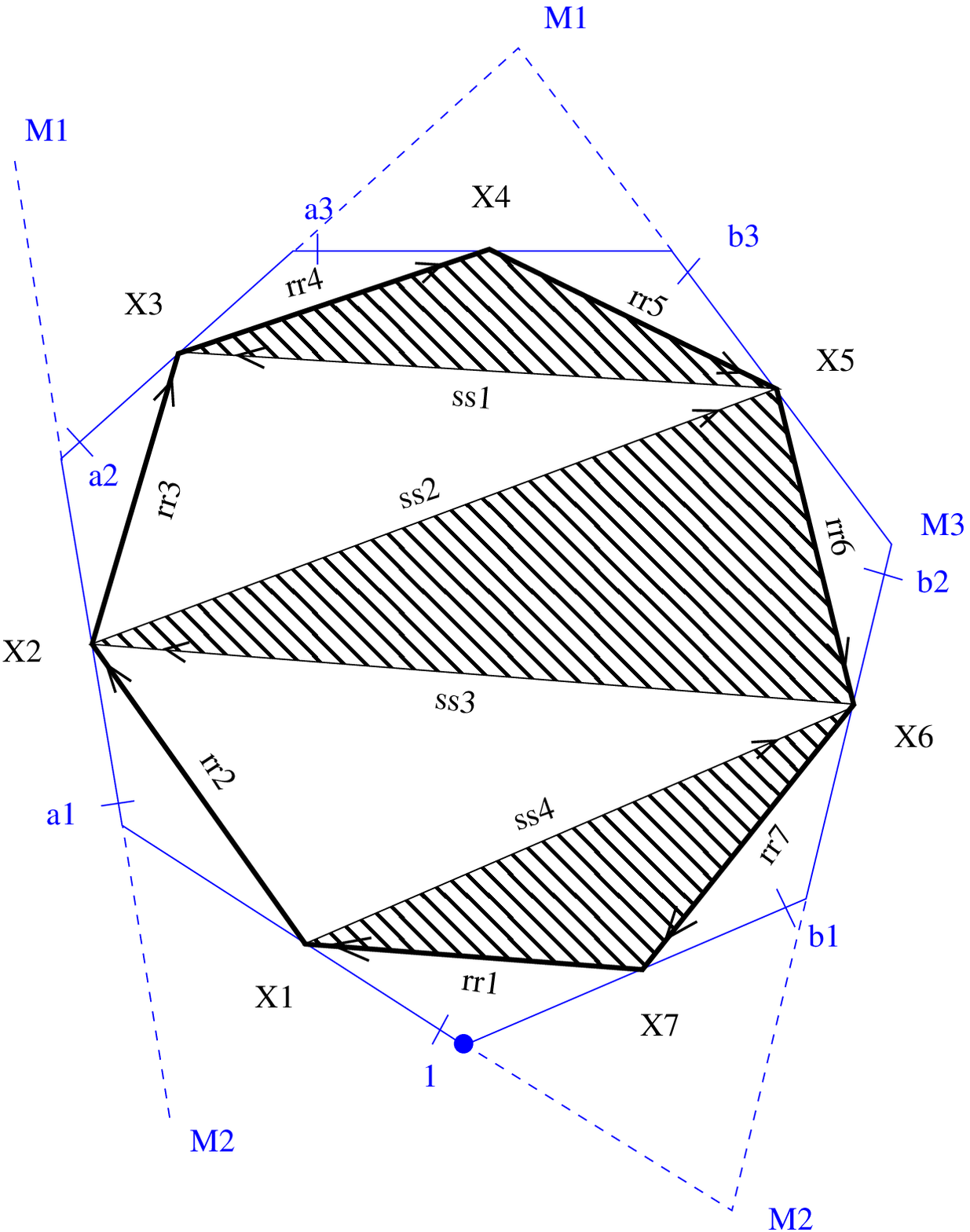}}
\psfrag{X1}[cc][cc]{$X_1$}
\psfrag{X2}[cc][cc]{$X_2$}
\psfrag{X3}[cc][cc]{$X_3$}
\psfrag{X4}[cc][cc]{$X_4$}
\psfrag{X5}[cc][cc]{$X_5$}
\psfrag{X6}[cc][cc]{$X_6$}
\psfrag{X7}[cc][cc]{$X_7$}
\psfrag{M1}[cc][cc]{$\scriptstyle \mathcal{O}_1$}
\psfrag{M2}[cc][cc]{$\scriptstyle \mathcal{O}_2$}
\psfrag{M3}[cc][cc]{$\scriptstyle \mathcal{O}_3$}
\hspace*{25mm} {\includegraphics[width=62mm]{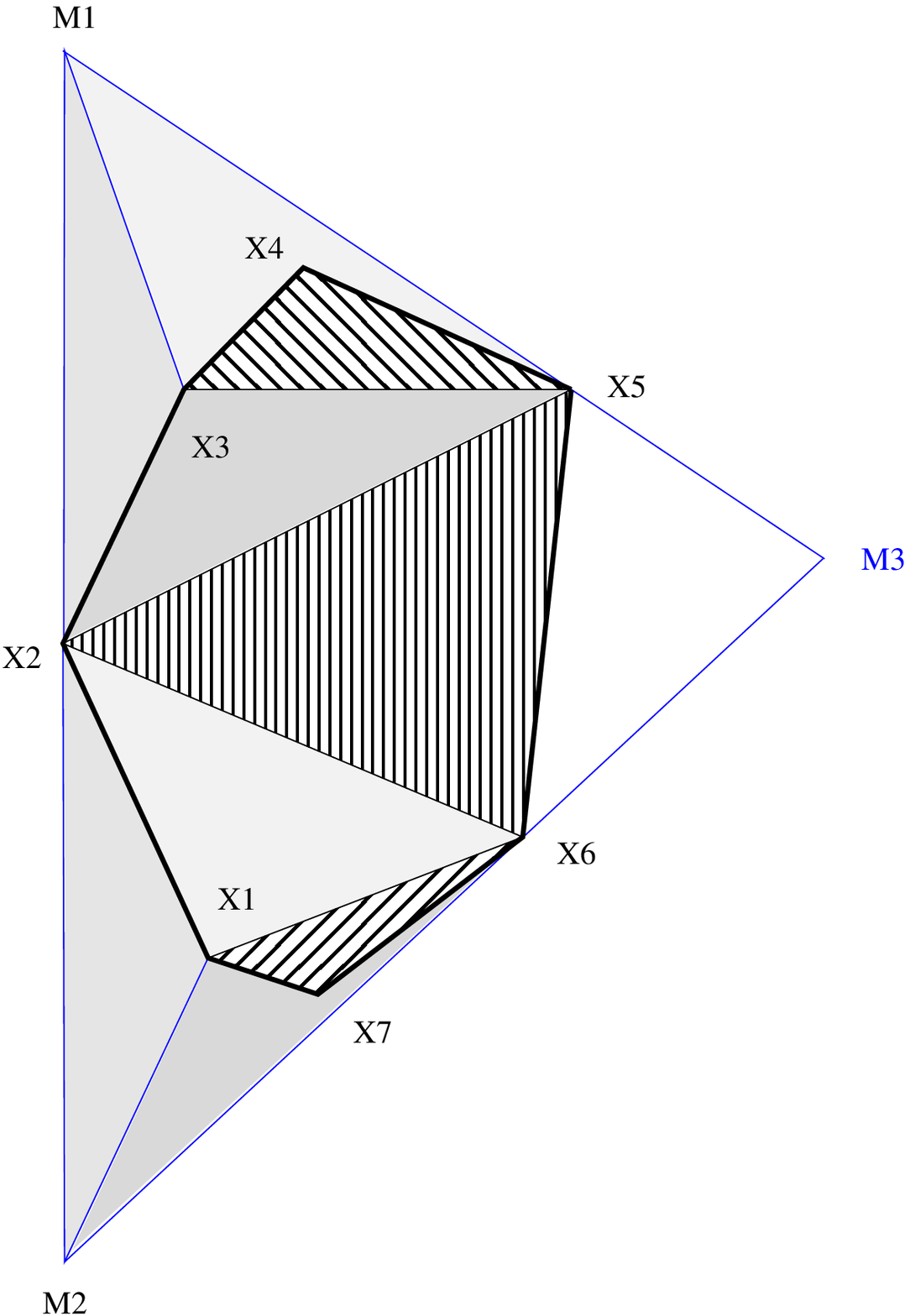}}
 \end{center}
\caption{\small N${}^3$MHV: heptagon (A1) in two dimensions
and in three dimensions }\label{A12d}
\end{figure}

\begin{figure}[!h]
\psfrag{X1}[cc][cc]{$X_1$}
\psfrag{X2}[cc][cc]{$X_2$}
\psfrag{X3}[cc][cc]{$X_3$}
\psfrag{X4}[cc][cc]{$X_4$}
\psfrag{X5}[cc][cc]{$X_5$}
\psfrag{X6}[cc][cc]{$X_6$}
\psfrag{X7}[cc][cc]{$X_7$}
\psfrag{a1}[cc][cc]{$\scriptstyle a_1$}
\psfrag{a2}[cc][cc]{$\scriptstyle a_2$}
\psfrag{a3}[cc][cc]{$\scriptstyle b_2$}
\psfrag{b1}[cc][cc]{$\scriptstyle b_1$}
\psfrag{b2}[cc][cc]{$\scriptstyle b_3$}
\psfrag{b3}[cc][cc]{$\scriptstyle a_3$}
\psfrag{1}[cc][cc]{$\scriptstyle 1$}\psfrag{n}[cc][cc]{$\scriptstyle n$}
\psfrag{rr1}[cc][cc]{$\scriptstyle \r_0\rt_1$}
\psfrag{rr2}[cc][cc]{$\scriptstyle \r_1(\rt_1+\rt_2+\rt_3)$}
\psfrag{rr3}[cc][cc]{$\scriptstyle \r_2 \rt_2$}
\psfrag{rr4}[cc][cc]{$\scriptstyle \sigma_2 \rt_2$}
\psfrag{rr5}[cc][cc]{$\scriptstyle \r_3\rt_3$}
\psfrag{rr6}[cc][cc]{$\scriptstyle \sigma_3\rt_3$}
\psfrag{rr7}[cc][cc]{$\scriptstyle \sigma_1\rt_1$}
\psfrag{ss1}[cc][cc]{$\scriptstyle \r_1\rt_2$}
\psfrag{ss2}[cc][cc]{$\scriptstyle \r_1\rt_3$}
\psfrag{ss3}[cc][cc]{$\scriptstyle \r_1\rt_1$}
\begin{center}
{\includegraphics[width=70mm]{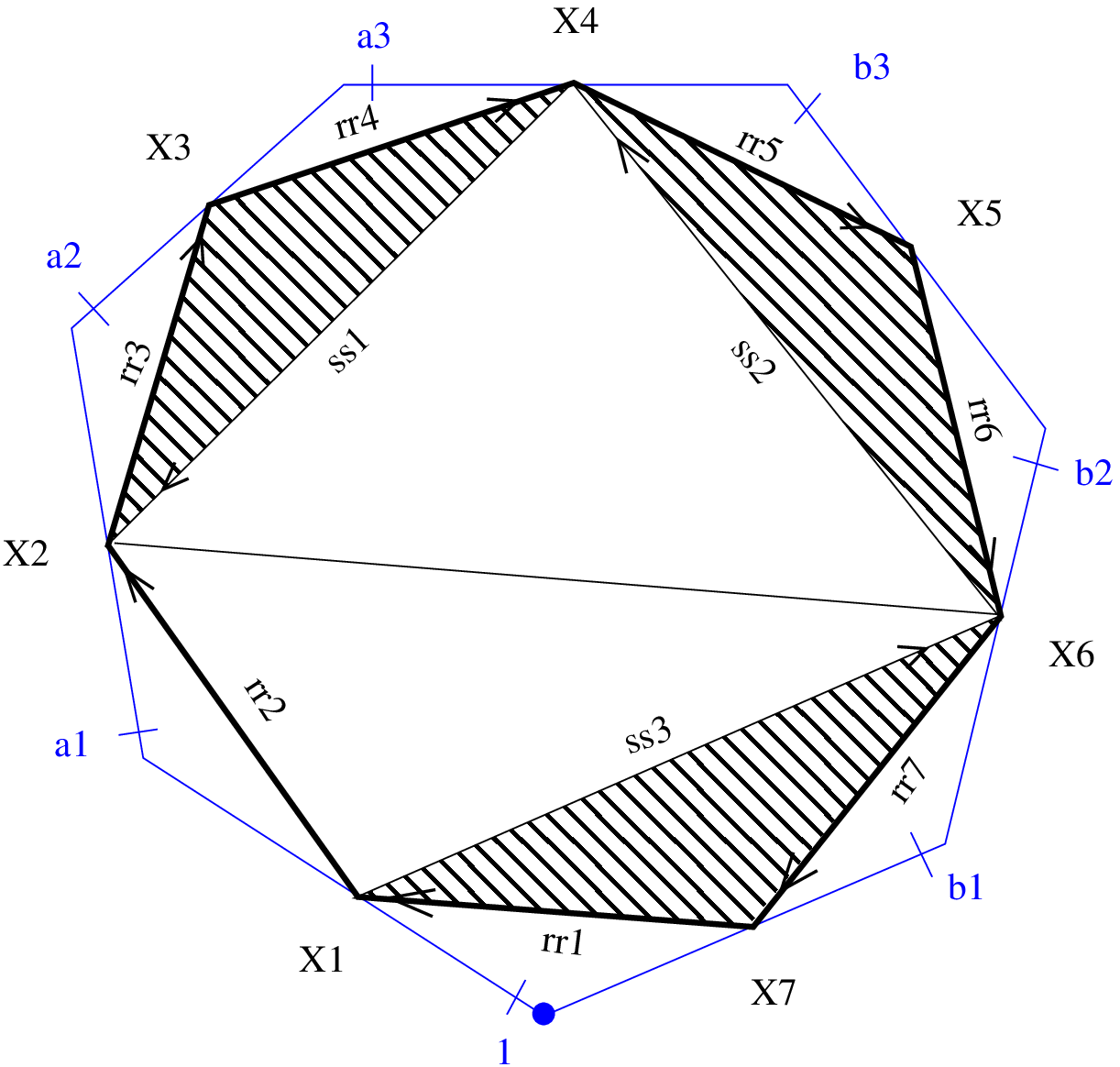}}
%
%
\psfrag{X1}[cc][cc]{$X_1$}
\psfrag{X2}[cc][cc]{$X_2$}
\psfrag{X3}[cc][cc]{$X_3$}
\psfrag{X4}[cc][cc]{$X_4$}
\psfrag{X5}[cc][cc]{$X_5$}
\psfrag{X6}[cc][cc]{$X_6$}
\psfrag{X7}[cc][cc]{$X_7$}
\psfrag{a1}[cc][cc]{$\scriptstyle a_1$}
\psfrag{a2}[cc][cc]{$\scriptstyle a_2$}
\psfrag{a3}[cc][cc]{$\scriptstyle a_3$}
\psfrag{b1}[cc][cc]{$\scriptstyle b_1$}
\psfrag{b2}[cc][cc]{$\scriptstyle b_2$}
\psfrag{b3}[cc][cc]{$\scriptstyle b_3$}
\psfrag{1}[cc][cc]{$\scriptstyle 1$}\psfrag{n}[cc][cc]{$\scriptstyle n$}
%
\hspace*{15mm}
{\includegraphics[height=90mm]{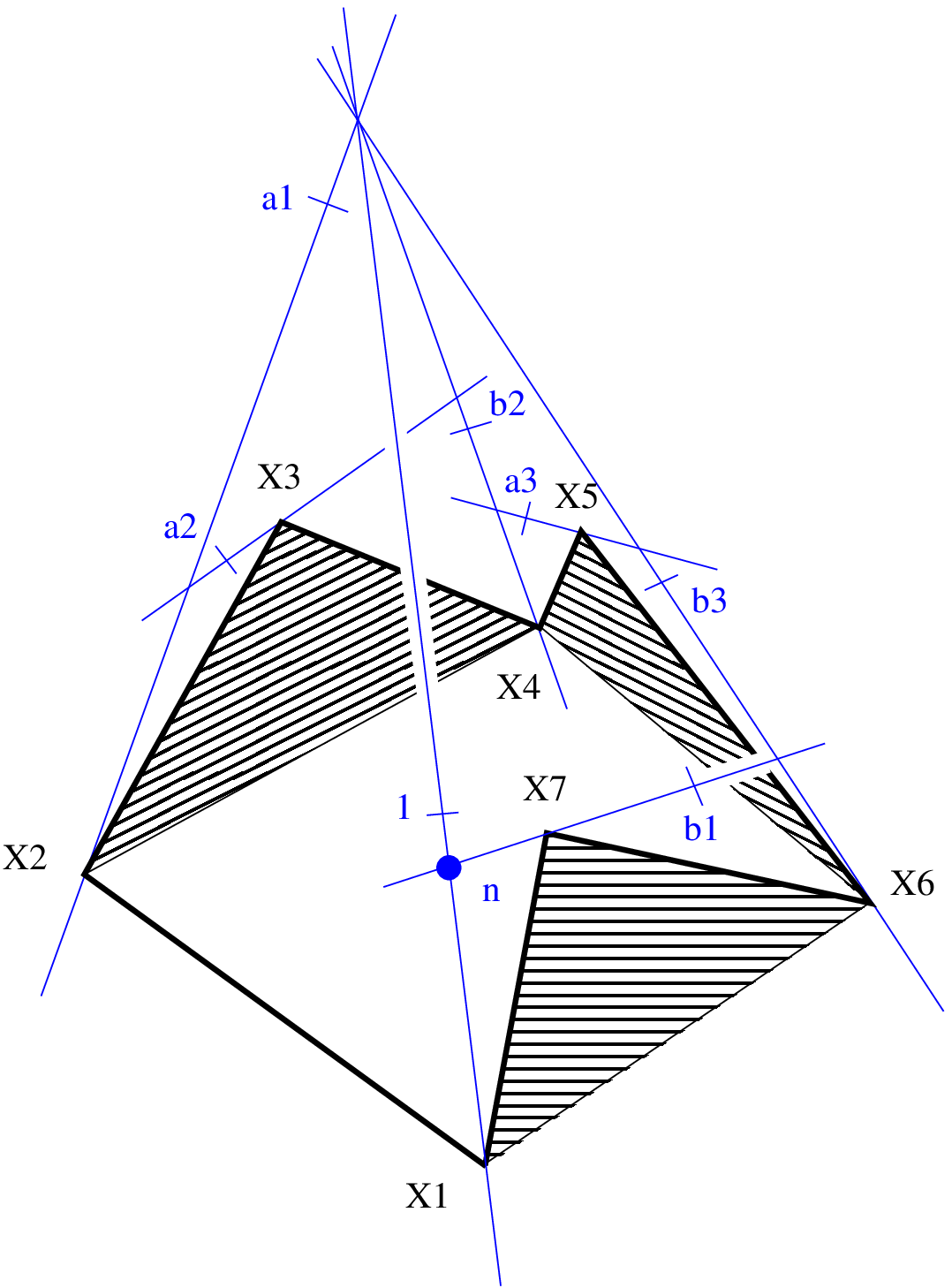}}
\end{center}
\caption{\small N${}^3$MHV: heptagon (A2) in two dimensions  and in
 three dimensions }\label{A23d}
\end{figure}

\begin{figure}[!h]
\begin{minipage}[t]{.4\linewidth}
    \begin{center}
\psfrag{X1}[cc][cc]{$X_1$}
\psfrag{X2}[cc][cc]{$X_2$}
\psfrag{X3}[cc][cc]{$X_3$}
\psfrag{X4}[cc][cc]{$X_4$}
\psfrag{X5}[cc][cc]{$X_5$}
\psfrag{X6}[cc][cc]{$X_6$}
\psfrag{X7}[cc][cc]{$X_7$}
\psfrag{a1}[cc][cc]{$\scriptstyle a_1$}
\psfrag{a2}[cc][cc]{$\scriptstyle a_2$}
\psfrag{a3}[cc][cc]{$\scriptstyle b_2$}
\psfrag{b3}[cc][cc]{$\scriptstyle b_1$}
\psfrag{b2}[cc][cc]{$\scriptstyle a_3$}
\psfrag{b1}[cc][cc]{$\scriptstyle b_3$}
\psfrag{1}[cc][cc]{$\scriptstyle 1$}\psfrag{n}[cc][cc]{$\scriptstyle n$}
\psfrag{rr1}[cc][cc]{$\scriptstyle \r_0(\rt_1+\rt_3)$}
\psfrag{rr2}[cc][cc]{$\scriptstyle \r_1(\rt_1+\rt_2)$}
\psfrag{rr3}[cc][cc]{$\scriptstyle \r_2 \rt_2$}
\psfrag{rr4}[cc][cc]{$\scriptstyle \sigma_2 \rt_2$}
\psfrag{rr5}[cc][cc]{$\scriptstyle \sigma_1\rt_1$}
\psfrag{rr6}[cc][cc]{$\scriptstyle \r_3\rt_3$}
\psfrag{rr7}[cc][cc]{$\scriptstyle \sigma_3\rt_3$}
\psfrag{ss1}[cc][cc]{$\scriptstyle \r_1\rt_2$}
\psfrag{ss2}[cc][cc]{$\scriptstyle \r_1\rt_1$}
\psfrag{ss3}[cc][cc]{$\scriptstyle \r_0\rt_1$}
\psfrag{ss4}[cc][cc]{$\scriptstyle \r_0\rt_3$}
       \includegraphics[width=70mm]{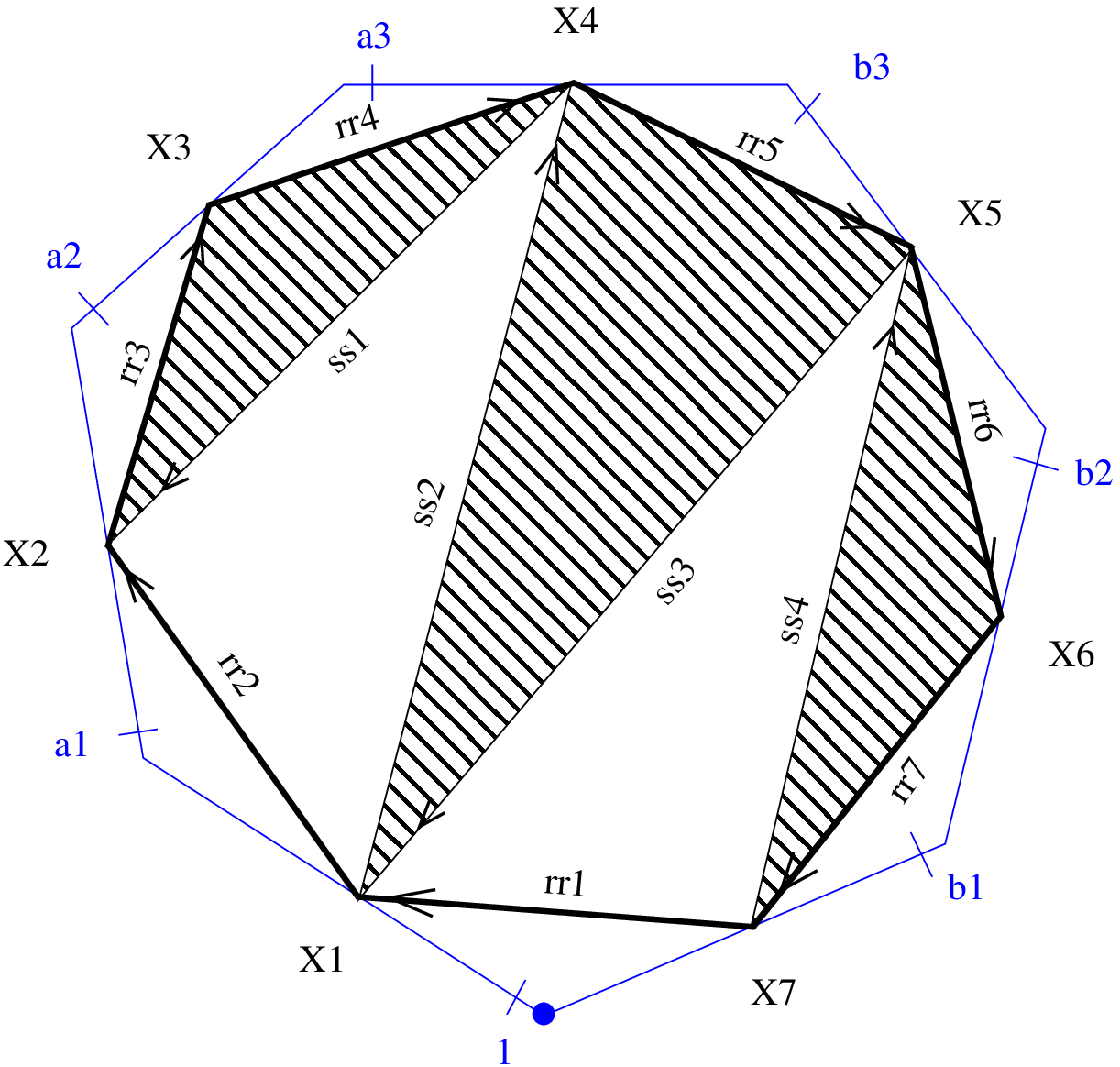}
       \caption{\small N${}^3$MHV: heptagon (A3)}\label{A3}
    \end{center}
\end{minipage}
\hfill
\begin{minipage}[t]{.4\linewidth}
    \begin{center}
\psfrag{X1}[cc][cc]{$X_1$}
\psfrag{X2}[cc][cc]{$X_2$}
\psfrag{X3}[cc][cc]{$X_3$}
\psfrag{X4}[cc][cc]{$X_4$}
\psfrag{X5}[cc][cc]{$X_5$}
\psfrag{X6}[cc][cc]{$X_6$}
\psfrag{X7}[cc][cc]{$X_7$}

\psfrag{a1}[cc][cc]{$\scriptstyle a_1$}
\psfrag{a2}[cc][cc]{$\scriptstyle b_1$}
\psfrag{a3}[cc][cc]{$\scriptstyle a_2$}
\psfrag{b3}[cc][cc]{$\scriptstyle b_2$}
\psfrag{b2}[cc][cc]{$\scriptstyle a_3$}
\psfrag{b1}[cc][cc]{$\scriptstyle b_3$}
\psfrag{1}[cc][cc]{$\scriptstyle 1$}\psfrag{n}[cc][cc]{$\scriptstyle n$}
\psfrag{rr1}[cc][cc]{$\scriptstyle \r_0(\rt_1+\rt_2+\rt_3)$}
\psfrag{rr2}[cc][cc]{$\scriptstyle \r_1\rt_1$}
\psfrag{rr3}[cc][cc]{$\scriptstyle \sigma_1 \rt_1$}
\psfrag{rr4}[cc][cc]{$\scriptstyle \r_2 \rt_2$}
\psfrag{rr5}[cc][cc]{$\scriptstyle \sigma_2\rt_2$}
\psfrag{rr6}[cc][cc]{$\scriptstyle \r_3\rt_3$}
\psfrag{rr7}[cc][cc]{$\scriptstyle \sigma_3\rt_3$}
\psfrag{ss1}[cc][cc]{$\scriptstyle \r_0\rt_1$}
\psfrag{ss2}[cc][cc]{$\scriptstyle \r_0\rt_2$}
\psfrag{ss3}[cc][cc]{$\scriptstyle \r_0\rt_3$}    
       \includegraphics[width=70mm]{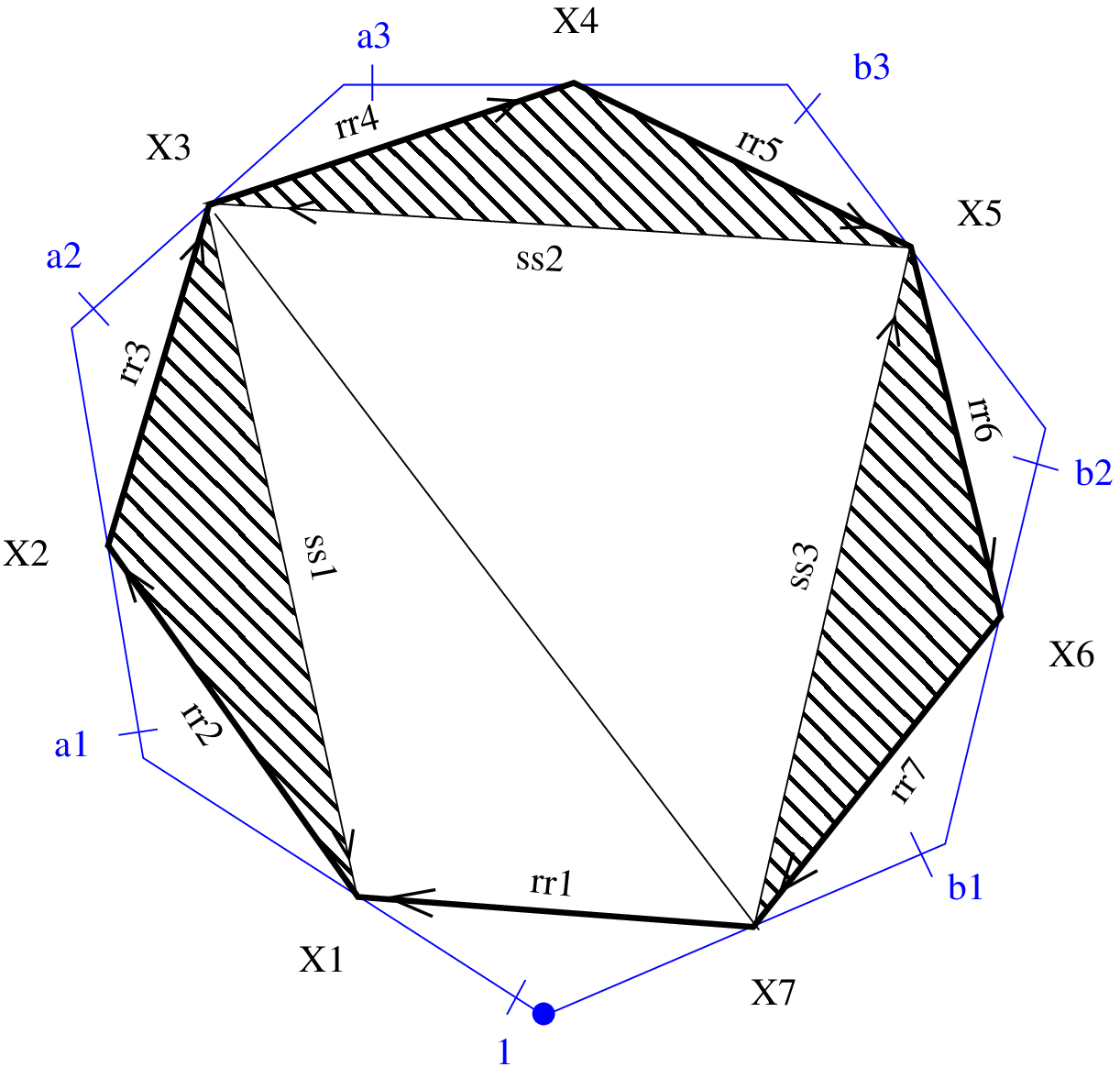}
       \caption{\small N${}^3$MHV: heptagon (B1)}\label{B1}
    \end{center}
\end{minipage}
\end{figure}

\begin{figure}[h]
\begin{minipage}[t]{.4\linewidth}
\psfrag{X1}[cc][cc]{$X_1$}
\psfrag{X2}[cc][cc]{$X_2$}
\psfrag{X3}[cc][cc]{$X_3$}
\psfrag{X4}[cc][cc]{$X_4$}
\psfrag{X5}[cc][cc]{$X_5$}
\psfrag{X6}[cc][cc]{$X_6$}
\psfrag{X7}[cc][cc]{$X_7$}
\psfrag{a1}[cc][cc]{$\scriptstyle a_1$}
\psfrag{a2}[cc][cc]{$\scriptstyle b_1$}
\psfrag{a3}[cc][cc]{$\scriptstyle a_2$}
\psfrag{b3}[cc][cc]{$\scriptstyle a_3$}
\psfrag{b2}[cc][cc]{$\scriptstyle b_3$}
\psfrag{b1}[cc][cc]{$\scriptstyle b_2$}
\psfrag{1}[cc][cc]{$\scriptstyle 1$}\psfrag{n}[cc][cc]{$\scriptstyle n$}
\psfrag{rr1}[cc][cc]{$\scriptstyle \r_0(\rt_1+\rt_2)$}
\psfrag{rr2}[cc][cc]{$\scriptstyle \r_1\rt_1$}
\psfrag{rr3}[cc][cc]{$\scriptstyle \sigma_1\rt_1$}
\psfrag{rr4}[cc][cc]{$\scriptstyle \r_2 (\rt_2+\rt_3)$}
\psfrag{rr5}[cc][cc]{$\scriptstyle \r_3\rt_3$}
\psfrag{rr6}[cc][cc]{$\scriptstyle \sigma_3\rt_3$}
\psfrag{rr7}[cc][cc]{$\scriptstyle \sigma_2\rt_2$}
\psfrag{ss1}[cc][cc]{$\scriptstyle \r_0\rt_1$}
\psfrag{ss2}[cc][cc]{$\scriptstyle \r_0\rt_2$}
\psfrag{ss3}[cc][cc]{$\scriptstyle \r_2\rt_2$}
\psfrag{ss4}[cc][cc]{$\scriptstyle \r_2\rt_3$}
\centerline{\includegraphics[width=70mm]{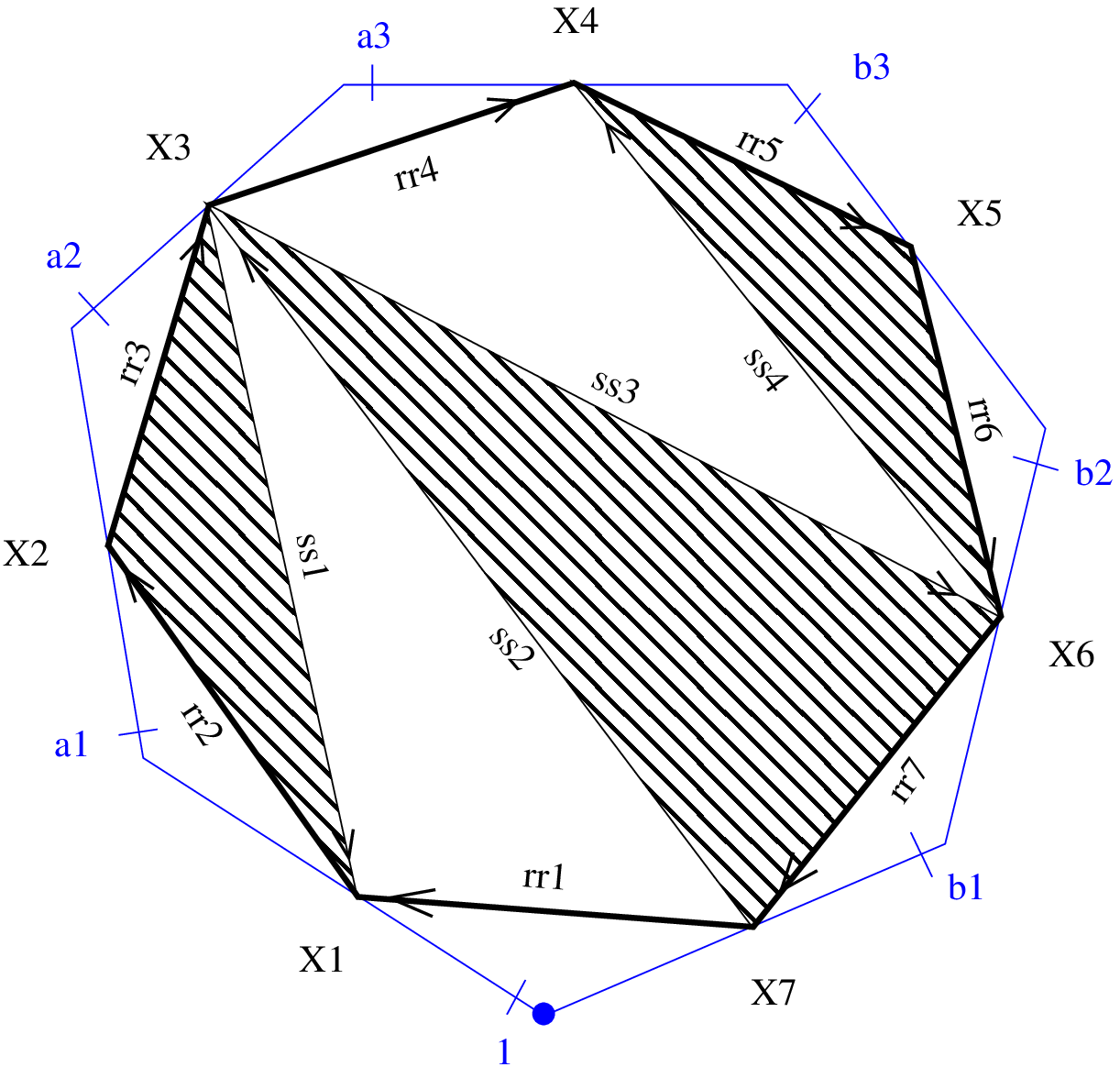}}
\caption{\small N${}^3$MHV: heptagon (B2)}\label{B2}
\end{minipage}
\hfill
\begin{minipage}[t]{.4\linewidth}
\psfrag{X1}[cc][cc]{$X_1$}
\psfrag{X2}[cc][cc]{$X_2$}
\psfrag{X3}[cc][cc]{$X_3$}
\psfrag{X4}[cc][cc]{$X_4$}
\psfrag{X5}[cc][cc]{$X_5$}
\psfrag{X6}[cc][cc]{$X_6$}
\psfrag{X7}[cc][cc]{$X_7$}
\psfrag{X8}[cc][cc]{$X_8$}
\psfrag{X9}[cc][cc]{$X_9$}
\psfrag{a1}[cc][cc]{$\scriptstyle a_1$}
\psfrag{a2}[cc][cc]{$\scriptstyle a_2$}
\psfrag{a3}[cc][cc]{$\scriptstyle a_3$}
\psfrag{a4}[cc][cc]{$\scriptstyle a_4$}
\psfrag{b4}[cc][cc]{$\scriptstyle b_4$}
\psfrag{b3}[cc][cc]{$\scriptstyle b_3$}
\psfrag{b2}[cc][cc]{$\scriptstyle b_2$}
\psfrag{b1}[cc][cc]{$\scriptstyle b_1$}
\psfrag{1}[cc][cc]{$\scriptstyle 1$}\psfrag{n}[cc][cc]{$\scriptstyle n$}
\psfrag{rr1}[cc][cc]{$\scriptstyle \r_0\rt_1$}
\psfrag{rr2}[cc][cc]{$\scriptstyle \r_1(\rt_1+\rt_2)$}
\psfrag{rr3}[cc][cc]{$\scriptstyle \r_2(\rt_2+\rt_3)$}
\psfrag{rr4}[cc][cc]{$\scriptstyle \r_3 (\rt_3+\rt_4)$}
\psfrag{rr5}[cc][cc]{$\scriptstyle \r_4\rt_4$}
\psfrag{rr6}[cc][cc]{$\scriptstyle \sigma_4\rt_4$}
\psfrag{rr7}[cc][cc]{$\scriptstyle \sigma_3\rt_3$}
\psfrag{rr8}[cc][cc]{$\scriptstyle \sigma_2\rt_2$}
\psfrag{rr9}[cc][cc]{$\scriptstyle \sigma_1\rt_1$}
\psfrag{ss1}[cc][cc]{$\scriptstyle \r_1\rt_1$}
\psfrag{ss2}[cc][cc]{$\scriptstyle \r_1\rt_2$}
\psfrag{ss3}[cc][cc]{$\scriptstyle \r_2\rt_2$}
\psfrag{ss4}[cc][cc]{$\scriptstyle \r_2\rt_3$}
\psfrag{ss5}[cc][cc]{$\scriptstyle \r_3\rt_3$}
\psfrag{ss6}[cc][cc]{$\scriptstyle \r_3\rt_4$}
\centerline{\includegraphics[width=70mm]{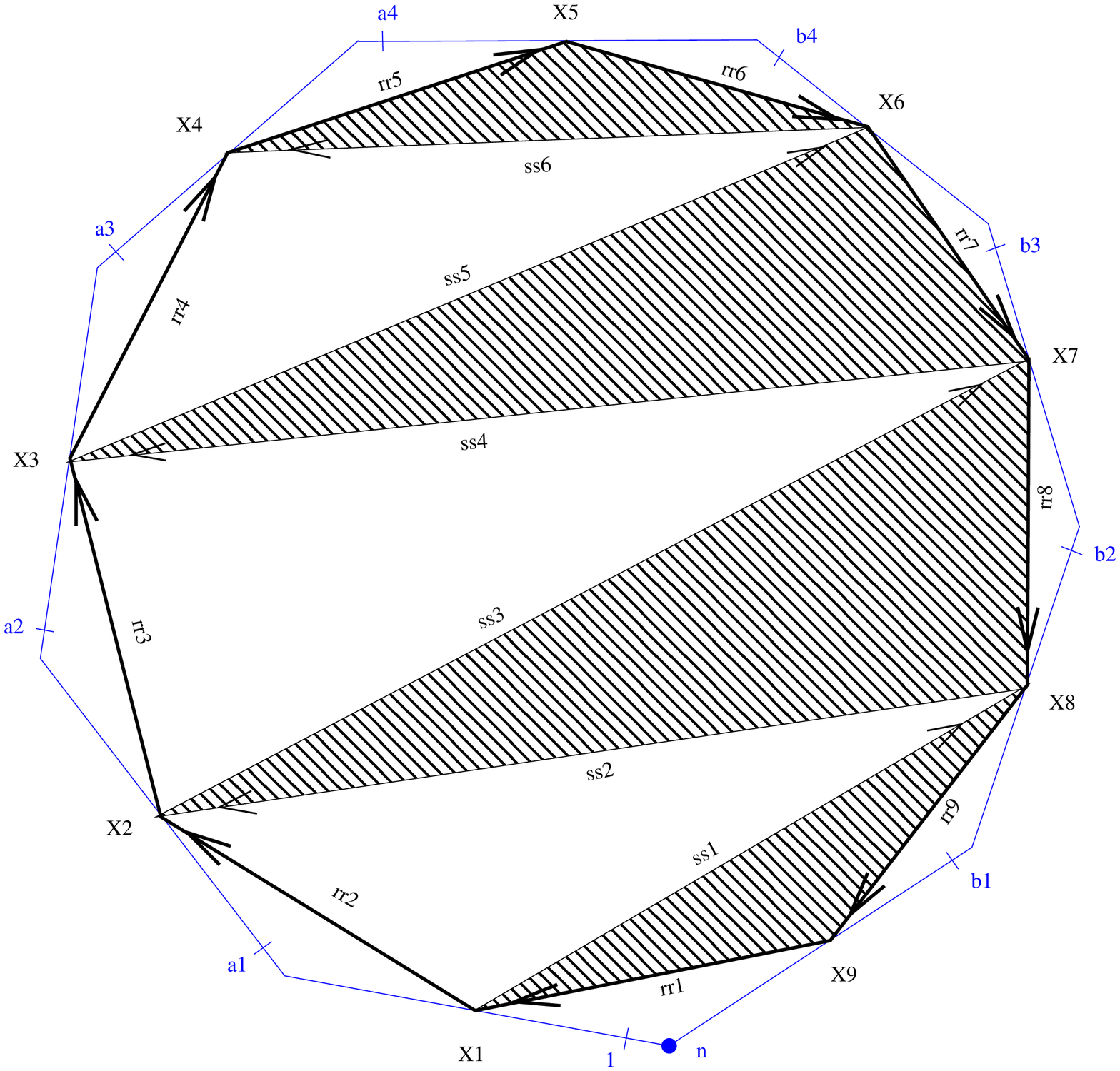}}
\caption{\small N${}^4$MHV: nonagon (A11)}\label{A11}
\end{minipage}
\end{figure}

\begin{figure}[h]
\begin{minipage}[t]{.4\linewidth}
\psfrag{X1}[cc][cc]{$X_1$}
\psfrag{X2}[cc][cc]{$X_2$}
\psfrag{X3}[cc][cc]{$X_3$}
\psfrag{X4}[cc][cc]{$X_4$}
\psfrag{X5}[cc][cc]{$X_5$}
\psfrag{X6}[cc][cc]{$X_6$}
\psfrag{X7}[cc][cc]{$X_7$}
\psfrag{X8}[cc][cc]{$X_8$}
\psfrag{X9}[cc][cc]{$X_9$}

\psfrag{a1}[cc][cc]{$\scriptstyle a_1$}
\psfrag{a2}[cc][cc]{$\scriptstyle a_2$}
\psfrag{a3}[cc][cc]{$\scriptstyle a_3$}
\psfrag{a4}[cc][cc]{$\scriptstyle b_3$}
\psfrag{b4}[cc][cc]{$\scriptstyle a_4$}
\psfrag{b3}[cc][cc]{$\scriptstyle b_4$}
\psfrag{b2}[cc][cc]{$\scriptstyle b_2$}
\psfrag{b1}[cc][cc]{$\scriptstyle b_1$}

\psfrag{1}[cc][cc]{$\scriptstyle 1$}\psfrag{n}[cc][cc]{$\scriptstyle n$}

\psfrag{rr1}[cc][cc]{$\scriptstyle \r_0\rt_1$}
\psfrag{rr2}[cc][cc]{$\scriptstyle \r_1(\rt_1+\rt_2)$}
\psfrag{rr3}[cc][cc]{$\scriptstyle \r_2(\rt_2+\rt_3+\rt_4)$}
\psfrag{rr4}[cc][cc]{$\scriptstyle \r_3 \rt_3$}
\psfrag{rr5}[cc][cc]{$\scriptstyle \sigma_3\rt_3$}
\psfrag{rr6}[cc][cc]{$\scriptstyle \r_4\rt_4$}
\psfrag{rr7}[cc][cc]{$\scriptstyle \sigma_4\rt_4$}
\psfrag{rr8}[cc][cc]{$\scriptstyle \sigma_2\rt_2$}
\psfrag{rr9}[cc][cc]{$\scriptstyle \sigma_1\rt_1$}

\psfrag{ss1}[cc][cc]{$\scriptstyle \r_1\rt_1$}
\psfrag{ss2}[cc][cc]{$\scriptstyle \r_1\rt_2$}
\psfrag{ss3}[cc][cc]{$\scriptstyle \r_2\rt_2$}
\psfrag{ss4}[cc][cc]{$\scriptstyle \r_2\rt_3$}
\psfrag{ss5}[cc][cc]{$\scriptstyle \r_2\rt_4$}

\centerline{\includegraphics[width=70mm]{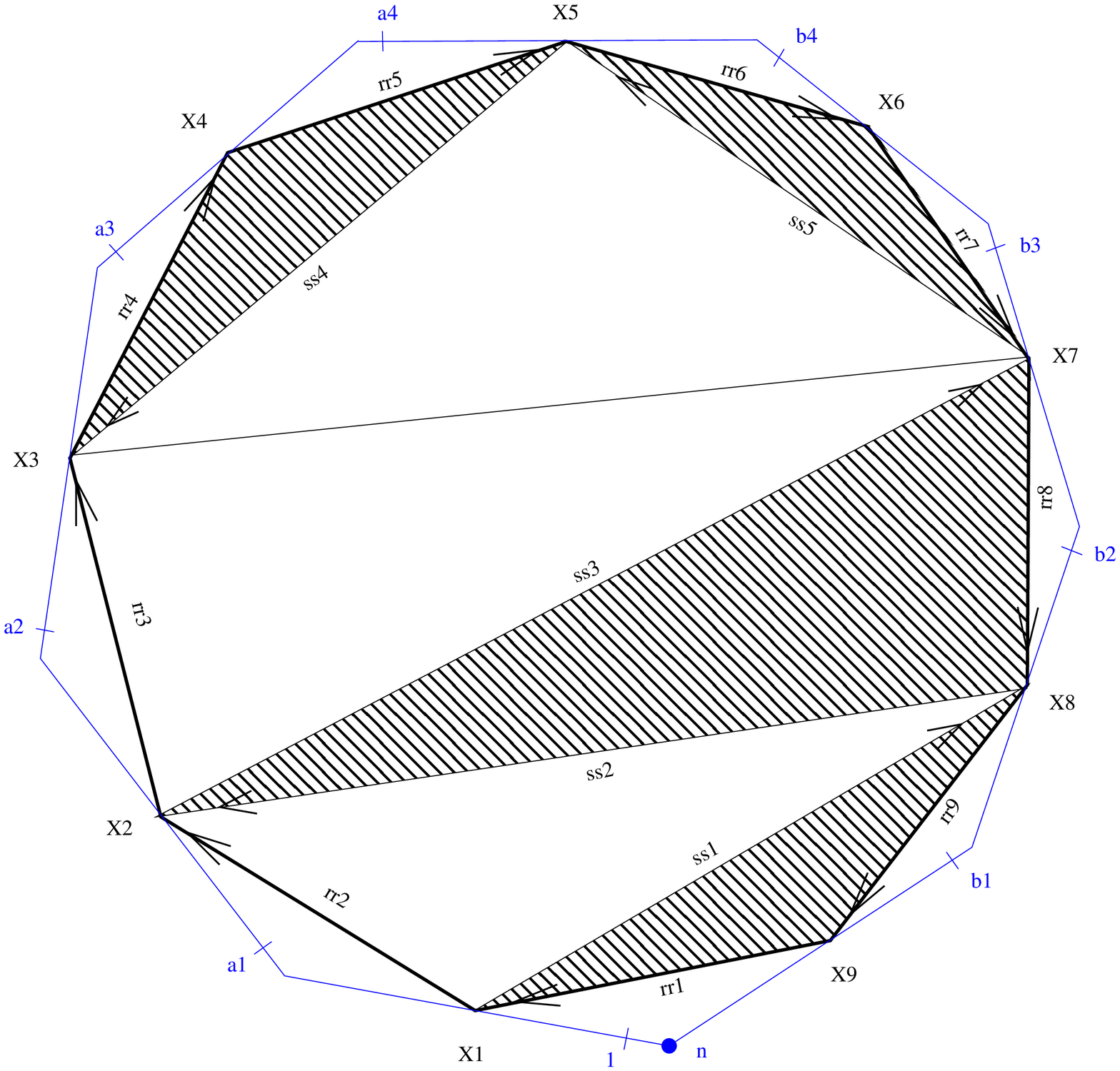}}

\caption{\small N${}^4$MHV: nonagon (A12)}\label{A12}
\end{minipage}
\hfill
\begin{minipage}[t]{.4\linewidth}
\psfrag{X1}[cc][cc]{$X_1$}
\psfrag{X2}[cc][cc]{$X_2$}
\psfrag{X3}[cc][cc]{$X_3$}
\psfrag{X4}[cc][cc]{$X_4$}
\psfrag{X5}[cc][cc]{$X_5$}
\psfrag{X6}[cc][cc]{$X_6$}
\psfrag{X7}[cc][cc]{$X_7$}
\psfrag{X8}[cc][cc]{$X_8$}
\psfrag{X9}[cc][cc]{$X_9$}
\psfrag{a1}[cc][cc]{$\scriptstyle a_1$}
\psfrag{a2}[cc][cc]{$\scriptstyle b_1$}
\psfrag{a3}[cc][cc]{$\scriptstyle a_2$}
\psfrag{a4}[cc][cc]{$\scriptstyle b_2$}
\psfrag{b4}[cc][cc]{$\scriptstyle a_3$}
\psfrag{b3}[cc][cc]{$\scriptstyle b_3$}
\psfrag{b2}[cc][cc]{$\scriptstyle a_4$}
\psfrag{b1}[cc][cc]{$\scriptstyle b_4$}
\psfrag{1}[cc][cc]{$\scriptstyle 1$}\psfrag{n}[cc][cc]{$\scriptstyle n$}
\psfrag{rr1}[cc][cc]{$\scriptstyle \r_0(\rt_1+\rt_2+\rt_3+\rt_4)$}
\psfrag{rr2}[cc][cc]{$\scriptstyle \r_1\rt_1$}
\psfrag{rr3}[cc][cc]{$\scriptstyle \sigma_1 \rt_1$}
\psfrag{rr4}[cc][cc]{$\scriptstyle \r_2 \rt_2$}
\psfrag{rr5}[cc][cc]{$\scriptstyle \sigma_2\rt_2$}
\psfrag{rr6}[cc][cc]{$\scriptstyle \r_3\rt_3$}
\psfrag{rr7}[cc][cc]{$\scriptstyle \sigma_3\rt_3$}
\psfrag{rr8}[cc][cc]{$\scriptstyle \r_4\rt_4$}
\psfrag{rr9}[cc][cc]{$\scriptstyle \sigma_4\rt_4$}
\psfrag{ss1}[cc][cc]{$\scriptstyle \r_0\rt_1$}
\psfrag{ss2}[cc][cc]{$\scriptstyle \r_0\rt_2$}
\psfrag{ss3}[cc][cc]{$\scriptstyle \r_0\rt_3$}
\psfrag{ss4}[cc][cc]{$\scriptstyle \r_0\rt_4$}
\centerline{\includegraphics[width=70mm]{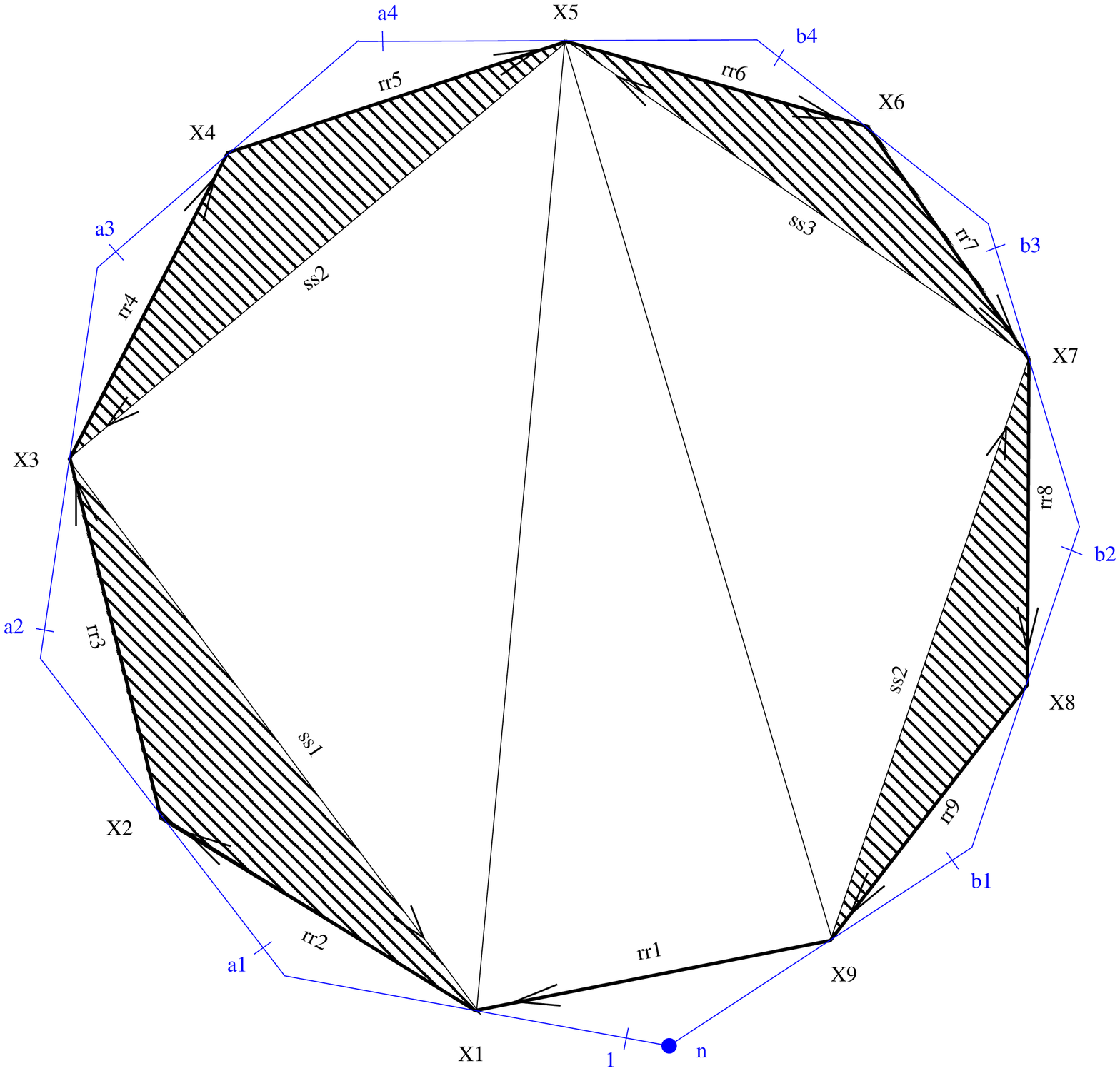}}
\caption{\small N${}^4$MHV: nonagon (B12)}\label{B12}
\end{minipage}
\end{figure}

{Let us illustrate the simple rules formulated in the previous subsection by the
examples of the 5 heptagon diagrams for the N${}^3$MHV case (Figs.~\ref{A12d} -- \ref{B2}) and some of the 14 nonagon diagrams for the N${}^4$MHV case (Figs.~\ref{A11} -- \ref{B12}).}

In the left-hand side part of Fig.~\ref{A12d} we have shown the two-dimensional diagram of the first heptagon configuration (A1). It is obtained from the pentagon (A) shown in Fig.~\ref{5a2d} by inserting the new segment $[a_3, b_3-1]$ into the segment  $[a_2, b_2-1]$. As a result, the number of shaded triangles goes from two to three, and that of the white triangles from one to two. The lines passing through the vertices of the white triangles $X_2 X_3 X_5$ and  $X_1 X_2 X_6$ intersect at the common points ${\cal O}_1$ and ${\cal O}_2$, respectively. In two dimensions, these points appear split up into two points each. The three-dimensional version of the same diagram is shown in the right-hand side part of Fig.~\ref{A12d}, where the surfaces extending into the third dimension are grayed.

The second heptagon configuration (A2), shown in Fig.~\ref{A23d}, is obtained from the same pentagon (A) (see Fig.~\ref{5a2d}), this time inserting $[a_3, b_3-1]$ after  $[a_2, b_2-1]$, but still inside  $[a_1, b_1-1]$. Now the two white triangles have a common side, thus forming a planar rectangle. The four sides of the rectangle, as well as its two diagonals, share the common chiral spinor $\r_1$. Consequently, the twistor lines passing through the vertices $X_1$, $X_2$, $X_4$ and $X_6$ join up at a common intersection point, which serves as the apex of a four-sided pyramid, as shown in the three-dimensional version of the same diagram.

The third heptagon configuration (A3), shown in Fig.~\ref{A3}, is obtained by inserting $[a_3, b_3-1]$ after  $[a_1, b_1-1]$. It resembles (A1) (see Fig.~\ref{A12d}), up to a rotation of the inscribed moduli space heptagon with respect to the twistor space one. The remaining two heptagons (B1) and (B2), shown in Figs.~\ref{B1} and \ref{B2}, respectively, are obtained from the pentagon (B) (see Fig.~\ref{5a2dB}).

{In the N${}^4$MHV case, we show in Figs.~\ref{A11} -- \ref{B12} examples of three different configurations, all obtained from the heptagon (A1) by various insertions.}
Apart from the four shaded triangles, the nonagon (A11) has three distinct white triangles. In the nonagon (A12) two of them merge together into a planar rectangle with the common spinor $\r_2$, and in (A13) the three of them form a planar pentagon with the common spinor $\r_0$.


\clearpage

\section{Discussion and conclusions}

Let us recall some of the main motivations for the recent attempts to reexamine the twistor transform of the tree-level superamplitudes in $\mathcal{N}=4$ SYM theory
undertaken in \cite{Mason:2009sa,ArkaniHamed:2009si}
and in the present paper:
\begin{itemize}
  \item  Make conformal invariance manifest;
  \item  Elucidate the nature of the twistor space support of the amplitude;
  \item  Reveal some hidden geometric structure of the amplitudes in twistor space.
\end{itemize}
To what extent have these goals been achieved?

Let us start with the conformal invariance of the amplitudes. The approaches of
Refs.~\cite{Mason:2009sa,ArkaniHamed:2009si} employ different twistor formalisms, but their common point is the linear realization of the superconformal group $SL(4|4)$ on the space of functions (amplitudes) depending on
(super)twistors  $\hat Z = (\la^\a, \mu_{\da}, \psi_A)$  and dual (super)twistors $\hat W = (\pi_\a, \tau^{\da}, \kappa^A)$, i.e., the fundamental and the anti-fundamental representations of $SL(4|4)$. {The approach of Ref.~\cite{Mason:2009sa} employs only fundamental supertwistors  $\hat Z$. It allows one to express the solution to the BCFW recursion relations in twistor space in terms of  conformally {invariant} three-point delta functions of the type
\begin{equation}\label{msdelta}
    \int \frac{ds}{s} \frac{dt}{t}\ \delta^{(4|4)}(\hat Z_1- s\hat Z_2 -t \hat Z_3)\,,
\end{equation}
in which the points $Z_1$, $Z_2$ and $Z_3$ are aligned on twistor lines.
}

The ambitwistor approach of Ref.~\cite{ArkaniHamed:2009si} uses both types of twistors
and, as a consequence, it offers the possibility to introduce the most natural $SL(4|4)$ {invariants}, the contractions of fundamental and anti-fundamental twistors $Z \cdot W$. The generic twistor transform in this approach is a multi-parameter integral of the type
\begin{equation}\label{aha}
    \int dc_{iJ} M(c_{iJ}) \e^{i \sum_{i,J} c_{iJ} (Z_i  \cdot W_J)}\,.
\end{equation}
Here the essential information about the amplitude is contained in the `link function' $M(c_{iJ})$ which can be determined, in principle, for any given amplitude using graphical rules. Note that in this approach one has to make a choice which particles are described by $Z$'s and which by $W$'s, thus generating a number of (equivalent) descriptions of the same amplitude.

{Although conformal symmetry has indeed been made manifest in both approaches, doing the twistor transform carefully, one discovers that the above simple expressions are plagued with various sign factors which, in fact, break  the conformal invariance
of the amplitude. In \cite{Mason:2009sa} such factors take the form of non-local differential operators, while in \cite{ArkaniHamed:2009si} they are of the type ${\rm sgn}(\vev{1 2})$. In both approaches, the appearance of the conformal symmetry breaking factors is an inherent feature of the twistor transform, which has to do with the presence of physical singularities in the scattering amplitudes. This effect is difficult to spot in  momentum space, but it becomes clearly visible after Fourier transforming the amplitude to twistor space (see \cite{Mason:2009sa} for a discussion of this point).}

Where does the original Witten's twistor transform, which we employ here, stand in this context? Unlike the two approaches above, where the moduli $s$ and $t$ entering \p{msdelta}, or $c_{iJ}$ in \p{aha}, are conformally invariant, the moduli $X$
and $\Theta$ in \re{-01} transform like the coordinates of a point in some kind of position superspace. Consequently, one needs to work out the conformal properties of the moduli space integration measure (see Appendix \ref{jacobi}). Thus, conformal invariance is slightly less manifest than in the other approaches. As to its breakdown due to sign factors, we have confirmed this phenomenon in both the MHV and NMHV cases. While in the former this breakdown only concerns global conformal transformations, in the latter the problem gets worse, affecting even infinitesimal conformal transformations, as we emphasize in Appendix \ref{dramatic}.

Let us now move to the second issue mentioned in the beginning of this section, the
support of the amplitudes in twistor space. In the approach of Ref.~\cite{Mason:2009sa},
the twistor transform is expressed in terms of delta functions \p{msdelta}, which impose linear relations among triplets of points in twistor space. This, as discussed in Sect.~\ref{cctsp}, implies the collinearity of such points. The twistor transform of the NMHV amplitude derived in  \cite{Mason:2009sa} has support on three intersecting lines,
in agreement with the previous findings of Refs.~\cite{Britto:2004tx,Bern:2004ky,Bern:2004bt}.
{In this paper, we have been able to explicitly identify the twistor line structure of all tree amplitudes in $\mathcal{N}=4$ SYM.} Indeed, we have demonstrated that the twistor transform is given by products of delta functions, whose arguments are the twistor line equations themselves. Moreover, the graphical representation we have developed allows us to visualize this structure, with the possibility to directly translate it into analytic expressions, both in twistor and in momentum space.~\footnote{We should mention in this context the `Hodges diagrams' \cite{Hodges:2005bf}, and their relationship with the `link representation' of \cite{ArkaniHamed:2009si}. They determine, in a graphical way, the kernel (moduli dependence) in the integral \p{aha}. It is an interesting question to find out if there is any relation between Hodges' diagrams and our moduli space diagrams.  }

However, the sign factors mentioned above, common to all twistor approaches, blur the nice and simple picture in twistor space. More precisely, the half-Fourier transform of the
explicit expressions for the scattering amplitude, which include such factors, has a rather
complicated form (see Appendix \ref{dramatic}). {Besides breaking the conformal
symmetry, the sign factors result in loosing some of the delta functions needed to align all particles on certain twistor lines. } A similar phenomenon was also observed in \cite{Mason:2009sa}. It is natural to expect that by removing the unwanted sign factors
from the amplitude, we can restore the conformal symmetry and at the same time significantly simplify the twistor transform. We should emphasize that  the resulting expression
in momentum space is {\it not} the true amplitude, but it coincides with the latter
in a restricted kinematic region, where the kinematic variables from the sign factors have fixed signs. Clearly, this restriction yields the loss of some important information about the behavior of the amplitudes near the physical singularities. The most conservative attitude towards this problem might be to consider the `modified' twistor transform as a convenient generating function for the momentum space amplitudes, which have then to be analytically continued to the full kinematic domain. A more optimistic point of view would be that a careful reformulation of quantum field theory in twistor space might resolve the problem, but {this goes beyond the scope of the present work}.

Finally, do the amplitudes in twistor space reveal some hidden geometric structure? We believe that our approach, with the particular moduli space associated to the twistor transform, is the {appropriate} framework for studying this question. As was already pointed out, the {emerging} moduli space looks very much like some fictitious configuration space. The basic objects in it are lightlike polygons, which are triangulated according to a specific pattern. They form surfaces in moduli space, whose topology determines the different contributions to an amplitude of a given kind. More work is needed to better understand the meaning of these triangulated surfaces and their possible hidden symmetries. Intriguingly, the recent proposal by Hodges \cite{Hodges:2009hk} for an alternative `momentum-twistor' formulation of the amplitudes, in the form of polytopes with polygonal faces, bears some resemblance with our picture. Although these polytopes live in a different space, it may be that there is an intimate link between the two formulations.

Another open question is how to make the twistor representation considered here, as well as the one of Ref.~\cite{Mason:2009sa}, compatible with PCT invariance.
Namely,  the MHV and $\widebar{\text{MHV}}$ amplitudes are related to each by PCT conjugation,  and the same is true for the N${}^k$MHV and N${}^{(n-4-k)}$MHV amplitudes. However, their descriptions in {\it chiral} on-shell superspace (and in the associated twistor space) seem radically different: The N${}^k$MHV amplitude has support
on $(2k+1)$ lines whereas for the N${}^{(n-4-k)}$MHV amplitude the number of lines is
$2(n-4-k)+1$. Yet, they must be equivalent, so there should exist a direct link between the two different line configurations in twistor space.

On the more speculative side, we may try to push the analogy between the lightlike polygons in moduli space with the lightlike $n-$gon contours
in the remarkable duality between Wilson loops and $n-$particle MHV amplitudes~\cite{Alday:2007hr,Drummond:2007aua,Brandhuber:2007yx}. The Wilson loop contours, defined in {\it configuration space}, have $n$ cusps, matching each of the scattered particles of the MHV amplitude. At the same time, the number $(2k+1)$ of vertices of the moduli space polygon is related to the type of N${}^k$MHV amplitude. Still, what will happen if we try to develop some kind of dual field theory of Wilson loops in moduli space? Could this lead to a solution of the problem of extending the Wilson loop/MHV amplitude duality to non-MHV amplitudes?
Answering these questions, one might be able to shed some light on another, major open question: How to apply twistor methods to loop amplitudes?
 
\section*{Acknowledgments}

We would like to thank Nima Arkani-Hamed, Iosif Bena, James Drummond, David Kosower, Lionel Mason and David Skinner for interesting discussions. ES would like to acknowledge the hospitality of the Galileo Galilei Institute (Florence), where part of this work was done. This work was supported
in part by the French Agence Nationale de la Recherche under grant
ANR-06-BLAN-0142, by the CNRS/RFFI grant 09-02-00308.

\appendix

\section{Appendix: Transformation of the integration measure under conformal inversion}\label{jacobi}

Here we show that the new moduli space measure $d^2\r d^2\rt d^4\xi$ in \p{sugg} has the right transformation properties under conformal inversion, as announced in Sect.~\ref{scpro}.
Let us start with $\rho$ and denote
\begin{align}
\rho_I \equiv I[\rho] = -\frac{X_3^2}{X_2^2} \bra{\rho} X_1= -\frac{X_3^2}{X_2^2} \bra{\rho} \lr{X_3-\ket{n}[\tilde\rho|}\,.
\end{align}
Then,
\begin{align}
d\rho_I = -\frac{X_3^2}{X_2^2}\bra{d\rho} X_1 +\frac{X_3^2}{X_2^4}\bra{\rho} X_1 d(X_2^2)= -\frac{X_3^2}{X_2^2}\bra{d\rho} \lr{1+ \frac{X_3}{X_2^2}X_{12} } X_1\,,
\end{align}
where we took into account that $X_2^2=(X_3+X_{23})^2 =
X_3^2+\bra{\sigma} X_3|\tilde\rho]$ with $\sigma=-\lambda_n-\rho$.
We conclude that
\begin{align}
d^2 \rho_I = \left| \det \left[ \frac{X_3^2}{X_2^2} \lr{1+ \frac{X_3}{X_2^2}X_{12} } X_1 \right] \right| d^2 \rho
= {\rm sgn}(X_2^2) {(X_3^2)^2 (X_1^2)^2}{(X_2^2)^{-3}} d^2 \rho \,.
\end{align}

The study of the measure $d^2\tilde\rho$ goes along the same lines. The transformation rule
\begin{align}
\tilde\rho_I \equiv I[\tilde\rho] = -(X^2_1 X^2_3)^{-1}\ X_3|\rt]
\end{align}
leads to
\begin{align}
d\tilde\rho_I = -(X^2_1 X^2_3)^{-1}\ X_3|d \rt] + (X^4_1 X^2_3)^{-1}\  (dX_1^2) X_3|\rt]
\end{align}
with $X_1=X_3-X_{31} = X_3 - \lambda_n \tilde\rho$ and
\begin{align}
d X_1^2 = d(X_3+X_{13})^2 = - \bra{n} X_3|d \tilde \rho|\,.
\end{align}
A short calculation gives
\begin{align}
d^2 \tilde\rho_I = {\rm sgn}(X_1^2) (X_1^2)^{-3}  d^2 \tilde\rho\,.
\end{align}

Finally,
\begin{align}
d^4 \xi_I = (X_1^2)^4 d^4 \xi\,.
\end{align}
Putting together all factors, we find
\begin{align}
d^2 \rho_I d^2 \tilde\rho_I d^4 \xi_I =  {\rm sgn}(X_1^2 X_2^2) {(X_1^2)^3 (X_3^2)^2}{(X_2^2)^{-3}}\, d^2 \rho d^2 \tilde\rho d^4 \xi\,.
\end{align}

We are now ready to verify the conformal invariance of \re{sugg}. Let us first rewrite it as follows,
\begin{align}
\int d^2\rho d^2\tilde\rho  d^4\xi\  &  \prod_1^{a-2}\frac{\delta^{(2)}(\mu_i + \lan{i} X_1)}{\vev{i i+1}}
  \prod_{a}^{b-2} \frac{ \delta^{(2)}(\mu_i + \lan{i} X_2)}{\vev{i i+1}}
 \prod_b^{n-1} \frac{ \delta^{(2)}(\mu_i + \lan{i} X_3)}{\vev{i i+1}}
\\
\times & {}
 \frac{\delta^{(2)}(\mu_{a-1}+ \lan{a-1} X_1)\delta^{(2)}(\mu_{b-1} + \lan{b-1} X_2)
 \delta^{(2)}(\mu_{n}+ \lan{n} X_3)}{  \vev{a-1\, \rho} \vev{\rho\, a}   \vev{b-1\, \sigma} \vev{\sigma\, b}  \vev{n1} }\,,
\end{align}
where we have dropped the conformally invariant integral $\int d^4X d^8\Theta$. Each factor inside the products in the first line brings in the weight ${\rm sgn}(X_i^2)$ (with $i=1,2,3$).
In the second line, each delta function gives a factor $|X_i^2|$, while in the denominator we have
\begin{align}
I[\vev{n1}] = [\mu_n \mu_1] =\vev{n|X_3X_1|1} = X_1^2 \vev{n1}\,,
\end{align}
in addition to \p{newcontr}.
Combining all these results, we get the overall factor
\begin{equation}\label{sigfac}
    [\text{sgn}(X_1^2)]^{a-2} [\text{sgn}(X_2^2)]^{b-a-1} [\text{sgn}(X_3^2)]^{n-b+1}\,.
\end{equation}

\section{Appendix: The role of ${\rm sgn}(x^2_{ab})$ and the breaking of conformal invariance}\label{dramatic}

When doing the twistor transform in Sect.~\ref{ttnmhv}, we pointed out that the factor $|x^2_{ab}|$ in the right-hand side of \p{trick} does not exactly cancel the similar factor $x^2_{ab}$ in the denominator in \p{mbubl}. If we wish to find the simple supertwistor transform \re{sugg}, with its clear line structure, we must start with the invariant \p{mbubl} where the factor $x^2_{ab}$ in the denominator is replaced by $|x^2_{ab}|$,
\begin{equation}\label{mbublmod}
    R^+_{nab} = \frac{\vev{a-1\, a} \vev{b-1\, b}\ \delta^{(4)}(\sum_1^{a-1} \lan{n} x_{nb} x^{-1}_{ba}\ran{i}  \eta_i + \sum_1^{b-1} \lan{n} x_{na} x^{-1}_{ab}\ran{i} \eta_i)}{|x^2_{ab}|\lan{n} x_{nb} x^{-1}_{ba}\ran{a-1}\lan{n} x_{nb} x^{-1}_{ba}\ran{a} \lan{n} x_{na} x^{-1}_{ab}\ran{b-1}\lan{n} x_{na} x^{-1}_{ab}\ran{b}}\,.
\end{equation}
Here $R^+$ indicates that this term only coincides with the true amplitude in the kinematic region where $x^2_{ab}>0$.  If instead we insist on the original definition \p{mbubl}, there will be an additional sign factor
\begin{equation}\label{sign}
    {\rm sgn}(x^2_{ab}) = \int^\infty_{-\infty} \frac{dt}{t} e^{it x^2_{ab}} = \int^\infty_{-\infty} \frac{dt}{t} \exp\left\{ it\sum_{a\leq i < j \leq b-1}\vev{ij} [ij] \right\}\ .
\end{equation}

Let us consider the simplest case $n=5$, where the only invariant is $R_{524}$, representing the googly $\overline{\rm MHV}$ amplitude. In this case the half-Fourier transform is reduced to
\begin{equation}\label{halff}
   \int \prod_1^5 d^2\bl_i \int^\infty_{-\infty} \frac{dt}{t} \exp\left\{ i\left(t\vev{23} [23] + \sum_1^5 [i \hat\mu_i] \right)\right\}\ ,
\end{equation}
where we have introduced the notation (recall \p{arr}) $\hat\mu_i = \mu_i + \lan{i}\hat X_i$ and
\begin{eqnarray}
  \hat X_1 &=& X - \la_5 \rt \nn \\
  \hat X_p &=& X +\sigma \rt \,, \quad p=2,3 \nn \\
  \hat X_q &=& X\,, \quad q=4,5\ . \label{mu}
\end{eqnarray}
The integrals over $\bl_{1,4,5}$ give the usual delta functions from the clusters $[1,a-1]$ and $[b,n]$ in \p{2}, while those over $\bl_{2,3}$ require special care:
\begin{eqnarray}
  && \int d^2\bl_2 d^2\bl_3 \int^\infty_{-\infty} \frac{dt}{t} \exp\left\{ i(t\vev{23} [23] + [2 \hat\mu_2]  + [3 \hat\mu_3])\right\} \nn\\
  &=& \int  d^2\bl_3 \int^\infty_{-\infty} \frac{dt}{t} \delta^{(2)}(\hat\mu_2 + t \vev{23} \bl_3) e^{i[3\hat\mu_3]} \nn\\
  &=& \int^\infty_{-\infty} \frac{dt}{t}  \frac{1}{t^2 \vev{23}^2} \exp\left\{ -i\frac{[\hat\mu_2 \hat\mu_3]}{t\vev{23}}\right\} = \delta'([\hat\mu_2 \hat\mu_3])\,.
\end{eqnarray}
Combining this result with the rest of the bosonic delta functions, as well as with all the fermionic ones (they are not affected by the sign factor), we obtain a modified version of \p{2},
\begin{eqnarray}
   (\prod)_{x^2_{ab}} &=&  \delta^{(2)}(\hat\mu_1 )\ \delta^{(4)}(\hat\psi_1)\nn\\
  &\times& \delta'([\hat\mu_2 \hat\mu_3])\ \delta^{(4)}(\hat\psi_2) \delta^{(4)}(\hat\psi_3)\nn\\
  &\times& \delta^{(2)}(\hat\mu_4 )\delta^{(2)}(\hat\mu_5 )\ \delta^{(4)}(\hat\psi_4) \delta^{(4)}(\hat\psi_5)\,, \label{2'''}
\end{eqnarray}
with $\hat\psi_i = \psi_i + \lan{i}\hat \Theta_i$ and
\begin{eqnarray}
  \hat\Theta_1 &=& \Theta - \la_5 \xi \nn \\
  \hat\Theta_p &=&  \Theta +\sigma \xi \,, \quad p=2,3 \nn \\
  \hat\Theta_q &=&   \Theta \,, \quad q=4,5 \ .\label{psi}
\end{eqnarray}

Let us now discuss the superconformal properties of this expression. Under the antichiral conformal supersymmetry with generator $\bar s$ we have $\delta_{\bar s}\psi_{A\, i} = [\mu_i\, \bar\kappa_A]$. Let us first apply this to  the third line of \p{2'''}:
\begin{equation}\label{delpsi}
    \delta_{\bar s}\hat\psi_q = [\mu_q\, \bar\kappa] + \vev{q\ \delta_{\bar s}\Theta}  = [\hat\mu_q\, \bar\kappa] + \vev{q\ (\delta_{\bar s}\Theta - X|\bar\kappa])}\,, \qquad q=4,5\ .
\end{equation}
The first term in this variation vanishes because of the bosonic delta function in the third line of \p{2'''}, while the second term defines the transformation $\delta_{\bar s}\Theta = X|\bar\kappa]$. Just as in Sect. \ref{mmhhvv}, this is the standard superconformal transformation of a point in {position} superspace.

Next, we move to the first line of \p{2'''}:
\begin{equation}\label{delpsi1}
    \delta_{\bar s}\hat\psi_1 = [\mu_1\, \bar\kappa] + \vev{1\ \delta_{\bar s}\Theta} - \vev{15} \delta_{\bar s}\xi  = [\hat\mu_1\, \bar\kappa] + \vev{15} ([\rt \bar\kappa] - \delta_{\bar s}\xi)\,.
\end{equation}
Once again, the first term in the last relation is annihilated by the bosonic delta function, while the second term defines $\delta_{\bar s}\xi = [\rt \bar\kappa]$.

Finally, let us try the second line of \p{2'''}. In the fermionic sector we again obtain $\delta_{\bar s}\hat\psi_p = [\hat\mu_p\, \bar\kappa]$, $p=2,3$. However, now we do not have the matching bosonic delta functions $\delta^{(2)}(\hat\mu_p)$ to annihilate the variations of the fermionic terms. Instead, we have $\delta'([\hat\mu_2 \hat\mu_3])$ which is unable to do the same job! We are led to the conclusion that the twistor transform of the `correct' amplitude with $x^2_{ab}$ is {\it not invariant under infinitesimal superconformal transformations} (and, as a corollary, under conformal transformations).

On the other hand, if we consider the `wrong' $R_{524}$ \p{mbublmod}, its twistor transform is
\begin{eqnarray}
   (\prod)_{|x^2_{ab}|} &=&  \delta^{(2)}(\hat\mu_1 )\ \delta^{(4)}(\hat\psi_1)\nn\\
  &\times& \delta^{(2)}(\hat\mu_2 )\delta^{(2)}(\hat\mu_3 )\ \delta^{(4)}(\hat\psi_2) \delta^{(4)}(\hat\psi_3)\nn\\
  &\times& \delta^{(2)}(\hat\mu_4 )\delta^{(2)}(\hat\mu_5 )\ \delta^{(4)}(\hat\psi_4) \delta^{(4)}(\hat\psi_5)\,, \label{2''''}
\end{eqnarray}
which is clearly invariant under the conformal supersymmetry transformation $\delta_{\bar s}\hat\psi_i = [\hat\mu_i\, \bar\kappa]$, $i=1,\ldots,5$.

In conclusion, we see that only the `wrong' twistor transform \p{2''''} is a superconformal invariant, even with respect to {\it infinitesimal} transformations.


\end{document}